\documentclass[notitlepage,twocolumn,prl,nobalancelastpage,amssymb,superscriptaddress,showpacs,longbibliography,nofootinbib]{revtex4-1}
\usepackage{graphicx}
\usepackage{amsmath}
\usepackage{amssymb}
\usepackage[colorlinks=true,citecolor=blue,linkcolor=blue]{hyperref}
\usepackage{float}
\usepackage{lipsum}
\usepackage{comment}
\usepackage{cancel}
\usepackage[dvipsnames]{xcolor}
\usepackage{physics}
\definecolor{IEcolor}{RGB}{255, 153, 51}

\definecolor{CYcolor}{RGB}{50, 100, 50}

\begin{document}

\title{Quantized Acoustoelectric Floquet Effect in Quantum Nanowires}

\author{Christopher Yang}
\address{Department of Physics, IQIM, California Institute of Technology, Pasadena, CA 91125, USA}

\author{Will Hunt}
\address{Department of Physics, University of Cambridge, Cambridge, United Kingdom}
\address{Department of Physics, IQIM, California Institute of Technology, Pasadena, CA 91125, USA}

\author{Gil Refael}
\address{Department of Physics, IQIM, California Institute of Technology, Pasadena, CA 91125, USA}

\author{Iliya Esin}
\address{Department of Physics, IQIM, California Institute of Technology, Pasadena, CA 91125, USA}

\begin{abstract}
External coherent fields can drive quantum materials into non-equilibrium states, revealing exotic properties that are unattainable under equilibrium conditions---an approach known as ``Floquet engineering.'' While optical lasers have commonly been used as the driving fields, recent advancements have introduced nontraditional sources, such as coherent phonon drives. Building on this progress, we demonstrate that driving a metallic quantum nanowire with a coherent wave of terahertz phonons can induce an electronic steady state characterized by a persistent quantized current along the wire. The quantization of the current is achieved due to the coupling of electrons to the nanowire's vibrational modes, providing the low-temperature heat bath and energy relaxation mechanisms. Our findings underscore the potential of using non-optical drives, such as coherent phonon sources, to induce non-equilibrium phenomena in materials. Furthermore, our approach suggests a new method for the high-precision detection of coherent phonon oscillations via transport measurements.
\end{abstract}

\maketitle

\maketitle

\textit{Introduction}---Recent advances in the creation of phonon sources have enabled on-demand access to coherent phonon beams across a broad spectrum of frequencies \cite{D2NR04100F,PhysRevA.54.943,phaser-1,PhysRevLett.121.043201,phaser-2,PhysRevLett.104.083901,PhysRevLett.113.053604,hyp-ph-po}. These phonon excitations in solids \cite{phononic-nonlinear, RevModPhys.84.1045, BALANDIN2012266, sound-heat-revolution, phononics-graphene, PhysRevB.89.220301, solids-nonlinear-phononics, PhysRevLett.118.054101} can induce new optical properties \cite{Trovatello,PhysRevX.11.021067}, strong correlation physics \cite{superconductivity-1,PhysRevB.94.214504,PhysRevB.96.014512,superconductivity-3,Mitrano2016,Li2019}, tunable magnetic properties \cite{PhysRevLett.118.197601, magnetic-1, D2NR04100F,phononics-graphene}, and acousto-electric effects \cite{PhysRev.89.990,PhysRev.106.1104,PhysRevLett.122.256801,PhysRevLett.85.2829,PhysRevLett.128.215902,Kawada2021}. The unique characteristics of coherent phonons---specifically their finite momentum, low energy, and coupling to the electrons---pave the way for Floquet engineering of non-equilibrium spatial-temporal electronic phenomena \cite{PhysRevLett.128.186802, PhysRevB.106.224311, phonon-floquet,PhysRevResearch.2.043431}. Notably, terahertz (THz) frequency phonons \cite{Esin2022GeneratingPhaser,PhysRevLett.104.085501,phaser-thz-1,aguilar2023} can potentially serve as tools for the dynamic manipulation of materials with narrow bandwidths, such as moiré systems \cite{Katz2020OpticallyGraphene,PhysRevLett.131.026901,Esin2022GeneratingPhaser}. Recent proposals have demonstrated that coherent phonon beams can be used for Floquet engineering of nontrivial band topology in trivial materials \cite{PhysRevResearch.2.043431,PhysRevLett.128.186802}.

\begin{figure} [h!]
    \centering
    \includegraphics[width=\linewidth]{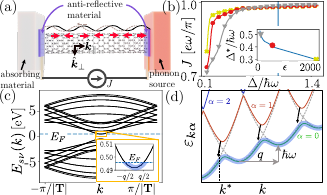}
    \caption{(a) Schematic experimental setup. A THz-frequency coherent phonon wave (with atomic displacements sketched by red arrows) of momentum $q$ and angular frequency $\omega$ induces a quantized current $J = e \omega/\pi$ in a CNT. A screening material of dielectric constant $\epsilon$ surrounds the CNT, and anti-reflective material minimize reflection of the phonon wave. (b) Current $J$ vs. Floquet gap $\Delta/\hbar\omega$ for different $\epsilon$ (see inset). Vertical line: gap below which the drive is non-adiabatic ($\Delta < \hbar\omega$). Inset: $\Delta^*$ vs. $\epsilon$, where $\Delta^*$ is the minimal Floquet gap at which $J = 0.96 e\omega/\pi$. (c) Band structure of a $(10, 0)$ armchair CNT. Inset: Fermi energy lies near the band bottom of the lowest positive-energy band, and the electronic density $n_e$ is chosen to be commensurate to $q$. (d) Quasienergy spectrum of the driven system. Blue shading on the $\alpha = 0$ band: optimal filling of the bands resulting in quantized current. Scattering transitions (black arrows) generated by incoherent phonons relax electrons into the $\alpha= 0$ band. 
    }\label{fig:intro}
\end{figure}

We demonstrate that a continuous propagating wave of coherent THz phonons can drive a quantum wire into a non-equilibrium Floquet steady state, resulting in a persistent charge current [see Fig. \ref{fig:intro}(a)]. Furthermore, under optimal conditions of doping and driving strength, this current can achieve a quantized value, $J=e \omega /\pi$, where $\omega$ represents the phonon driving angular frequency and $e$ is the electron charge [see Fig. \ref{fig:intro}(b)]. In the adiabatic limit, $\omega \to 0$, the quantization of current aligns with the principles of a topological pump \cite{thouless-pump,thouless-topo-ins,thouless-ultracold,PhysRevLett.111.060802,Citro2023}. Remarkably, we find that quantized transport can be maintained over a wide range of THz frequencies, even beyond the adiabatic limit. This suggests a robust mechanism underpinning the quantization, stabilized by the non-equilibrium electronic steady state, which is set by the coupling to a bath of low-temperature thermal phonons and electron-electron interactions \cite{PhysRevB.97.195411,PhysRevResearch.6.013094,PhysRevB.98.115147,Walter2023,PhysRevLett.120.106601}.

A device for robust generation of quantized current on demand through coherent phonon illumination has numerous potential applications in metrology \cite{PhysRevLett.64.1812,RevModPhys.85.1421,Kaneko2016,Stein2015,PhysRevApplied.8.044021,Giblin2019}, electronics, and quantum computing \cite{PhysRevLett.97.096602,PhysRevLett.128.244302,Buerle2018,Feve2007}. Furthermore, this device can be used as a tool for the characterization and detection of coherent phonons through transport measurement. Traditionally, the detection of coherent phonons has relied on optical methods such as reflectivity measurements \cite{PhysRevB.84.064307,10.1063/1.90855,pump-probe-1,https://doi.org/10.48550/arxiv.2310.04939,Yoon2023}. Leveraging current quantization as a probe of the phonon field offers a more direct and sensitive method of coherent phonon detection.

In this work, we focus on a concrete experimental setup of carbon nanotube (CNT) \cite{PhysRevB.79.205434, cnt-review, PhysRevLett.78.1932,PhysRevB.46.1804,PhysRevLett.68.631,PhysRevLett.68.1579} coupled to a continuous source of coherent phonons. The phonon waves propagate from the right to the left end of the CNT, where they are absorbed by the absorbing material, see Fig. \ref{fig:intro}(a). Anti-reflective material with appropriately chosen thickness on both ends of the CNT [see blue slabs in Fig. \ref{fig:intro}(a)] suppress reflected phonon waves that destructively interfere with the coherent phonon mode. The electronic steady state in the CNT is formed from the balance between the interaction with the coherent phonon wave, as well as interactions with incoherent phonons of the CNT and other electrons. In our model, we consider a detailed microscopic description of the phononic spectrum of the CNT that serves as the low-temperature heat bath for the phonons and the electron-electron interactions.

\textit{Phonon-driven carbon nanotube.}---To analyze the steady state properties of this setup, we use an effective model of the CNT. We define $\hat{\boldsymbol{\psi}}_{k,s}^{\dagger} \equiv \begin{pmatrix} \hat{\psi}_{{k}, A,s}^{\dagger} & \hat{\psi}_{{k}, B,s}^{\dagger} \end{pmatrix}$ as the creation operator of a Bloch state on the CNT, where $\hat{\psi}_{{k}, j,s}^\dagger$ creates an electron of crystal momentum ${k}$ on sublattice $j = A,B$, and ${k}$ is the Bloch momentum in the direction $\boldsymbol{\hat{k}_{\parallel}}$ along the tube axis. The index $s = 0, \hdots, N-1$ enumerates the electronic momentum $s \boldsymbol{k}_{\perp}$ along the direction $\hat{\boldsymbol{k}}_{\perp}$ around the circumference of the tube. Here, $\boldsymbol{k}_{\perp} = (2\pi / P) \hat{\boldsymbol{k}}_{\perp}$, $P$ is the perimeter of the CNT, $N$ is the number of graphene unit cells contained in a length $|\boldsymbol{T}|$ along the tube axis, and the CNT is periodic along the tube axis by translations of vector $\boldsymbol{T}$. The electronic momentum $k \in [-\pi / |\boldsymbol{T}|, \pi / |\boldsymbol{T}|]$ along $\hat{\boldsymbol{k}}_{\parallel}$ is approximated as continuous for a long tube. Corresponding eigenenergies of the electronic states are given by $E_{s\nu}(k)$, where $\nu = +, -$ denote the conduction and valence bands, respectively, see Fig. \ref{fig:intro}(c). 

We focus on a semiconducting CNT whose Fermi surface in equilibrium lies near the bottom of the lowest conduction band [see inset of Fig. \ref{fig:intro}(c)]. The lowest conduction and highest valence bands are described by the effective Hamiltonian $\hat{H}_{\text{e}} = \int {dk}/{2\pi} \ \hat{\boldsymbol{\psi}}_{{k}}^{\dagger} H_{\text{e}}(k) \hat{\boldsymbol{\psi}}_{{k}}$, where $H_{\text{e}}(k) = \hbar v_F (\delta k \hat{\boldsymbol{k}}_{\perp} + k \hat{\boldsymbol{k}}_{\parallel}) \cdot \boldsymbol{\sigma}$, $\delta k$ is a constant dependent on the chirality indices (see the Supplemental Materials \cite{SupplementalMaterials} for details), $v_F$ is the Fermi velocity of graphene, $\boldsymbol{\sigma} = (\sigma_x, \sigma_y)$ is a vector of Pauli matrices acting in the graphene sublattice basis, and the $x$ axis is aligned along a bond between
carbon atoms. In what follows, we omit the index $s$, assuming only the two bands described by $\nu=+,-$ with eigenergies $E_{\nu}(k)$ and eigenstates $|\nu k\rangle$.

To describe electron-phonon interactions in the CNT, we consider the Hamiltonian $\hat{H}_{\text{ep}}({t}) = \sum_{\lambda} \hat{H}_{\text{ep}}^{(\lambda)}({t})$, where $\hat{H}_{\text{ep}}^{(\lambda)}({t}) = \int d^2\boldsymbol{r} \  \hat{\boldsymbol{\psi}}_{\boldsymbol{r}}^{\dagger} \hat{V}^{(\lambda)}(\boldsymbol{r},t) \hat{\boldsymbol{\psi}}_{\boldsymbol{r}}$ and 
$\hat{V}^{(\lambda)}(\boldsymbol{r},t) = \hbar v_F \hat{\boldsymbol{\mathcal{A}}}_{\text{ph}}^{(\lambda)}(\boldsymbol{r},t) \cdot \boldsymbol{\sigma} + \hat{\phi}_{\text{ph}}^{(\lambda)}(\boldsymbol{r},t)$ describe electronic coupling to a phonon mode indicated by $\lambda$, $\boldsymbol{r} = (x, y)$ is the spatial coordinate along the tube, and $\hat{\boldsymbol{{\psi}}}_{\boldsymbol{r}}^{\dagger} = P^{-1} \int d{k}/(2\pi) e^{-i(\delta k \hat{\boldsymbol{k}}_{\perp}+k\hat{\boldsymbol{k}}_{\parallel} ) \cdot \boldsymbol{r}} \hat{{\boldsymbol{\psi}}}^{\dagger}_{k}$. The electrons interact with the phonons through the effective vector potential $\hat{\boldsymbol{\mathcal{A}}}_{\text{ph}}^{(\lambda)}(\boldsymbol{r},t) = \sqrt{3} \beta / (2a) (\hat{u}^{(\lambda)}_{xx}(\boldsymbol{r},t) - \hat{u}^{(\lambda)}_{yy}(\boldsymbol{r},t), 2\hat{u}^{(\lambda)}_{xy}(\boldsymbol{r},t))$
and through the local scalar potential $\hat{\phi}_{\text{ph}}^{(\lambda)}(\boldsymbol{r},t) = D[\hat{u}^{(\lambda)}_{xx}(\boldsymbol{r},t) + \hat{u}^{(\lambda)}_{yy}(\boldsymbol{r},t)] I$, where $a = 0.246 \ \mathrm{nm}$, $\beta \approx 3.14$, and deformation potential $D = 15 \ \mathrm{eV}$ \cite{Esin2022GeneratingPhaser, PhysRevB.80.075420}. Here, $\hat{\boldsymbol{u}}^{(\lambda)}(\boldsymbol{r},t)$ is the displacement operator of the phonon mode, $\hat{u}^{(\lambda)}_b(\boldsymbol{r},t)$ is its $b$-th component, and $\hat{u}^{(\lambda)}_{bc}(\boldsymbol{r},t) = [\partial_b \hat{u}^{(\lambda)}_c(\boldsymbol{r},t) + \partial_c \hat{u}^{(\lambda)}_b(\boldsymbol{r},t)] / 2$.  

We assume that a source of phonons [see Fig. \ref{fig:intro}(a)] generates a coherent sound mode with index $\lambda = \lambda_0$ that propagates through the CNT, while the rest of the phonons are in low-temperature thermal equilibrium. The phonon mode $\lambda_0$ has momentum $\boldsymbol{q} = q\hat{\boldsymbol{k}}_{\parallel}$, angular frequency $\omega$, and finite displacement expectation value $\langle\hat{\boldsymbol{u}}^{(\lambda_0)}(\boldsymbol{r},t) \rangle  = u_0 \cos(\boldsymbol{q}\cdot \boldsymbol{r} - \omega t) \hat{\boldsymbol{k}}_{\parallel}$, where $u_0$ is the displacement amplitude \cite{Longitudinal}.

The electronic dynamics can be divided to coherent components described by the time- and spatially-periodic Hamiltonian $\hat{H}_0(t) = \hat{H}_{\text{e}} + \hat{H}_{\text{ep}}^{(\lambda_0)}(t)$ and incoherent components due to coupling to thermal phonon modes, $\hat{H}_{\text{b}}(t) = \sum_{\lambda \neq \lambda_0} \hat{H}_{\text{ep}}^{(\lambda)} (t)$ \cite{Commensurate}. The single-particle Hamiltonian $\hat{H}_0(t)$ can be diagonalized by the Floquet-Bloch states, $|\psi_{k\alpha} (\boldsymbol{r}, t)\rangle = e^{-i(\boldsymbol{k} \cdot \boldsymbol{r}+\varepsilon_{k\alpha} t/\hbar)} \sum_{n \in \mathbb{Z}} e^{-in(\boldsymbol{q}\cdot \boldsymbol{r} - \omega t)} |\phi^{(n)}_{k\alpha}\rangle$ \cite{floquethandbook,PhysRevLett.128.186802, PhysRevB.106.224311, phonon-floquet}. Here, $\varepsilon_{k\alpha}$ is the quasienergy satisfying
\begin{equation}\label{eq:floquet-se}
    (\varepsilon_{k\alpha} + m\hbar\omega) |\phi^{(m)}_{k\alpha}\rangle = {H}_{\text{e}}({{k}+m{q}}) + \sum_{m'\neq 0}  V_{m-m'} |\phi^{(m')}_{k\alpha}\rangle
\end{equation}
where $\alpha$ enumerates the Floquet bands and $V_{n}$ are the Fourier harmonics of $\langle \hat{V}^{(\lambda_0)}(\boldsymbol{r},t) \rangle$, i.e., $\langle \hat{V}^{(\lambda_0)}(\boldsymbol{r},t) \rangle = \sum_{n\neq 0} e^{-in(\boldsymbol{q}\cdot \boldsymbol{r} - \omega t)} V_{n}$. The quasienergy spectrum $\varepsilon_{k\alpha}$ arises from replicas of the original energy bands $E_{\nu}(k)$ shifted in energy and momentum by $m \hbar\omega$ and $mq$, respectively, where $m \in \mathbb{Z}$ [see light grey, dashed curves in Fig. \ref{fig:intro}(d)]. Rabi-like gaps of size $\Delta \approx qu_0 [D  + \hat{k}_{\parallel}^y \hbar v_F \sqrt{3} \beta / 2a]$ open at momenta $k = k^* + mq$ where the Floquet replicas cross, resulting in the quasienergy spectrum sketched with solid curves in Fig. \ref{fig:intro}(d) (see the Supplemental Materials \cite{SupplementalMaterials} for details). Remarkably, the quasienergy satisfies the periodicity condition $\varepsilon_{k\alpha} = \varepsilon_{k+q,\alpha} -\hbar\omega$, which is the basis for the quantized current presented in this work. Specifically, the current is given by \cite{transport}
\begin{equation} \label{eq:current-floquet}
    J = \frac{2e}{\hbar}\sum_{\alpha} \int_0^q \frac{dk}{2\pi} \ \frac{d\varepsilon_{k\alpha}}{dk} F_{k\alpha}.
\end{equation}
Here, $F_{k\alpha}(t) = \langle \hat{f}^{\dagger}_{k\alpha} (t) \hat{f}_{k\alpha}(t) \rangle$ is the occupation of the Floquet state $|\psi_{k\alpha}(\boldsymbol{r}, t) \rangle$ created by operator $\hat{f}_{k\alpha}^{\dagger}(t)$ \cite{SpinDegen}. When only the $\alpha = 0$ band is fully-occupied, $J = e\omega/\pi$, resulting in quantized current.

\textit{Floquet-Boltzmann Equation.}---To model the dynamics under electron-phonon interactions, we consider the microscopic Hamiltonian $\hat{H}_{\text{b}}(t)$ for electronic coupling to incoherent bath phonons. The incoherent low-energy longitudinal acoustic phonons of speed $c_{\text{ph}}$, momentum $p$, and energy $\hbar \omega_l(p) = \hbar c_{\text{ph}}|s \boldsymbol{k}_{\perp} + p \boldsymbol{\hat{k}}_{\parallel}|$ dominate the electron-phonon scattering near the Fermi surface of the CNT. Additionally, we consider electron-electron coupling:
\begin{equation} \label{eq:h-el-el}
\begin{split}
    \hat{H}_{\text{ee}} = &\int \frac{dk_1dk_2dp}{(2\pi)^3}\mathcal{V}_{{k}_1,{k}_2}(p) \hat{c}_{{k}_1+{p},+}^{\dagger} \hat{c}_{{k}_2 -{p},+}^{\dagger} \hat{c}_{{k}_2,+}\hat{c}_{{k}_1,+},
\end{split}
\end{equation}
where $\mathcal{V}_{{k}_1,{k}_2}(p) = U(p)  \mathcal{W}_{{k}_1,{p}} \mathcal{W}_{{k}_2,-{p}} / (2\epsilon)$, $\mathcal{W}_{k,p} \equiv \langle  +,k+p | +, k\rangle$ is the form-factor, $U(p)$ is the Coulomb potential, $\epsilon$ is the dielectric constant of the surrounding screening medium, and $\hat{c}_{k\nu}$ creates an electron in the eigenstate $|\nu k \rangle$ of $\hat{H}_{\text{e}}$. We consider scattering within the $\nu = +$ band only, because the Fermi level [see inset of Fig. \ref{fig:intro}(c)] is energetically well-separated from other bands, restricting scattering to small momentum transfers $p \lesssim q$ near the Fermi surface. For small $p$, we approximate $U(p) \approx 1 \ \mathrm{eV}$, consistent with estimates in Refs. \cite{PhysRevLett.93.157402, coulomb-cnt}.

We assume that the CNT incoherent phonons are coupled to an external heat bath and remain in thermal equilibrium at temperature $T$. Under these conditions, the electrons form a steady state distribution with occupations $F_{k\alpha}(t)$ determined by solving the Floquet-Boltzmann equation (FBE) \cite{fbe_adv, fbe_orig,PhysRevA.92.062108,PhysRevE.79.051129}, $\dot F_{{k}\alpha}(t) = I^{\text{b}}_{{k}\alpha}[\{F_{{k}\alpha}(t)\}] + I^{\text{ee}}_{{k}\alpha}[\{F_{{k}\alpha}(t)\}]$, for $\dot F_{{k}\alpha}(t) = 0$. The FBE is valid when $\hat{H}_{\text{ee}}$ and $\hat{H}_{\text{b}}$ weakly scatter electrons between single-particle Floquet eigenstates. Here, we use the Fermi golden rule modified for transitions between Floquet states \cite{fbe_adv, fbe_orig} to calculate the electron-phonon and electron-electron collision integrals $I^{\text{b}}_{{k}\alpha}[\{F_{{k}\alpha}(t)\}]$ and $I^{\text{ee}}_{{k}\alpha}[\{F_{{k}\alpha}(t)\}]$, respectively. (For details, see the Supplemental Materials \cite{SupplementalMaterials}.) The coherences $\langle \hat{f}^{\dagger}_{k\alpha}(t) \hat{f}_{k\alpha'}(t) \rangle$ for $\alpha \neq \alpha'$ are suppressed when $1/ \tau^{\text{ph}}_{k\alpha \alpha'} + 1/ \tau^{\text{el}}_{k\alpha \alpha'} \ll |\varepsilon_{k\alpha} - \varepsilon_{k\alpha'}|/ \hbar$, where $1/\tau^{\text{ph}}_{k\alpha \alpha'}$ and $1/ \tau^{\text{el}}_{k\alpha \alpha'}$ are respectively the electron-phonon and electron-electron interband scattering rates between bands $\alpha$ and $\alpha' \neq \alpha$ \cite{Kohn2001PeriodicThermodynamics, fbe_orig, fbe_adv}. Interband scattering transitions also broaden the electronic spectral function by an energy scale of roughly $\delta \varepsilon \approx \hbar / \tau^{\text{tot}}_{k\alpha\alpha'}$, relaxing energy conservation in the FBE. 

\begin{figure}
    \centering    \includegraphics[width=0.98\linewidth]{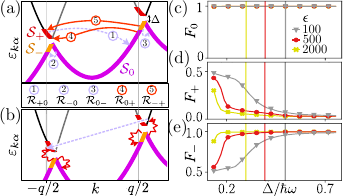}
    \caption{(a) Floquet bands upon driving by a coherent phonon wave. Dashed light purple arrows (1-3): dominant electron-phonon intraband and cooling processes relaxing electrons into the $\alpha = 0$ band. Solid red arrows (4-5): electron-phonon heating processes exciting electrons into the $\alpha = 1$ band, which are kinematically suppressed when $\Delta \gg \delta \varepsilon$. (b) Pairs of zigzag and squiggly red arrows: electron-electron scattering exciting electrons into the $\alpha = 1$ band. Dashed light purple arrow: electron-phonon process relaxing excited electrons into $\mathcal{S}_+$. (c)-(e) Occupation $F_i$ of the patches $\mathcal{S}_j$ for $i = +, -, 0$ [see panel (a)] vs $\Delta$ for various dielectric constants $\epsilon$. Vertical lines: $\Delta = \Delta^*$. Note $F_+ \to 0, F_- \to 1$ as $\Delta \to \Delta^*$, resulting in quantized current.}
    \label{fig:phenom}
\end{figure}

\textit{Phenomenological rate equation.}---The current in Eq. (\ref{eq:current-floquet}) attains a quantized value when $F_{k\alpha}=F^{\text{opt}}_{k\alpha}$, where $F^{\text{opt}}_{k0}=1$, and $F^{\text{opt}}_{k,\alpha\neq 0}=0$. Our goal in this section is to find the conditions on the scattering rates that lead to $F_{k\alpha}\approx F^{\text{opt}}_{k\alpha}$ in the steady state. To this end, we identify the key scattering processes that contribute to the steady state, indicated by arrows in Fig. \ref{fig:phenom}(a)-(b). These processes connect three patches of Floquet states denoted by $\mathcal{S}_{+}$, $\mathcal{S}_{-}$, and $\mathcal{S}_{0}$, with approximately uniform electronic occupation $F_{+}$, $F_{-}$, and $F_0$ respectively. The patch $\mathcal{S}_0$ includes Floquet states with momentum $k^* \leq k < q/2$ in the $\alpha =0$ band, while $\mathcal{S}_{+}$ and  $\mathcal{S}_{-}$ encloses those with momentum $-q/2 \leq k < k^*$ in the $\alpha = 1$ and $\alpha = 0$ bands, respectively [see Fig. \ref{fig:phenom}(a)].

To estimate the electronic occupations of the patches, let us first consider the limit $\epsilon \to \infty$ and $I^{\text{ee}}_{k\alpha} \to 0$ in which scattering is mediated by acoustic phonons. Averaging the FBE over the patches, we obtain rate equations for their occupations, $\dot{F}_{i} = \sum_{j} [\mathcal{R}_{ji} F_j (1-F_i) - \mathcal{R}_{ij} F_i (1-F_j)]$, where $i, j = +, -, 0$ and $\mathcal{R}_{ij}$ denotes the average scattering rate of an electron from patch $\mathcal{S}_i$ to $\mathcal{S}_j$. We begin by assuming that the system is optimally doped, i.e., $n_e=q$, where $n_e$ is the density of the electrons. The optimal steady state distribution $F^{\text{opt}}_{k\alpha}$, corresponding to $F_{-}=F_0=1$, and $F_+=0$, is obtained when the ``Floquet-cooling'' processes $\mathcal{R}_{+0}$, $\mathcal{R}_{0-}$ and $\mathcal{R}_{-0}$ [dashed, light purple arrows numbered 1-3 in Fig. \ref{fig:phenom}(a)] dominate the scattering rates. The rest of the scattering rates $\mathcal{R}_{ij}$ create excitations in $\mathcal{S}_+$, giving rise to deviations from $F^{\text{opt}}_{k\alpha}$ and therefore are dubbed ``Floquet-heating'' rates [solid red arrows numbered 4-5 in Fig. \ref{fig:phenom}(a)].

When the incoherent phonons remain at temperature $T = 0$, all ``Floquet-heating'' processes mediated by acoustic phonons require a small energy and large momentum transfer. This kinematically constrains the rates $\mathcal{R}_{+-}$ and $\mathcal{R}_{0+}$ at high drive intensities, i.e., when $\Delta > \delta\varepsilon$, disabling all ``Floquet-heating'' processes. Simultaneously, the processes described by the rates  $\mathcal{R}_{0-}$ and $\mathcal{R}_{-0}$ are kinematically allowed. These processes are of the Floquet-Umklapp (FU) type and therefore increase with the drive intensity as $(\Delta/\hbar \omega)^2$ yielding $F_{k\alpha} \to F^{\text{opt}}_{k\alpha}$ \cite{floquethandbook,gyro,fbe_orig}. This is a remarkable result that shows that the coupling to acoustic phonons can stabilize a quantized current in the non-adiabatic regime $\delta \varepsilon < \Delta < \hbar\omega$. In contrast, in the low intensity limit ($\Delta\to 0$) of the drive, the rates $\mathcal{R}_{0-}$ and $\mathcal{R}_{-0}$ vanish, whereas the rates $\mathcal{R}_{-+}$ and $\mathcal{R}_{0+}$ become kinematically enabled. This situation leads to $F_+=F_0=1$ and $F_-=0$, recovering the equilibrium Fermi-Dirac distribution. 

A finite incoherent phonon temperature $k_B T < \hbar\omega$ and electron-electron interactions (finite $\epsilon$) cause deviations of the steady state from $F^{\text{opt}}_{k\alpha}$ in the regime $\Delta > \delta\varepsilon$. Absorption of incoherent phonons yields a finite but weak electron-phonon heating rate $\mathcal{R}_{0+}$, generating small electron and hole densities in $\mathcal{S}_+$ and $\mathcal{S}_-$ respectively. Heating processes due to electron-electron interactions excite electrons into the patch $\mathcal{S}_+$ in the $\alpha = 1$ band [pairs of squiggly red arrows in Fig. \ref{fig:phenom}(b)]. Other electron-electron scattering processes [pairs of zigzag arrows in Fig. \ref{fig:phenom}(b)] excite electrons to states elsewhere in the $\alpha = 1$ band, which are relaxed to $\mathcal{S}_+$ by electron-phonon cooling [dashed, light purple arrow in Fig. \ref{fig:phenom}(b)]. These processes result in a net increase in $F_+$ and reduced current response. The electron-electron heating can be suppressed by increasing the drive intensity, since the phase space for such scattering processes is constrained to small energy and momentum transfers near the Floquet gap [see Fig. \ref{fig:phenom}(b)] and is reduced as $\Delta$ is increased. In Fig. \ref{fig:phenom}(c)-(e), we show the average occupations of the patches as a function of $\Delta$ for various dielectric constants $\epsilon$. We define $\Delta^*$ as the minimal Floquet gap at which $J = 0.96 e\omega / \pi$. The equilibrium Fermi-Dirac distribution ($F_+ = F_0 = 1-F_- = 1$) transitions to $F^{\text{opt}}_{k\alpha}$ as $\Delta \to \Delta^*$.  

\begin{figure}
    \centering    \includegraphics[width=1\linewidth]{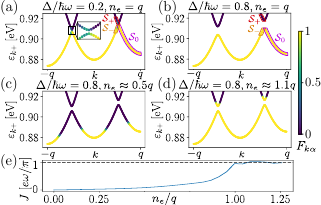}
    \caption{(a) Steady state occupation of the phonon-driven CNT for a weak phonon drive ($\Delta < \Delta^*$), dielectric constant $\epsilon = 80$, and optimal doping $n_e = q$. Inset shows excitations in the $\alpha = 1$ band near the Floquet gap. (b) Same as (a) but for a strong drive amplitude ($\Delta > \Delta^*$) where the occupation of the $\alpha = 1$ band is negligible, and the $\alpha = 0$ band is fully occupied. (c)-(d) Same as (b), but with two different electronic densities $n_e$ away from optimal doping. (e) Steady state current vs $n_e$ evaluated at $\Delta = 0.8 \hbar\omega > \Delta^*$ for $\epsilon = 80$.}
    \label{fig:numer}
\end{figure}
\textit{Numerical analysis.}---To test our prediction, we solve the FBE numerically in the steady state. We consider a coherent acoustic phonon mode of angular frequency $\omega = 6 \ \mathrm{meV} / \hbar$ with speed of sound $c_{\text{ph}} = 20 \ \mathrm{km/s}$, incoherent phonon modes at temperature $T = 7 \ \mathrm{K}$, and optimal electronic density $n_e = q$. Fig. \ref{fig:numer}(a)-(b) compares the steady state distributions for weak ($\delta \varepsilon < \Delta < \Delta^*$) and strong ($\Delta^* < \Delta < \hbar\omega$) non-adiabatic drives. As predicted using the phenomenological model, the density of excitations in the $\alpha=1$ band is suppressed in the strong drive limit, approaching the optimal distribution $F^{\text{opt}}_{k\alpha}$. 

Fig. \ref{fig:intro}(b) shows the current $J$ as a function of ${\Delta}/\hbar\omega$ for different dielectric constants $\epsilon$. The vertical line indicates $\Delta = \hbar \omega$, the boundary between the non-adiabatic and adiabatic drive regimes. The current approaches quantization in the non-adiabatic regime $\Delta < \hbar \omega$ when $\epsilon$ is sufficiently large, verifying the phenomenological model. The inset of Fig. \ref{fig:intro}(b) shows that the optimal Floquet gap $\Delta^*$ decreases as a function of $\epsilon$. 

Finally, we study the current as a function of the doping. The optimal distribution, $F^{\text{opt}}_{k\alpha}$, is obtained at optimal doping $n_e = q$. Fig. \ref{fig:numer}(c)-(d) compares the steady state distributions for electronic densities below ($n_e < q$) and above ($n_e > q$) optimal doping, where the steady state deviates significantly from $F^{\text{opt}}_{k\alpha}$. Fig. \ref{fig:numer}(e) shows $J$ as a function of $n_e$ for $\Delta \approx 0.8 \hbar\omega > \Delta^*$ and $\epsilon = 80$. The quantized current is reached at optimal doping. 

\textit{Experimental realization.}---Lastly, let us comment on the experimental setup needed to detect a phonon-induced quantized current. In Fig. \ref{fig:intro}(a), we show a CNT surrounded by a dielectric material of relative permittivity $\epsilon$. The source of coherent phonons is placed on the right end of the CNT, and the CNT is enclosed on both ends with a conducting anti-reflective material of speed of sound $c_{\text{ph}}^r$ and thickness of $2 \pi c_{\text{ph}}^r / (4\omega)$ tuned to minimize reflection of the phonon wave. The current is measured using gold leads attached to the anti-reflective material. Lastly, the phonon drive amplitude $u_0$ must be much weaker than that which causes melting, i.e., $u_0 \ll 0.1 q^{-1}$ as predicted by the Lindemann criterion \cite{PhysRevResearch.2.012040, lindemann-orig}. The inset of Fig. \ref{fig:intro}(b) shows that, for experimentally accessible values of $\epsilon$, $\Delta^* /\hbar\omega \sim 0.5$, corresponding to the coherent phonon amplitude $u^* \sim 0.001 a \ll 0.1q^{-1}$.

\textit{Conclusion.}---Phonon-driven systems exhibit exciting non-equilibrium physics. Our work shows that screened quantum wires can exhibit quantized transport when driven non-adiabatically by coherent phonons. Low-temperature incoherent phonons relax electronic occupation to a topological Floquet energy band [see Fig. \ref{fig:intro}(c)] that hosts a quantized current response. Because the topological transport does not rely on adiabaticity, the quantized current is realized for short-wavelength phonons with weak amplitudes ($\Delta < \hbar\omega$). This phenomenon differs from the non-interacting and adiabatic drive limit realizing a Thouless pump \cite{PhysRevB.27.6083,thouless-pump,thouless-topo-ins,thouless-ultracold,PhysRevLett.111.060802}, or the quantized acousto-electric effect induced by GHz-frequency phonons \cite{PhysRevLett.87.276802,PhysRevLett.95.256802,Buitelaar2006,Ahlers2004,PhysRevB.70.233401,PhysRevB.68.245310}. While the adiabatic quantized acousto-electric effect induces nano-Ampere quantized currents, our proposal may be realized for higher, THz-frequency phonons which result in much stronger quantized currents on the order of $0.1 \ \mu \mathrm{A}$. It also differs from proposals to realize quantized transport in systems far from equilibrium via strong disorder \cite{PhysRevB.101.041403}, or one-way transport without guaranteed quantized current in driven open systems \cite{one-way-transport}. 

The setup can serve as a new detector of coherent phonons relying only on transport measurements, in contrast to conventional methods utilizing optical reflectivity measurements \cite{PhysRevB.84.064307,10.1063/1.90855,pump-probe-1,https://doi.org/10.48550/arxiv.2310.04939}. Carbon nanotubes [see Fig. \ref{fig:intro}(a)] could provide a suitable experimental platform to realize such phonon detection. The detection would be sensitive even to weak coherent phonon amplitudes [see Fig. \ref{fig:intro}(b)] because adiabaticity is not required. Topological transport in higher-dimensional phonon-driven systems is a subject of future work, in which the direction of the transport could be sensitively controlled by the direction of the coherent phonon propagation. 

\begin{acknowledgments}
We thank Yang Peng, Michael Kolodrubetz, and Erez Berg for valuable discussions. C.Y. gratefully acknowledges support from the DOE NNSA Stewardship Science Graduate Fellowship program, which is provided under cooperative agreement number DE-NA0003960. W.H. gratefully acknowledges support from the Caltech Cambridge Scholars Exchange Program. G.R. and I.E. are grateful for support from the Simons Foundation and the Institute of Quantum Information and Matter, as well as support from the NSF DMR grant number 1839271. 
This work is supported by ARO MURI Grant No. W911NF-16-1-0361, and was performed in part at Aspen Center for Physics, which is supported by National Science Foundation grant PHY-1607611.
\end{acknowledgments}

\bibliography{references_use.bib}

\begin{thebibliography}{101}%
\makeatletter
\providecommand \@ifxundefined [1]{%
 \@ifx{#1\undefined}
}%
\providecommand \@ifnum [1]{%
 \ifnum #1\expandafter \@firstoftwo
 \else \expandafter \@secondoftwo
 \fi
}%
\providecommand \@ifx [1]{%
 \ifx #1\expandafter \@firstoftwo
 \else \expandafter \@secondoftwo
 \fi
}%
\providecommand \natexlab [1]{#1}%
\providecommand \enquote  [1]{``#1''}%
\providecommand \bibnamefont  [1]{#1}%
\providecommand \bibfnamefont [1]{#1}%
\providecommand \citenamefont [1]{#1}%
\providecommand \href@noop [0]{\@secondoftwo}%
\providecommand \href [0]{\begingroup \@sanitize@url \@href}%
\providecommand \@href[1]{\@@startlink{#1}\@@href}%
\providecommand \@@href[1]{\endgroup#1\@@endlink}%
\providecommand \@sanitize@url [0]{\catcode `\\12\catcode `\$12\catcode `\&12\catcode `\#12\catcode `\^12\catcode `\_12\catcode `\%12\relax}%
\providecommand \@@startlink[1]{}%
\providecommand \@@endlink[0]{}%
\providecommand \url  [0]{\begingroup\@sanitize@url \@url }%
\providecommand \@url [1]{\endgroup\@href {#1}{\urlprefix }}%
\providecommand \urlprefix  [0]{URL }%
\providecommand \Eprint [0]{\href }%
\providecommand \doibase [0]{http://dx.doi.org/}%
\providecommand \selectlanguage [0]{\@gobble}%
\providecommand \bibinfo  [0]{\@secondoftwo}%
\providecommand \bibfield  [0]{\@secondoftwo}%
\providecommand \translation [1]{[#1]}%
\providecommand \BibitemOpen [0]{}%
\providecommand \bibitemStop [0]{}%
\providecommand \bibitemNoStop [0]{.\EOS\space}%
\providecommand \EOS [0]{\spacefactor3000\relax}%
\providecommand \BibitemShut  [1]{\csname bibitem#1\endcsname}%
\let\auto@bib@innerbib\@empty
\bibitem [{\citenamefont {Ng}\ \emph {et~al.}(2022)\citenamefont {Ng}, \citenamefont {El~Sachat}, \citenamefont {Cespedes}, \citenamefont {Poblet}, \citenamefont {Madiot}, \citenamefont {Jaramillo-Fernandez}, \citenamefont {Florez}, \citenamefont {Xiao}, \citenamefont {Sledzinska}, \citenamefont {Sotomayor-Torres},\ and\ \citenamefont {Chavez-Angel}}]{D2NR04100F}%
  \BibitemOpen
  \bibfield  {author} {\bibinfo {author} {\bibfnamefont {R.~C.}\ \bibnamefont {Ng}}, \bibinfo {author} {\bibfnamefont {A.}~\bibnamefont {El~Sachat}}, \bibinfo {author} {\bibfnamefont {F.}~\bibnamefont {Cespedes}}, \bibinfo {author} {\bibfnamefont {M.}~\bibnamefont {Poblet}}, \bibinfo {author} {\bibfnamefont {G.}~\bibnamefont {Madiot}}, \bibinfo {author} {\bibfnamefont {J.}~\bibnamefont {Jaramillo-Fernandez}}, \bibinfo {author} {\bibfnamefont {O.}~\bibnamefont {Florez}}, \bibinfo {author} {\bibfnamefont {P.}~\bibnamefont {Xiao}}, \bibinfo {author} {\bibfnamefont {M.}~\bibnamefont {Sledzinska}}, \bibinfo {author} {\bibfnamefont {C.~M.}\ \bibnamefont {Sotomayor-Torres}}, \ and\ \bibinfo {author} {\bibfnamefont {E.}~\bibnamefont {Chavez-Angel}},\ }\href {\doibase 10.1039/D2NR04100F} {\bibfield  {journal} {\bibinfo  {journal} {Nanoscale}\ }\textbf {\bibinfo {volume} {14}},\ \bibinfo {pages} {13428} (\bibinfo {year} {2022})}\BibitemShut {NoStop}%
\bibitem [{\citenamefont {Wallentowitz}\ \emph {et~al.}(1996)\citenamefont {Wallentowitz}, \citenamefont {Vogel}, \citenamefont {Siemers},\ and\ \citenamefont {Toschek}}]{PhysRevA.54.943}%
  \BibitemOpen
  \bibfield  {author} {\bibinfo {author} {\bibfnamefont {S.}~\bibnamefont {Wallentowitz}}, \bibinfo {author} {\bibfnamefont {W.}~\bibnamefont {Vogel}}, \bibinfo {author} {\bibfnamefont {I.}~\bibnamefont {Siemers}}, \ and\ \bibinfo {author} {\bibfnamefont {P.~E.}\ \bibnamefont {Toschek}},\ }\href {\doibase 10.1103/PhysRevA.54.943} {\bibfield  {journal} {\bibinfo  {journal} {Phys. Rev. A}\ }\textbf {\bibinfo {volume} {54}},\ \bibinfo {pages} {943} (\bibinfo {year} {1996})}\BibitemShut {NoStop}%
\bibitem [{\citenamefont {Vahala}\ \emph {et~al.}(2009)\citenamefont {Vahala}, \citenamefont {Herrmann}, \citenamefont {Kn|[uuml]|nz}, \citenamefont {Batteiger}, \citenamefont {Saathoff}, \citenamefont {H|[auml]|nsch},\ and\ \citenamefont {Udem}}]{phaser-1}%
  \BibitemOpen
  \bibfield  {author} {\bibinfo {author} {\bibfnamefont {K.}~\bibnamefont {Vahala}}, \bibinfo {author} {\bibfnamefont {M.}~\bibnamefont {Herrmann}}, \bibinfo {author} {\bibfnamefont {S.}~\bibnamefont {Kn|[uuml]|nz}}, \bibinfo {author} {\bibfnamefont {V.}~\bibnamefont {Batteiger}}, \bibinfo {author} {\bibfnamefont {G.}~\bibnamefont {Saathoff}}, \bibinfo {author} {\bibfnamefont {T.}~\bibnamefont {H|[auml]|nsch}}, \ and\ \bibinfo {author} {\bibfnamefont {T.}~\bibnamefont {Udem}},\ }\href {\doibase 10.1038/nphys1367} {\bibfield  {journal} {\bibinfo  {journal} {Nature Physics}\ }\textbf {\bibinfo {volume} {5}},\ \bibinfo {pages} {682} (\bibinfo {year} {2009})}\BibitemShut {NoStop}%
\bibitem [{\citenamefont {Ip}\ \emph {et~al.}(2018)\citenamefont {Ip}, \citenamefont {Ransford}, \citenamefont {Jayich}, \citenamefont {Long}, \citenamefont {Roman},\ and\ \citenamefont {Campbell}}]{PhysRevLett.121.043201}%
  \BibitemOpen
  \bibfield  {author} {\bibinfo {author} {\bibfnamefont {M.}~\bibnamefont {Ip}}, \bibinfo {author} {\bibfnamefont {A.}~\bibnamefont {Ransford}}, \bibinfo {author} {\bibfnamefont {A.~M.}\ \bibnamefont {Jayich}}, \bibinfo {author} {\bibfnamefont {X.}~\bibnamefont {Long}}, \bibinfo {author} {\bibfnamefont {C.}~\bibnamefont {Roman}}, \ and\ \bibinfo {author} {\bibfnamefont {W.~C.}\ \bibnamefont {Campbell}},\ }\href {\doibase 10.1103/PhysRevLett.121.043201} {\bibfield  {journal} {\bibinfo  {journal} {Phys. Rev. Lett.}\ }\textbf {\bibinfo {volume} {121}},\ \bibinfo {pages} {043201} (\bibinfo {year} {2018})}\BibitemShut {NoStop}%
\bibitem [{\citenamefont {Pettit}\ \emph {et~al.}(2019)\citenamefont {Pettit}, \citenamefont {Ge}, \citenamefont {Kumar}, \citenamefont {Luntz-Martin}, \citenamefont {Schultz}, \citenamefont {Neukirch}, \citenamefont {Bhattacharya},\ and\ \citenamefont {Vamivakas}}]{phaser-2}%
  \BibitemOpen
  \bibfield  {author} {\bibinfo {author} {\bibfnamefont {R.}~\bibnamefont {Pettit}}, \bibinfo {author} {\bibfnamefont {W.}~\bibnamefont {Ge}}, \bibinfo {author} {\bibfnamefont {P.}~\bibnamefont {Kumar}}, \bibinfo {author} {\bibfnamefont {D.}~\bibnamefont {Luntz-Martin}}, \bibinfo {author} {\bibfnamefont {J.}~\bibnamefont {Schultz}}, \bibinfo {author} {\bibfnamefont {L.}~\bibnamefont {Neukirch}}, \bibinfo {author} {\bibfnamefont {M.}~\bibnamefont {Bhattacharya}}, \ and\ \bibinfo {author} {\bibfnamefont {A.}~\bibnamefont {Vamivakas}},\ }\href {\doibase 10.1038/s41566-019-0395-5} {\bibfield  {journal} {\bibinfo  {journal} {Nature Photonics}\ }\textbf {\bibinfo {volume} {13}},\ \bibinfo {pages} {1} (\bibinfo {year} {2019})}\BibitemShut {NoStop}%
\bibitem [{\citenamefont {Grudinin}\ \emph {et~al.}(2010)\citenamefont {Grudinin}, \citenamefont {Lee}, \citenamefont {Painter},\ and\ \citenamefont {Vahala}}]{PhysRevLett.104.083901}%
  \BibitemOpen
  \bibfield  {author} {\bibinfo {author} {\bibfnamefont {I.~S.}\ \bibnamefont {Grudinin}}, \bibinfo {author} {\bibfnamefont {H.}~\bibnamefont {Lee}}, \bibinfo {author} {\bibfnamefont {O.}~\bibnamefont {Painter}}, \ and\ \bibinfo {author} {\bibfnamefont {K.~J.}\ \bibnamefont {Vahala}},\ }\href {\doibase 10.1103/PhysRevLett.104.083901} {\bibfield  {journal} {\bibinfo  {journal} {Phys. Rev. Lett.}\ }\textbf {\bibinfo {volume} {104}},\ \bibinfo {pages} {083901} (\bibinfo {year} {2010})}\BibitemShut {NoStop}%
\bibitem [{\citenamefont {Jing}\ \emph {et~al.}(2014)\citenamefont {Jing}, \citenamefont {\"Ozdemir}, \citenamefont {L\"u}, \citenamefont {Zhang}, \citenamefont {Yang},\ and\ \citenamefont {Nori}}]{PhysRevLett.113.053604}%
  \BibitemOpen
  \bibfield  {author} {\bibinfo {author} {\bibfnamefont {H.}~\bibnamefont {Jing}}, \bibinfo {author} {\bibfnamefont {S.~K.}\ \bibnamefont {\"Ozdemir}}, \bibinfo {author} {\bibfnamefont {X.-Y.}\ \bibnamefont {L\"u}}, \bibinfo {author} {\bibfnamefont {J.}~\bibnamefont {Zhang}}, \bibinfo {author} {\bibfnamefont {L.}~\bibnamefont {Yang}}, \ and\ \bibinfo {author} {\bibfnamefont {F.}~\bibnamefont {Nori}},\ }\href {\doibase 10.1103/PhysRevLett.113.053604} {\bibfield  {journal} {\bibinfo  {journal} {Phys. Rev. Lett.}\ }\textbf {\bibinfo {volume} {113}},\ \bibinfo {pages} {053604} (\bibinfo {year} {2014})}\BibitemShut {NoStop}%
\bibitem [{\citenamefont {Guo}\ \emph {et~al.}(2023)\citenamefont {Guo}, \citenamefont {Esin}, \citenamefont {Li}, \citenamefont {Chen}, \citenamefont {Liu}, \citenamefont {Edgar}, \citenamefont {Zhou}, \citenamefont {Demler}, \citenamefont {Refael},\ and\ \citenamefont {Xia}}]{hyp-ph-po}%
  \BibitemOpen
  \bibfield  {author} {\bibinfo {author} {\bibfnamefont {Q.}~\bibnamefont {Guo}}, \bibinfo {author} {\bibfnamefont {I.}~\bibnamefont {Esin}}, \bibinfo {author} {\bibfnamefont {C.}~\bibnamefont {Li}}, \bibinfo {author} {\bibfnamefont {C.}~\bibnamefont {Chen}}, \bibinfo {author} {\bibfnamefont {S.}~\bibnamefont {Liu}}, \bibinfo {author} {\bibfnamefont {J.~H.}\ \bibnamefont {Edgar}}, \bibinfo {author} {\bibfnamefont {S.}~\bibnamefont {Zhou}}, \bibinfo {author} {\bibfnamefont {E.}~\bibnamefont {Demler}}, \bibinfo {author} {\bibfnamefont {G.}~\bibnamefont {Refael}}, \ and\ \bibinfo {author} {\bibfnamefont {F.}~\bibnamefont {Xia}},\ }\href {\doibase 10.48550/ARXIV.2310.03926} {\enquote {\bibinfo {title} {Hyperbolic phonon-polariton electroluminescence in graphene-hbn van der waals heterostructures},}\ } (\bibinfo {year} {2023})\BibitemShut {NoStop}%
\bibitem [{\citenamefont {Först}\ \emph {et~al.}(2011)\citenamefont {Först}, \citenamefont {Manzoni}, \citenamefont {Kaiser}, \citenamefont {Tomioka}, \citenamefont {Tokura}, \citenamefont {Merlin},\ and\ \citenamefont {Cavalleri}}]{phononic-nonlinear}%
  \BibitemOpen
  \bibfield  {author} {\bibinfo {author} {\bibfnamefont {M.}~\bibnamefont {Först}}, \bibinfo {author} {\bibfnamefont {C.}~\bibnamefont {Manzoni}}, \bibinfo {author} {\bibfnamefont {S.}~\bibnamefont {Kaiser}}, \bibinfo {author} {\bibfnamefont {Y.}~\bibnamefont {Tomioka}}, \bibinfo {author} {\bibfnamefont {Y.}~\bibnamefont {Tokura}}, \bibinfo {author} {\bibfnamefont {R.}~\bibnamefont {Merlin}}, \ and\ \bibinfo {author} {\bibfnamefont {A.}~\bibnamefont {Cavalleri}},\ }\href {\doibase 10.1038/NPHYS2055} {\bibfield  {journal} {\bibinfo  {journal} {Nature Physics}\ }\textbf {\bibinfo {volume} {7}} (\bibinfo {year} {2011}),\ 10.1038/NPHYS2055}\BibitemShut {NoStop}%
\bibitem [{\citenamefont {Li}\ \emph {et~al.}(2012)\citenamefont {Li}, \citenamefont {Ren}, \citenamefont {Wang}, \citenamefont {Zhang}, \citenamefont {H\"anggi},\ and\ \citenamefont {Li}}]{RevModPhys.84.1045}%
  \BibitemOpen
  \bibfield  {author} {\bibinfo {author} {\bibfnamefont {N.}~\bibnamefont {Li}}, \bibinfo {author} {\bibfnamefont {J.}~\bibnamefont {Ren}}, \bibinfo {author} {\bibfnamefont {L.}~\bibnamefont {Wang}}, \bibinfo {author} {\bibfnamefont {G.}~\bibnamefont {Zhang}}, \bibinfo {author} {\bibfnamefont {P.}~\bibnamefont {H\"anggi}}, \ and\ \bibinfo {author} {\bibfnamefont {B.}~\bibnamefont {Li}},\ }\href {\doibase 10.1103/RevModPhys.84.1045} {\bibfield  {journal} {\bibinfo  {journal} {Rev. Mod. Phys.}\ }\textbf {\bibinfo {volume} {84}},\ \bibinfo {pages} {1045} (\bibinfo {year} {2012})}\BibitemShut {NoStop}%
\bibitem [{\citenamefont {Balandin}\ and\ \citenamefont {Nika}(2012)}]{BALANDIN2012266}%
  \BibitemOpen
  \bibfield  {author} {\bibinfo {author} {\bibfnamefont {A.~A.}\ \bibnamefont {Balandin}}\ and\ \bibinfo {author} {\bibfnamefont {D.~L.}\ \bibnamefont {Nika}},\ }\href {\doibase https://doi.org/10.1016/S1369-7021(12)70117-7} {\bibfield  {journal} {\bibinfo  {journal} {Materials Today}\ }\textbf {\bibinfo {volume} {15}},\ \bibinfo {pages} {266} (\bibinfo {year} {2012})}\BibitemShut {NoStop}%
\bibitem [{\citenamefont {Maldovan}(2013)}]{sound-heat-revolution}%
  \BibitemOpen
  \bibfield  {author} {\bibinfo {author} {\bibfnamefont {M.}~\bibnamefont {Maldovan}},\ }\href {\doibase 10.1038/nature12608} {\bibfield  {journal} {\bibinfo  {journal} {Nature}\ }\textbf {\bibinfo {volume} {503}},\ \bibinfo {pages} {209} (\bibinfo {year} {2013})}\BibitemShut {NoStop}%
\bibitem [{\citenamefont {Balandin}(2020)}]{phononics-graphene}%
  \BibitemOpen
  \bibfield  {author} {\bibinfo {author} {\bibfnamefont {A.}~\bibnamefont {Balandin}},\ }\href {\doibase 10.1021/acsnano.0c02718} {\bibfield  {journal} {\bibinfo  {journal} {ACS Nano}\ }\textbf {\bibinfo {volume} {14}} (\bibinfo {year} {2020}),\ 10.1021/acsnano.0c02718}\BibitemShut {NoStop}%
\bibitem [{\citenamefont {Subedi}\ \emph {et~al.}(2014)\citenamefont {Subedi}, \citenamefont {Cavalleri},\ and\ \citenamefont {Georges}}]{PhysRevB.89.220301}%
  \BibitemOpen
  \bibfield  {author} {\bibinfo {author} {\bibfnamefont {A.}~\bibnamefont {Subedi}}, \bibinfo {author} {\bibfnamefont {A.}~\bibnamefont {Cavalleri}}, \ and\ \bibinfo {author} {\bibfnamefont {A.}~\bibnamefont {Georges}},\ }\href {\doibase 10.1103/PhysRevB.89.220301} {\bibfield  {journal} {\bibinfo  {journal} {Phys. Rev. B}\ }\textbf {\bibinfo {volume} {89}},\ \bibinfo {pages} {220301} (\bibinfo {year} {2014})}\BibitemShut {NoStop}%
\bibitem [{\citenamefont {Mankowsky}\ \emph {et~al.}(2016)\citenamefont {Mankowsky}, \citenamefont {Först},\ and\ \citenamefont {Cavalleri}}]{solids-nonlinear-phononics}%
  \BibitemOpen
  \bibfield  {author} {\bibinfo {author} {\bibfnamefont {R.}~\bibnamefont {Mankowsky}}, \bibinfo {author} {\bibfnamefont {M.}~\bibnamefont {Först}}, \ and\ \bibinfo {author} {\bibfnamefont {A.}~\bibnamefont {Cavalleri}},\ }\href {\doibase 10.1088/0034-4885/79/6/064503} {\bibfield  {journal} {\bibinfo  {journal} {Reports on Progress in Physics}\ }\textbf {\bibinfo {volume} {79}},\ \bibinfo {pages} {064503} (\bibinfo {year} {2016})}\BibitemShut {NoStop}%
\bibitem [{\citenamefont {Juraschek}\ \emph {et~al.}(2017)\citenamefont {Juraschek}, \citenamefont {Fechner},\ and\ \citenamefont {Spaldin}}]{PhysRevLett.118.054101}%
  \BibitemOpen
  \bibfield  {author} {\bibinfo {author} {\bibfnamefont {D.~M.}\ \bibnamefont {Juraschek}}, \bibinfo {author} {\bibfnamefont {M.}~\bibnamefont {Fechner}}, \ and\ \bibinfo {author} {\bibfnamefont {N.~A.}\ \bibnamefont {Spaldin}},\ }\href {\doibase 10.1103/PhysRevLett.118.054101} {\bibfield  {journal} {\bibinfo  {journal} {Phys. Rev. Lett.}\ }\textbf {\bibinfo {volume} {118}},\ \bibinfo {pages} {054101} (\bibinfo {year} {2017})}\BibitemShut {NoStop}%
\bibitem [{\citenamefont {Trovatello}\ \emph {et~al.}(2020)\citenamefont {Trovatello}, \citenamefont {Miranda}, \citenamefont {Molina-S{\'{a}}nchez}, \citenamefont {Borrego-Varillas}, \citenamefont {Manzoni}, \citenamefont {Moretti}, \citenamefont {Ganzer}, \citenamefont {Maiuri}, \citenamefont {Wang}, \citenamefont {Dumcenco}, \citenamefont {Kis}, \citenamefont {Wirtz}, \citenamefont {Marini}, \citenamefont {Soavi}, \citenamefont {Ferrari}, \citenamefont {Cerullo}, \citenamefont {Sangalli},\ and\ \citenamefont {Conte}}]{Trovatello}%
  \BibitemOpen
  \bibfield  {author} {\bibinfo {author} {\bibfnamefont {C.}~\bibnamefont {Trovatello}}, \bibinfo {author} {\bibfnamefont {H.~P.~C.}\ \bibnamefont {Miranda}}, \bibinfo {author} {\bibfnamefont {A.}~\bibnamefont {Molina-S{\'{a}}nchez}}, \bibinfo {author} {\bibfnamefont {R.}~\bibnamefont {Borrego-Varillas}}, \bibinfo {author} {\bibfnamefont {C.}~\bibnamefont {Manzoni}}, \bibinfo {author} {\bibfnamefont {L.}~\bibnamefont {Moretti}}, \bibinfo {author} {\bibfnamefont {L.}~\bibnamefont {Ganzer}}, \bibinfo {author} {\bibfnamefont {M.}~\bibnamefont {Maiuri}}, \bibinfo {author} {\bibfnamefont {J.}~\bibnamefont {Wang}}, \bibinfo {author} {\bibfnamefont {D.}~\bibnamefont {Dumcenco}}, \bibinfo {author} {\bibfnamefont {A.}~\bibnamefont {Kis}}, \bibinfo {author} {\bibfnamefont {L.}~\bibnamefont {Wirtz}}, \bibinfo {author} {\bibfnamefont {A.}~\bibnamefont {Marini}}, \bibinfo {author} {\bibfnamefont {G.}~\bibnamefont {Soavi}}, \bibinfo {author} {\bibfnamefont {A.~C.}\ \bibnamefont {Ferrari}}, \bibinfo {author} {\bibfnamefont
  {G.}~\bibnamefont {Cerullo}}, \bibinfo {author} {\bibfnamefont {D.}~\bibnamefont {Sangalli}}, \ and\ \bibinfo {author} {\bibfnamefont {S.~D.}\ \bibnamefont {Conte}},\ }\href {\doibase 10.1021/acsnano.0c00309} {\bibfield  {journal} {\bibinfo  {journal} {{ACS} Nano}\ }\textbf {\bibinfo {volume} {14}},\ \bibinfo {pages} {5700} (\bibinfo {year} {2020})}\BibitemShut {NoStop}%
\bibitem [{\citenamefont {Khalsa}\ \emph {et~al.}(2021)\citenamefont {Khalsa}, \citenamefont {Benedek},\ and\ \citenamefont {Moses}}]{PhysRevX.11.021067}%
  \BibitemOpen
  \bibfield  {author} {\bibinfo {author} {\bibfnamefont {G.}~\bibnamefont {Khalsa}}, \bibinfo {author} {\bibfnamefont {N.~A.}\ \bibnamefont {Benedek}}, \ and\ \bibinfo {author} {\bibfnamefont {J.}~\bibnamefont {Moses}},\ }\href {\doibase 10.1103/PhysRevX.11.021067} {\bibfield  {journal} {\bibinfo  {journal} {Phys. Rev. X}\ }\textbf {\bibinfo {volume} {11}},\ \bibinfo {pages} {021067} (\bibinfo {year} {2021})}\BibitemShut {NoStop}%
\bibitem [{\citenamefont {Mankowsky}\ \emph {et~al.}(2014)\citenamefont {Mankowsky}, \citenamefont {Subedi}, \citenamefont {Först}, \citenamefont {Mariager}, \citenamefont {Chollet}, \citenamefont {Lemke}, \citenamefont {Robinson}, \citenamefont {Glownia}, \citenamefont {Minitti}, \citenamefont {Frano}, \citenamefont {Fechner}, \citenamefont {Spaldin}, \citenamefont {Loew}, \citenamefont {Keimer}, \citenamefont {Georges},\ and\ \citenamefont {Cavalleri}}]{superconductivity-1}%
  \BibitemOpen
  \bibfield  {author} {\bibinfo {author} {\bibfnamefont {R.}~\bibnamefont {Mankowsky}}, \bibinfo {author} {\bibfnamefont {A.}~\bibnamefont {Subedi}}, \bibinfo {author} {\bibfnamefont {M.}~\bibnamefont {Först}}, \bibinfo {author} {\bibfnamefont {S.}~\bibnamefont {Mariager}}, \bibinfo {author} {\bibfnamefont {M.}~\bibnamefont {Chollet}}, \bibinfo {author} {\bibfnamefont {H.}~\bibnamefont {Lemke}}, \bibinfo {author} {\bibfnamefont {J.}~\bibnamefont {Robinson}}, \bibinfo {author} {\bibfnamefont {J.}~\bibnamefont {Glownia}}, \bibinfo {author} {\bibfnamefont {M.}~\bibnamefont {Minitti}}, \bibinfo {author} {\bibfnamefont {A.}~\bibnamefont {Frano}}, \bibinfo {author} {\bibfnamefont {M.}~\bibnamefont {Fechner}}, \bibinfo {author} {\bibfnamefont {N.}~\bibnamefont {Spaldin}}, \bibinfo {author} {\bibfnamefont {T.}~\bibnamefont {Loew}}, \bibinfo {author} {\bibfnamefont {B.}~\bibnamefont {Keimer}}, \bibinfo {author} {\bibfnamefont {A.}~\bibnamefont {Georges}}, \ and\ \bibinfo {author} {\bibfnamefont {A.}~\bibnamefont
  {Cavalleri}},\ }\href {\doibase 10.1038/nature13875} {\bibfield  {journal} {\bibinfo  {journal} {Nature}\ }\textbf {\bibinfo {volume} {516}},\ \bibinfo {pages} {71} (\bibinfo {year} {2014})}\BibitemShut {NoStop}%
\bibitem [{\citenamefont {Knap}\ \emph {et~al.}(2016)\citenamefont {Knap}, \citenamefont {Babadi}, \citenamefont {Refael}, \citenamefont {Martin},\ and\ \citenamefont {Demler}}]{PhysRevB.94.214504}%
  \BibitemOpen
  \bibfield  {author} {\bibinfo {author} {\bibfnamefont {M.}~\bibnamefont {Knap}}, \bibinfo {author} {\bibfnamefont {M.}~\bibnamefont {Babadi}}, \bibinfo {author} {\bibfnamefont {G.}~\bibnamefont {Refael}}, \bibinfo {author} {\bibfnamefont {I.}~\bibnamefont {Martin}}, \ and\ \bibinfo {author} {\bibfnamefont {E.}~\bibnamefont {Demler}},\ }\href {\doibase 10.1103/PhysRevB.94.214504} {\bibfield  {journal} {\bibinfo  {journal} {Phys. Rev. B}\ }\textbf {\bibinfo {volume} {94}},\ \bibinfo {pages} {214504} (\bibinfo {year} {2016})}\BibitemShut {NoStop}%
\bibitem [{\citenamefont {Babadi}\ \emph {et~al.}(2017)\citenamefont {Babadi}, \citenamefont {Knap}, \citenamefont {Martin}, \citenamefont {Refael},\ and\ \citenamefont {Demler}}]{PhysRevB.96.014512}%
  \BibitemOpen
  \bibfield  {author} {\bibinfo {author} {\bibfnamefont {M.}~\bibnamefont {Babadi}}, \bibinfo {author} {\bibfnamefont {M.}~\bibnamefont {Knap}}, \bibinfo {author} {\bibfnamefont {I.}~\bibnamefont {Martin}}, \bibinfo {author} {\bibfnamefont {G.}~\bibnamefont {Refael}}, \ and\ \bibinfo {author} {\bibfnamefont {E.}~\bibnamefont {Demler}},\ }\href {\doibase 10.1103/PhysRevB.96.014512} {\bibfield  {journal} {\bibinfo  {journal} {Phys. Rev. B}\ }\textbf {\bibinfo {volume} {96}},\ \bibinfo {pages} {014512} (\bibinfo {year} {2017})}\BibitemShut {NoStop}%
\bibitem [{\citenamefont {Kennes}\ \emph {et~al.}(2017)\citenamefont {Kennes}, \citenamefont {Wilner}, \citenamefont {Reichman},\ and\ \citenamefont {Millis}}]{superconductivity-3}%
  \BibitemOpen
  \bibfield  {author} {\bibinfo {author} {\bibfnamefont {D.}~\bibnamefont {Kennes}}, \bibinfo {author} {\bibfnamefont {E.}~\bibnamefont {Wilner}}, \bibinfo {author} {\bibfnamefont {D.}~\bibnamefont {Reichman}}, \ and\ \bibinfo {author} {\bibfnamefont {A.}~\bibnamefont {Millis}},\ }\href {\doibase 10.1038/nphys4024} {\bibfield  {journal} {\bibinfo  {journal} {Nature Physics}\ }\textbf {\bibinfo {volume} {13}} (\bibinfo {year} {2017}),\ 10.1038/nphys4024}\BibitemShut {NoStop}%
\bibitem [{\citenamefont {Mitrano}\ \emph {et~al.}(2016)\citenamefont {Mitrano}, \citenamefont {Cantaluppi}, \citenamefont {Nicoletti}, \citenamefont {Kaiser}, \citenamefont {Perucchi}, \citenamefont {Lupi}, \citenamefont {Di~Pietro}, \citenamefont {Pontiroli}, \citenamefont {Riccò}, \citenamefont {Clark}, \citenamefont {Jaksch},\ and\ \citenamefont {Cavalleri}}]{Mitrano2016}%
  \BibitemOpen
  \bibfield  {author} {\bibinfo {author} {\bibfnamefont {M.}~\bibnamefont {Mitrano}}, \bibinfo {author} {\bibfnamefont {A.}~\bibnamefont {Cantaluppi}}, \bibinfo {author} {\bibfnamefont {D.}~\bibnamefont {Nicoletti}}, \bibinfo {author} {\bibfnamefont {S.}~\bibnamefont {Kaiser}}, \bibinfo {author} {\bibfnamefont {A.}~\bibnamefont {Perucchi}}, \bibinfo {author} {\bibfnamefont {S.}~\bibnamefont {Lupi}}, \bibinfo {author} {\bibfnamefont {P.}~\bibnamefont {Di~Pietro}}, \bibinfo {author} {\bibfnamefont {D.}~\bibnamefont {Pontiroli}}, \bibinfo {author} {\bibfnamefont {M.}~\bibnamefont {Riccò}}, \bibinfo {author} {\bibfnamefont {S.~R.}\ \bibnamefont {Clark}}, \bibinfo {author} {\bibfnamefont {D.}~\bibnamefont {Jaksch}}, \ and\ \bibinfo {author} {\bibfnamefont {A.}~\bibnamefont {Cavalleri}},\ }\href {\doibase 10.1038/nature16522} {\bibfield  {journal} {\bibinfo  {journal} {Nature}\ }\textbf {\bibinfo {volume} {530}},\ \bibinfo {pages} {461–464} (\bibinfo {year} {2016})}\BibitemShut {NoStop}%
\bibitem [{\citenamefont {Li}\ \emph {et~al.}(2019)\citenamefont {Li}, \citenamefont {Qiu}, \citenamefont {Zhang}, \citenamefont {Baldini}, \citenamefont {Lu}, \citenamefont {Rappe},\ and\ \citenamefont {Nelson}}]{Li2019}%
  \BibitemOpen
  \bibfield  {author} {\bibinfo {author} {\bibfnamefont {X.}~\bibnamefont {Li}}, \bibinfo {author} {\bibfnamefont {T.}~\bibnamefont {Qiu}}, \bibinfo {author} {\bibfnamefont {J.}~\bibnamefont {Zhang}}, \bibinfo {author} {\bibfnamefont {E.}~\bibnamefont {Baldini}}, \bibinfo {author} {\bibfnamefont {J.}~\bibnamefont {Lu}}, \bibinfo {author} {\bibfnamefont {A.~M.}\ \bibnamefont {Rappe}}, \ and\ \bibinfo {author} {\bibfnamefont {K.~A.}\ \bibnamefont {Nelson}},\ }\href {\doibase 10.1126/science.aaw4913} {\bibfield  {journal} {\bibinfo  {journal} {Science}\ }\textbf {\bibinfo {volume} {364}},\ \bibinfo {pages} {1079–1082} (\bibinfo {year} {2019})}\BibitemShut {NoStop}%
\bibitem [{\citenamefont {Mankowsky}\ \emph {et~al.}(2017)\citenamefont {Mankowsky}, \citenamefont {von Hoegen}, \citenamefont {F\"orst},\ and\ \citenamefont {Cavalleri}}]{PhysRevLett.118.197601}%
  \BibitemOpen
  \bibfield  {author} {\bibinfo {author} {\bibfnamefont {R.}~\bibnamefont {Mankowsky}}, \bibinfo {author} {\bibfnamefont {A.}~\bibnamefont {von Hoegen}}, \bibinfo {author} {\bibfnamefont {M.}~\bibnamefont {F\"orst}}, \ and\ \bibinfo {author} {\bibfnamefont {A.}~\bibnamefont {Cavalleri}},\ }\href {\doibase 10.1103/PhysRevLett.118.197601} {\bibfield  {journal} {\bibinfo  {journal} {Phys. Rev. Lett.}\ }\textbf {\bibinfo {volume} {118}},\ \bibinfo {pages} {197601} (\bibinfo {year} {2017})}\BibitemShut {NoStop}%
\bibitem [{\citenamefont {Nova}\ \emph {et~al.}(2019)\citenamefont {Nova}, \citenamefont {Disa}, \citenamefont {Fechner},\ and\ \citenamefont {Cavalleri}}]{magnetic-1}%
  \BibitemOpen
  \bibfield  {author} {\bibinfo {author} {\bibfnamefont {T.}~\bibnamefont {Nova}}, \bibinfo {author} {\bibfnamefont {A.}~\bibnamefont {Disa}}, \bibinfo {author} {\bibfnamefont {M.}~\bibnamefont {Fechner}}, \ and\ \bibinfo {author} {\bibfnamefont {A.}~\bibnamefont {Cavalleri}},\ }\href {\doibase 10.1126/science.aaw4911} {\bibfield  {journal} {\bibinfo  {journal} {Science}\ }\textbf {\bibinfo {volume} {364}},\ \bibinfo {pages} {1075} (\bibinfo {year} {2019})}\BibitemShut {NoStop}%
\bibitem [{\citenamefont {Parmenter}(1953)}]{PhysRev.89.990}%
  \BibitemOpen
  \bibfield  {author} {\bibinfo {author} {\bibfnamefont {R.~H.}\ \bibnamefont {Parmenter}},\ }\href {\doibase 10.1103/PhysRev.89.990} {\bibfield  {journal} {\bibinfo  {journal} {Phys. Rev.}\ }\textbf {\bibinfo {volume} {89}},\ \bibinfo {pages} {990} (\bibinfo {year} {1953})}\BibitemShut {NoStop}%
\bibitem [{\citenamefont {Weinreich}\ and\ \citenamefont {White}(1957)}]{PhysRev.106.1104}%
  \BibitemOpen
  \bibfield  {author} {\bibinfo {author} {\bibfnamefont {G.}~\bibnamefont {Weinreich}}\ and\ \bibinfo {author} {\bibfnamefont {H.~G.}\ \bibnamefont {White}},\ }\href {\doibase 10.1103/PhysRev.106.1104} {\bibfield  {journal} {\bibinfo  {journal} {Phys. Rev.}\ }\textbf {\bibinfo {volume} {106}},\ \bibinfo {pages} {1104} (\bibinfo {year} {1957})}\BibitemShut {NoStop}%
\bibitem [{\citenamefont {Kalameitsev}\ \emph {et~al.}(2019)\citenamefont {Kalameitsev}, \citenamefont {Kovalev},\ and\ \citenamefont {Savenko}}]{PhysRevLett.122.256801}%
  \BibitemOpen
  \bibfield  {author} {\bibinfo {author} {\bibfnamefont {A.~V.}\ \bibnamefont {Kalameitsev}}, \bibinfo {author} {\bibfnamefont {V.~M.}\ \bibnamefont {Kovalev}}, \ and\ \bibinfo {author} {\bibfnamefont {I.~G.}\ \bibnamefont {Savenko}},\ }\href {\doibase 10.1103/PhysRevLett.122.256801} {\bibfield  {journal} {\bibinfo  {journal} {Phys. Rev. Lett.}\ }\textbf {\bibinfo {volume} {122}},\ \bibinfo {pages} {256801} (\bibinfo {year} {2019})}\BibitemShut {NoStop}%
\bibitem [{\citenamefont {Reulet}\ \emph {et~al.}(2000)\citenamefont {Reulet}, \citenamefont {Kasumov}, \citenamefont {Kociak}, \citenamefont {Deblock}, \citenamefont {Khodos}, \citenamefont {Gorbatov}, \citenamefont {Volkov}, \citenamefont {Journet},\ and\ \citenamefont {Bouchiat}}]{PhysRevLett.85.2829}%
  \BibitemOpen
  \bibfield  {author} {\bibinfo {author} {\bibfnamefont {B.}~\bibnamefont {Reulet}}, \bibinfo {author} {\bibfnamefont {A.~Y.}\ \bibnamefont {Kasumov}}, \bibinfo {author} {\bibfnamefont {M.}~\bibnamefont {Kociak}}, \bibinfo {author} {\bibfnamefont {R.}~\bibnamefont {Deblock}}, \bibinfo {author} {\bibfnamefont {I.~I.}\ \bibnamefont {Khodos}}, \bibinfo {author} {\bibfnamefont {Y.~B.}\ \bibnamefont {Gorbatov}}, \bibinfo {author} {\bibfnamefont {V.~T.}\ \bibnamefont {Volkov}}, \bibinfo {author} {\bibfnamefont {C.}~\bibnamefont {Journet}}, \ and\ \bibinfo {author} {\bibfnamefont {H.}~\bibnamefont {Bouchiat}},\ }\href {\doibase 10.1103/PhysRevLett.85.2829} {\bibfield  {journal} {\bibinfo  {journal} {Phys. Rev. Lett.}\ }\textbf {\bibinfo {volume} {85}},\ \bibinfo {pages} {2829} (\bibinfo {year} {2000})}\BibitemShut {NoStop}%
\bibitem [{\citenamefont {Mahfouzi}\ and\ \citenamefont {Kioussis}(2022)}]{PhysRevLett.128.215902}%
  \BibitemOpen
  \bibfield  {author} {\bibinfo {author} {\bibfnamefont {F.}~\bibnamefont {Mahfouzi}}\ and\ \bibinfo {author} {\bibfnamefont {N.}~\bibnamefont {Kioussis}},\ }\href {\doibase 10.1103/PhysRevLett.128.215902} {\bibfield  {journal} {\bibinfo  {journal} {Phys. Rev. Lett.}\ }\textbf {\bibinfo {volume} {128}},\ \bibinfo {pages} {215902} (\bibinfo {year} {2022})}\BibitemShut {NoStop}%
\bibitem [{\citenamefont {Kawada}\ \emph {et~al.}(2021)\citenamefont {Kawada}, \citenamefont {Kawaguchi}, \citenamefont {Funato}, \citenamefont {Kohno},\ and\ \citenamefont {Hayashi}}]{Kawada2021}%
  \BibitemOpen
  \bibfield  {author} {\bibinfo {author} {\bibfnamefont {T.}~\bibnamefont {Kawada}}, \bibinfo {author} {\bibfnamefont {M.}~\bibnamefont {Kawaguchi}}, \bibinfo {author} {\bibfnamefont {T.}~\bibnamefont {Funato}}, \bibinfo {author} {\bibfnamefont {H.}~\bibnamefont {Kohno}}, \ and\ \bibinfo {author} {\bibfnamefont {M.}~\bibnamefont {Hayashi}},\ }\href {\doibase 10.1126/sciadv.abd9697} {\bibfield  {journal} {\bibinfo  {journal} {Science Advances}\ }\textbf {\bibinfo {volume} {7}} (\bibinfo {year} {2021}),\ 10.1126/sciadv.abd9697}\BibitemShut {NoStop}%
\bibitem [{\citenamefont {Peng}(2022)}]{PhysRevLett.128.186802}%
  \BibitemOpen
  \bibfield  {author} {\bibinfo {author} {\bibfnamefont {Y.}~\bibnamefont {Peng}},\ }\href {\doibase 10.1103/PhysRevLett.128.186802} {\bibfield  {journal} {\bibinfo  {journal} {Phys. Rev. Lett.}\ }\textbf {\bibinfo {volume} {128}},\ \bibinfo {pages} {186802} (\bibinfo {year} {2022})}\BibitemShut {NoStop}%
\bibitem [{\citenamefont {Gao}\ and\ \citenamefont {Niu}(2022)}]{PhysRevB.106.224311}%
  \BibitemOpen
  \bibfield  {author} {\bibinfo {author} {\bibfnamefont {Q.}~\bibnamefont {Gao}}\ and\ \bibinfo {author} {\bibfnamefont {Q.}~\bibnamefont {Niu}},\ }\href {\doibase 10.1103/PhysRevB.106.224311} {\bibfield  {journal} {\bibinfo  {journal} {Phys. Rev. B}\ }\textbf {\bibinfo {volume} {106}},\ \bibinfo {pages} {224311} (\bibinfo {year} {2022})}\BibitemShut {NoStop}%
\bibitem [{\citenamefont {Hübener}\ \emph {et~al.}(2018)\citenamefont {Hübener}, \citenamefont {De~Giovannini},\ and\ \citenamefont {Rubio}}]{phonon-floquet}%
  \BibitemOpen
  \bibfield  {author} {\bibinfo {author} {\bibfnamefont {H.}~\bibnamefont {Hübener}}, \bibinfo {author} {\bibfnamefont {U.}~\bibnamefont {De~Giovannini}}, \ and\ \bibinfo {author} {\bibfnamefont {A.}~\bibnamefont {Rubio}},\ }\href {\doibase 10.1021/acs.nanolett.7b05391} {\bibfield  {journal} {\bibinfo  {journal} {Nano Letters}\ }\textbf {\bibinfo {volume} {18}} (\bibinfo {year} {2018}),\ 10.1021/acs.nanolett.7b05391}\BibitemShut {NoStop}%
\bibitem [{\citenamefont {Chaudhary}\ \emph {et~al.}(2020)\citenamefont {Chaudhary}, \citenamefont {Haim}, \citenamefont {Peng},\ and\ \citenamefont {Refael}}]{PhysRevResearch.2.043431}%
  \BibitemOpen
  \bibfield  {author} {\bibinfo {author} {\bibfnamefont {S.}~\bibnamefont {Chaudhary}}, \bibinfo {author} {\bibfnamefont {A.}~\bibnamefont {Haim}}, \bibinfo {author} {\bibfnamefont {Y.}~\bibnamefont {Peng}}, \ and\ \bibinfo {author} {\bibfnamefont {G.}~\bibnamefont {Refael}},\ }\href {\doibase 10.1103/PhysRevResearch.2.043431} {\bibfield  {journal} {\bibinfo  {journal} {Phys. Rev. Res.}\ }\textbf {\bibinfo {volume} {2}},\ \bibinfo {pages} {043431} (\bibinfo {year} {2020})}\BibitemShut {NoStop}%
\bibitem [{\citenamefont {Esin}\ \emph {et~al.}(2023)\citenamefont {Esin}, \citenamefont {Esterlis}, \citenamefont {Demler},\ and\ \citenamefont {Refael}}]{Esin2022GeneratingPhaser}%
  \BibitemOpen
  \bibfield  {author} {\bibinfo {author} {\bibfnamefont {I.}~\bibnamefont {Esin}}, \bibinfo {author} {\bibfnamefont {I.}~\bibnamefont {Esterlis}}, \bibinfo {author} {\bibfnamefont {E.}~\bibnamefont {Demler}}, \ and\ \bibinfo {author} {\bibfnamefont {G.}~\bibnamefont {Refael}},\ }\href {\doibase 10.1103/PhysRevLett.130.147001} {\bibfield  {journal} {\bibinfo  {journal} {Phys. Rev. Lett.}\ }\textbf {\bibinfo {volume} {130}},\ \bibinfo {pages} {147001} (\bibinfo {year} {2023})}\BibitemShut {NoStop}%
\bibitem [{\citenamefont {Beardsley}\ \emph {et~al.}(2010)\citenamefont {Beardsley}, \citenamefont {Akimov}, \citenamefont {Henini},\ and\ \citenamefont {Kent}}]{PhysRevLett.104.085501}%
  \BibitemOpen
  \bibfield  {author} {\bibinfo {author} {\bibfnamefont {R.~P.}\ \bibnamefont {Beardsley}}, \bibinfo {author} {\bibfnamefont {A.~V.}\ \bibnamefont {Akimov}}, \bibinfo {author} {\bibfnamefont {M.}~\bibnamefont {Henini}}, \ and\ \bibinfo {author} {\bibfnamefont {A.~J.}\ \bibnamefont {Kent}},\ }\href {\doibase 10.1103/PhysRevLett.104.085501} {\bibfield  {journal} {\bibinfo  {journal} {Phys. Rev. Lett.}\ }\textbf {\bibinfo {volume} {104}},\ \bibinfo {pages} {085501} (\bibinfo {year} {2010})}\BibitemShut {NoStop}%
\bibitem [{\citenamefont {Cartella}\ \emph {et~al.}(2017)\citenamefont {Cartella}, \citenamefont {Nova}, \citenamefont {Fechner}, \citenamefont {Merlin},\ and\ \citenamefont {Cavalleri}}]{phaser-thz-1}%
  \BibitemOpen
  \bibfield  {author} {\bibinfo {author} {\bibfnamefont {A.}~\bibnamefont {Cartella}}, \bibinfo {author} {\bibfnamefont {T.}~\bibnamefont {Nova}}, \bibinfo {author} {\bibfnamefont {M.}~\bibnamefont {Fechner}}, \bibinfo {author} {\bibfnamefont {R.}~\bibnamefont {Merlin}}, \ and\ \bibinfo {author} {\bibfnamefont {A.}~\bibnamefont {Cavalleri}},\ }\href {\doibase 10.1073/pnas.1809725115} {\bibfield  {journal} {\bibinfo  {journal} {Proceedings of the National Academy of Sciences}\ }\textbf {\bibinfo {volume} {115}} (\bibinfo {year} {2017}),\ 10.1073/pnas.1809725115}\BibitemShut {NoStop}%
\bibitem [{\citenamefont {Barajas-Aguilar}\ \emph {et~al.}(2023)\citenamefont {Barajas-Aguilar}, \citenamefont {Zion}, \citenamefont {Sequeira}, \citenamefont {Barabas}, \citenamefont {Taniguchi}, \citenamefont {Watanabe}, \citenamefont {Barrett}, \citenamefont {Scaffidi},\ and\ \citenamefont {Sanchez-Yamagishi}}]{aguilar2023}%
  \BibitemOpen
  \bibfield  {author} {\bibinfo {author} {\bibfnamefont {A.~H.}\ \bibnamefont {Barajas-Aguilar}}, \bibinfo {author} {\bibfnamefont {J.}~\bibnamefont {Zion}}, \bibinfo {author} {\bibfnamefont {I.}~\bibnamefont {Sequeira}}, \bibinfo {author} {\bibfnamefont {A.~Z.}\ \bibnamefont {Barabas}}, \bibinfo {author} {\bibfnamefont {T.}~\bibnamefont {Taniguchi}}, \bibinfo {author} {\bibfnamefont {K.}~\bibnamefont {Watanabe}}, \bibinfo {author} {\bibfnamefont {E.}~\bibnamefont {Barrett}}, \bibinfo {author} {\bibfnamefont {T.}~\bibnamefont {Scaffidi}}, \ and\ \bibinfo {author} {\bibfnamefont {J.~D.}\ \bibnamefont {Sanchez-Yamagishi}},\ }\href {\doibase 10.48550/ARXIV.2310.12225} {\enquote {\bibinfo {title} {Electrically-driven amplification of terahertz acoustic waves in graphene},}\ } (\bibinfo {year} {2023})\BibitemShut {NoStop}%
\bibitem [{\citenamefont {Katz}\ \emph {et~al.}(2020)\citenamefont {Katz}, \citenamefont {Refael},\ and\ \citenamefont {Lindner}}]{Katz2020OpticallyGraphene}%
  \BibitemOpen
  \bibfield  {author} {\bibinfo {author} {\bibfnamefont {O.}~\bibnamefont {Katz}}, \bibinfo {author} {\bibfnamefont {G.}~\bibnamefont {Refael}}, \ and\ \bibinfo {author} {\bibfnamefont {N.~H.}\ \bibnamefont {Lindner}},\ }\href {\doibase 10.1103/PhysRevB.102.155123} {\bibfield  {journal} {\bibinfo  {journal} {Phys. Rev. B}\ }\textbf {\bibinfo {volume} {102}},\ \bibinfo {pages} {155123} (\bibinfo {year} {2020})}\BibitemShut {NoStop}%
\bibitem [{\citenamefont {Yang}\ \emph {et~al.}(2023)\citenamefont {Yang}, \citenamefont {Esin}, \citenamefont {Lewandowski},\ and\ \citenamefont {Refael}}]{PhysRevLett.131.026901}%
  \BibitemOpen
  \bibfield  {author} {\bibinfo {author} {\bibfnamefont {C.}~\bibnamefont {Yang}}, \bibinfo {author} {\bibfnamefont {I.}~\bibnamefont {Esin}}, \bibinfo {author} {\bibfnamefont {C.}~\bibnamefont {Lewandowski}}, \ and\ \bibinfo {author} {\bibfnamefont {G.}~\bibnamefont {Refael}},\ }\href {\doibase 10.1103/PhysRevLett.131.026901} {\bibfield  {journal} {\bibinfo  {journal} {Phys. Rev. Lett.}\ }\textbf {\bibinfo {volume} {131}},\ \bibinfo {pages} {026901} (\bibinfo {year} {2023})}\BibitemShut {NoStop}%
\bibitem [{\citenamefont {Nakajima}\ \emph {et~al.}(2015{\natexlab{a}})\citenamefont {Nakajima}, \citenamefont {Tomita}, \citenamefont {Taie}, \citenamefont {Ichinose}, \citenamefont {Ozawa}, \citenamefont {Wang}, \citenamefont {Troyer},\ and\ \citenamefont {Takahashi}}]{thouless-pump}%
  \BibitemOpen
  \bibfield  {author} {\bibinfo {author} {\bibfnamefont {S.}~\bibnamefont {Nakajima}}, \bibinfo {author} {\bibfnamefont {T.}~\bibnamefont {Tomita}}, \bibinfo {author} {\bibfnamefont {S.}~\bibnamefont {Taie}}, \bibinfo {author} {\bibfnamefont {T.}~\bibnamefont {Ichinose}}, \bibinfo {author} {\bibfnamefont {H.}~\bibnamefont {Ozawa}}, \bibinfo {author} {\bibfnamefont {L.}~\bibnamefont {Wang}}, \bibinfo {author} {\bibfnamefont {M.}~\bibnamefont {Troyer}}, \ and\ \bibinfo {author} {\bibfnamefont {Y.}~\bibnamefont {Takahashi}},\ }\href {\doibase 10.1038/nphys3622} {\bibfield  {journal} {\bibinfo  {journal} {Nature Physics}\ }\textbf {\bibinfo {volume} {12}} (\bibinfo {year} {2015}{\natexlab{a}}),\ 10.1038/nphys3622}\BibitemShut {NoStop}%
\bibitem [{\citenamefont {Hotzen~Grinberg}\ \emph {et~al.}(2020)\citenamefont {Hotzen~Grinberg}, \citenamefont {Lin}, \citenamefont {Harris}, \citenamefont {Benalcazar}, \citenamefont {Peterson}, \citenamefont {Hughes},\ and\ \citenamefont {Bahl}}]{thouless-topo-ins}%
  \BibitemOpen
  \bibfield  {author} {\bibinfo {author} {\bibfnamefont {I.}~\bibnamefont {Hotzen~Grinberg}}, \bibinfo {author} {\bibfnamefont {M.}~\bibnamefont {Lin}}, \bibinfo {author} {\bibfnamefont {C.}~\bibnamefont {Harris}}, \bibinfo {author} {\bibfnamefont {W.}~\bibnamefont {Benalcazar}}, \bibinfo {author} {\bibfnamefont {C.}~\bibnamefont {Peterson}}, \bibinfo {author} {\bibfnamefont {T.}~\bibnamefont {Hughes}}, \ and\ \bibinfo {author} {\bibfnamefont {G.}~\bibnamefont {Bahl}},\ }\href {\doibase 10.1038/s41467-020-14804-0} {\bibfield  {journal} {\bibinfo  {journal} {Nature Communications}\ }\textbf {\bibinfo {volume} {11}} (\bibinfo {year} {2020}),\ 10.1038/s41467-020-14804-0}\BibitemShut {NoStop}%
\bibitem [{\citenamefont {Nakajima}\ \emph {et~al.}(2015{\natexlab{b}})\citenamefont {Nakajima}, \citenamefont {Tomita}, \citenamefont {Taie}, \citenamefont {Ichinose}, \citenamefont {Ozawa}, \citenamefont {Wang}, \citenamefont {Troyer},\ and\ \citenamefont {Takahashi}}]{thouless-ultracold}%
  \BibitemOpen
  \bibfield  {author} {\bibinfo {author} {\bibfnamefont {S.}~\bibnamefont {Nakajima}}, \bibinfo {author} {\bibfnamefont {T.}~\bibnamefont {Tomita}}, \bibinfo {author} {\bibfnamefont {S.}~\bibnamefont {Taie}}, \bibinfo {author} {\bibfnamefont {T.}~\bibnamefont {Ichinose}}, \bibinfo {author} {\bibfnamefont {H.}~\bibnamefont {Ozawa}}, \bibinfo {author} {\bibfnamefont {L.}~\bibnamefont {Wang}}, \bibinfo {author} {\bibfnamefont {M.}~\bibnamefont {Troyer}}, \ and\ \bibinfo {author} {\bibfnamefont {Y.}~\bibnamefont {Takahashi}},\ }\href {\doibase 10.1038/nphys3622} {\bibfield  {journal} {\bibinfo  {journal} {Nature Physics}\ }\textbf {\bibinfo {volume} {12}} (\bibinfo {year} {2015}{\natexlab{b}}),\ 10.1038/nphys3622}\BibitemShut {NoStop}%
\bibitem [{\citenamefont {Bustos-Mar\'un}\ \emph {et~al.}(2013)\citenamefont {Bustos-Mar\'un}, \citenamefont {Refael},\ and\ \citenamefont {von Oppen}}]{PhysRevLett.111.060802}%
  \BibitemOpen
  \bibfield  {author} {\bibinfo {author} {\bibfnamefont {R.}~\bibnamefont {Bustos-Mar\'un}}, \bibinfo {author} {\bibfnamefont {G.}~\bibnamefont {Refael}}, \ and\ \bibinfo {author} {\bibfnamefont {F.}~\bibnamefont {von Oppen}},\ }\href {\doibase 10.1103/PhysRevLett.111.060802} {\bibfield  {journal} {\bibinfo  {journal} {Phys. Rev. Lett.}\ }\textbf {\bibinfo {volume} {111}},\ \bibinfo {pages} {060802} (\bibinfo {year} {2013})}\BibitemShut {NoStop}%
\bibitem [{\citenamefont {Citro}\ and\ \citenamefont {Aidelsburger}(2023)}]{Citro2023}%
  \BibitemOpen
  \bibfield  {author} {\bibinfo {author} {\bibfnamefont {R.}~\bibnamefont {Citro}}\ and\ \bibinfo {author} {\bibfnamefont {M.}~\bibnamefont {Aidelsburger}},\ }\href {\doibase 10.1038/s42254-022-00545-0} {\bibfield  {journal} {\bibinfo  {journal} {Nature Reviews Physics}\ }\textbf {\bibinfo {volume} {5}},\ \bibinfo {pages} {87–101} (\bibinfo {year} {2023})}\BibitemShut {NoStop}%
\bibitem [{\citenamefont {Bruch}\ \emph {et~al.}(2018)\citenamefont {Bruch}, \citenamefont {Kusminskiy}, \citenamefont {Refael},\ and\ \citenamefont {von Oppen}}]{PhysRevB.97.195411}%
  \BibitemOpen
  \bibfield  {author} {\bibinfo {author} {\bibfnamefont {A.}~\bibnamefont {Bruch}}, \bibinfo {author} {\bibfnamefont {S.~V.}\ \bibnamefont {Kusminskiy}}, \bibinfo {author} {\bibfnamefont {G.}~\bibnamefont {Refael}}, \ and\ \bibinfo {author} {\bibfnamefont {F.}~\bibnamefont {von Oppen}},\ }\href {\doibase 10.1103/PhysRevB.97.195411} {\bibfield  {journal} {\bibinfo  {journal} {Phys. Rev. B}\ }\textbf {\bibinfo {volume} {97}},\ \bibinfo {pages} {195411} (\bibinfo {year} {2018})}\BibitemShut {NoStop}%
\bibitem [{\citenamefont {Esin}\ \emph {et~al.}(2024)\citenamefont {Esin}, \citenamefont {Kuhlenkamp}, \citenamefont {Refael}, \citenamefont {Berg}, \citenamefont {Rudner},\ and\ \citenamefont {Lindner}}]{PhysRevResearch.6.013094}%
  \BibitemOpen
  \bibfield  {author} {\bibinfo {author} {\bibfnamefont {I.}~\bibnamefont {Esin}}, \bibinfo {author} {\bibfnamefont {C.}~\bibnamefont {Kuhlenkamp}}, \bibinfo {author} {\bibfnamefont {G.}~\bibnamefont {Refael}}, \bibinfo {author} {\bibfnamefont {E.}~\bibnamefont {Berg}}, \bibinfo {author} {\bibfnamefont {M.~S.}\ \bibnamefont {Rudner}}, \ and\ \bibinfo {author} {\bibfnamefont {N.~H.}\ \bibnamefont {Lindner}},\ }\href {\doibase 10.1103/PhysRevResearch.6.013094} {\bibfield  {journal} {\bibinfo  {journal} {Phys. Rev. Res.}\ }\textbf {\bibinfo {volume} {6}},\ \bibinfo {pages} {013094} (\bibinfo {year} {2024})}\BibitemShut {NoStop}%
\bibitem [{\citenamefont {Nakagawa}\ \emph {et~al.}(2018)\citenamefont {Nakagawa}, \citenamefont {Yoshida}, \citenamefont {Peters},\ and\ \citenamefont {Kawakami}}]{PhysRevB.98.115147}%
  \BibitemOpen
  \bibfield  {author} {\bibinfo {author} {\bibfnamefont {M.}~\bibnamefont {Nakagawa}}, \bibinfo {author} {\bibfnamefont {T.}~\bibnamefont {Yoshida}}, \bibinfo {author} {\bibfnamefont {R.}~\bibnamefont {Peters}}, \ and\ \bibinfo {author} {\bibfnamefont {N.}~\bibnamefont {Kawakami}},\ }\href {\doibase 10.1103/PhysRevB.98.115147} {\bibfield  {journal} {\bibinfo  {journal} {Phys. Rev. B}\ }\textbf {\bibinfo {volume} {98}},\ \bibinfo {pages} {115147} (\bibinfo {year} {2018})}\BibitemShut {NoStop}%
\bibitem [{\citenamefont {Walter}\ \emph {et~al.}(2023)\citenamefont {Walter}, \citenamefont {Zhu}, \citenamefont {G\"{a}chter}, \citenamefont {Minguzzi}, \citenamefont {Roschinski}, \citenamefont {Sandholzer}, \citenamefont {Viebahn},\ and\ \citenamefont {Esslinger}}]{Walter2023}%
  \BibitemOpen
  \bibfield  {author} {\bibinfo {author} {\bibfnamefont {A.-S.}\ \bibnamefont {Walter}}, \bibinfo {author} {\bibfnamefont {Z.}~\bibnamefont {Zhu}}, \bibinfo {author} {\bibfnamefont {M.}~\bibnamefont {G\"{a}chter}}, \bibinfo {author} {\bibfnamefont {J.}~\bibnamefont {Minguzzi}}, \bibinfo {author} {\bibfnamefont {S.}~\bibnamefont {Roschinski}}, \bibinfo {author} {\bibfnamefont {K.}~\bibnamefont {Sandholzer}}, \bibinfo {author} {\bibfnamefont {K.}~\bibnamefont {Viebahn}}, \ and\ \bibinfo {author} {\bibfnamefont {T.}~\bibnamefont {Esslinger}},\ }\href {\doibase 10.1038/s41567-023-02145-w} {\bibfield  {journal} {\bibinfo  {journal} {Nature Physics}\ }\textbf {\bibinfo {volume} {19}},\ \bibinfo {pages} {1471–1475} (\bibinfo {year} {2023})}\BibitemShut {NoStop}%
\bibitem [{\citenamefont {Privitera}\ \emph {et~al.}(2018)\citenamefont {Privitera}, \citenamefont {Russomanno}, \citenamefont {Citro},\ and\ \citenamefont {Santoro}}]{PhysRevLett.120.106601}%
  \BibitemOpen
  \bibfield  {author} {\bibinfo {author} {\bibfnamefont {L.}~\bibnamefont {Privitera}}, \bibinfo {author} {\bibfnamefont {A.}~\bibnamefont {Russomanno}}, \bibinfo {author} {\bibfnamefont {R.}~\bibnamefont {Citro}}, \ and\ \bibinfo {author} {\bibfnamefont {G.~E.}\ \bibnamefont {Santoro}},\ }\href {\doibase 10.1103/PhysRevLett.120.106601} {\bibfield  {journal} {\bibinfo  {journal} {Phys. Rev. Lett.}\ }\textbf {\bibinfo {volume} {120}},\ \bibinfo {pages} {106601} (\bibinfo {year} {2018})}\BibitemShut {NoStop}%
\bibitem [{\citenamefont {Niu}(1990)}]{PhysRevLett.64.1812}%
  \BibitemOpen
  \bibfield  {author} {\bibinfo {author} {\bibfnamefont {Q.}~\bibnamefont {Niu}},\ }\href {\doibase 10.1103/PhysRevLett.64.1812} {\bibfield  {journal} {\bibinfo  {journal} {Phys. Rev. Lett.}\ }\textbf {\bibinfo {volume} {64}},\ \bibinfo {pages} {1812} (\bibinfo {year} {1990})}\BibitemShut {NoStop}%
\bibitem [{\citenamefont {Pekola}\ \emph {et~al.}(2013)\citenamefont {Pekola}, \citenamefont {Saira}, \citenamefont {Maisi}, \citenamefont {Kemppinen}, \citenamefont {M\"ott\"onen}, \citenamefont {Pashkin},\ and\ \citenamefont {Averin}}]{RevModPhys.85.1421}%
  \BibitemOpen
  \bibfield  {author} {\bibinfo {author} {\bibfnamefont {J.~P.}\ \bibnamefont {Pekola}}, \bibinfo {author} {\bibfnamefont {O.-P.}\ \bibnamefont {Saira}}, \bibinfo {author} {\bibfnamefont {V.~F.}\ \bibnamefont {Maisi}}, \bibinfo {author} {\bibfnamefont {A.}~\bibnamefont {Kemppinen}}, \bibinfo {author} {\bibfnamefont {M.}~\bibnamefont {M\"ott\"onen}}, \bibinfo {author} {\bibfnamefont {Y.~A.}\ \bibnamefont {Pashkin}}, \ and\ \bibinfo {author} {\bibfnamefont {D.~V.}\ \bibnamefont {Averin}},\ }\href {\doibase 10.1103/RevModPhys.85.1421} {\bibfield  {journal} {\bibinfo  {journal} {Rev. Mod. Phys.}\ }\textbf {\bibinfo {volume} {85}},\ \bibinfo {pages} {1421} (\bibinfo {year} {2013})}\BibitemShut {NoStop}%
\bibitem [{\citenamefont {Kaneko}\ \emph {et~al.}(2016)\citenamefont {Kaneko}, \citenamefont {Nakamura},\ and\ \citenamefont {Okazaki}}]{Kaneko2016}%
  \BibitemOpen
  \bibfield  {author} {\bibinfo {author} {\bibfnamefont {N.-H.}\ \bibnamefont {Kaneko}}, \bibinfo {author} {\bibfnamefont {S.}~\bibnamefont {Nakamura}}, \ and\ \bibinfo {author} {\bibfnamefont {Y.}~\bibnamefont {Okazaki}},\ }\href {\doibase 10.1088/0957-0233/27/3/032001} {\bibfield  {journal} {\bibinfo  {journal} {Measurement Science and Technology}\ }\textbf {\bibinfo {volume} {27}},\ \bibinfo {pages} {032001} (\bibinfo {year} {2016})}\BibitemShut {NoStop}%
\bibitem [{\citenamefont {Stein}\ \emph {et~al.}(2015)\citenamefont {Stein}, \citenamefont {Drung}, \citenamefont {Fricke}, \citenamefont {Scherer}, \citenamefont {Hohls}, \citenamefont {Leicht}, \citenamefont {G\"{o}tz}, \citenamefont {Krause}, \citenamefont {Behr}, \citenamefont {Pesel}, \citenamefont {Pierz}, \citenamefont {Siegner}, \citenamefont {Ahlers},\ and\ \citenamefont {Schumacher}}]{Stein2015}%
  \BibitemOpen
  \bibfield  {author} {\bibinfo {author} {\bibfnamefont {F.}~\bibnamefont {Stein}}, \bibinfo {author} {\bibfnamefont {D.}~\bibnamefont {Drung}}, \bibinfo {author} {\bibfnamefont {L.}~\bibnamefont {Fricke}}, \bibinfo {author} {\bibfnamefont {H.}~\bibnamefont {Scherer}}, \bibinfo {author} {\bibfnamefont {F.}~\bibnamefont {Hohls}}, \bibinfo {author} {\bibfnamefont {C.}~\bibnamefont {Leicht}}, \bibinfo {author} {\bibfnamefont {M.}~\bibnamefont {G\"{o}tz}}, \bibinfo {author} {\bibfnamefont {C.}~\bibnamefont {Krause}}, \bibinfo {author} {\bibfnamefont {R.}~\bibnamefont {Behr}}, \bibinfo {author} {\bibfnamefont {E.}~\bibnamefont {Pesel}}, \bibinfo {author} {\bibfnamefont {K.}~\bibnamefont {Pierz}}, \bibinfo {author} {\bibfnamefont {U.}~\bibnamefont {Siegner}}, \bibinfo {author} {\bibfnamefont {F.~J.}\ \bibnamefont {Ahlers}}, \ and\ \bibinfo {author} {\bibfnamefont {H.~W.}\ \bibnamefont {Schumacher}},\ }\href {\doibase 10.1063/1.4930142} {\bibfield  {journal} {\bibinfo  {journal} {Applied Physics Letters}\ }\textbf
  {\bibinfo {volume} {107}} (\bibinfo {year} {2015}),\ 10.1063/1.4930142}\BibitemShut {NoStop}%
\bibitem [{\citenamefont {Zhao}\ \emph {et~al.}(2017)\citenamefont {Zhao}, \citenamefont {Rossi}, \citenamefont {Giblin}, \citenamefont {Fletcher}, \citenamefont {Hudson}, \citenamefont {M\"ott\"onen}, \citenamefont {Kataoka},\ and\ \citenamefont {Dzurak}}]{PhysRevApplied.8.044021}%
  \BibitemOpen
  \bibfield  {author} {\bibinfo {author} {\bibfnamefont {R.}~\bibnamefont {Zhao}}, \bibinfo {author} {\bibfnamefont {A.}~\bibnamefont {Rossi}}, \bibinfo {author} {\bibfnamefont {S.~P.}\ \bibnamefont {Giblin}}, \bibinfo {author} {\bibfnamefont {J.~D.}\ \bibnamefont {Fletcher}}, \bibinfo {author} {\bibfnamefont {F.~E.}\ \bibnamefont {Hudson}}, \bibinfo {author} {\bibfnamefont {M.}~\bibnamefont {M\"ott\"onen}}, \bibinfo {author} {\bibfnamefont {M.}~\bibnamefont {Kataoka}}, \ and\ \bibinfo {author} {\bibfnamefont {A.~S.}\ \bibnamefont {Dzurak}},\ }\href {\doibase 10.1103/PhysRevApplied.8.044021} {\bibfield  {journal} {\bibinfo  {journal} {Phys. Rev. Appl.}\ }\textbf {\bibinfo {volume} {8}},\ \bibinfo {pages} {044021} (\bibinfo {year} {2017})}\BibitemShut {NoStop}%
\bibitem [{\citenamefont {Giblin}\ \emph {et~al.}(2019)\citenamefont {Giblin}, \citenamefont {Fujiwara}, \citenamefont {Yamahata}, \citenamefont {Bae}, \citenamefont {Kim}, \citenamefont {Rossi}, \citenamefont {M\"{o}tt\"{o}nen},\ and\ \citenamefont {Kataoka}}]{Giblin2019}%
  \BibitemOpen
  \bibfield  {author} {\bibinfo {author} {\bibfnamefont {S.~P.}\ \bibnamefont {Giblin}}, \bibinfo {author} {\bibfnamefont {A.}~\bibnamefont {Fujiwara}}, \bibinfo {author} {\bibfnamefont {G.}~\bibnamefont {Yamahata}}, \bibinfo {author} {\bibfnamefont {M.-H.}\ \bibnamefont {Bae}}, \bibinfo {author} {\bibfnamefont {N.}~\bibnamefont {Kim}}, \bibinfo {author} {\bibfnamefont {A.}~\bibnamefont {Rossi}}, \bibinfo {author} {\bibfnamefont {M.}~\bibnamefont {M\"{o}tt\"{o}nen}}, \ and\ \bibinfo {author} {\bibfnamefont {M.}~\bibnamefont {Kataoka}},\ }\href {\doibase 10.1088/1681-7575/ab29a5} {\bibfield  {journal} {\bibinfo  {journal} {Metrologia}\ }\textbf {\bibinfo {volume} {56}},\ \bibinfo {pages} {044004} (\bibinfo {year} {2019})}\BibitemShut {NoStop}%
\bibitem [{\citenamefont {Das}\ \emph {et~al.}(2006)\citenamefont {Das}, \citenamefont {Kim},\ and\ \citenamefont {Mizel}}]{PhysRevLett.97.096602}%
  \BibitemOpen
  \bibfield  {author} {\bibinfo {author} {\bibfnamefont {K.~K.}\ \bibnamefont {Das}}, \bibinfo {author} {\bibfnamefont {S.}~\bibnamefont {Kim}}, \ and\ \bibinfo {author} {\bibfnamefont {A.}~\bibnamefont {Mizel}},\ }\href {\doibase 10.1103/PhysRevLett.97.096602} {\bibfield  {journal} {\bibinfo  {journal} {Phys. Rev. Lett.}\ }\textbf {\bibinfo {volume} {97}},\ \bibinfo {pages} {096602} (\bibinfo {year} {2006})}\BibitemShut {NoStop}%
\bibitem [{\citenamefont {You}\ \emph {et~al.}(2022)\citenamefont {You}, \citenamefont {Liang}, \citenamefont {Xie}, \citenamefont {Gao}, \citenamefont {Ye}, \citenamefont {Zhu},\ and\ \citenamefont {Zhang}}]{PhysRevLett.128.244302}%
  \BibitemOpen
  \bibfield  {author} {\bibinfo {author} {\bibfnamefont {O.}~\bibnamefont {You}}, \bibinfo {author} {\bibfnamefont {S.}~\bibnamefont {Liang}}, \bibinfo {author} {\bibfnamefont {B.}~\bibnamefont {Xie}}, \bibinfo {author} {\bibfnamefont {W.}~\bibnamefont {Gao}}, \bibinfo {author} {\bibfnamefont {W.}~\bibnamefont {Ye}}, \bibinfo {author} {\bibfnamefont {J.}~\bibnamefont {Zhu}}, \ and\ \bibinfo {author} {\bibfnamefont {S.}~\bibnamefont {Zhang}},\ }\href {\doibase 10.1103/PhysRevLett.128.244302} {\bibfield  {journal} {\bibinfo  {journal} {Phys. Rev. Lett.}\ }\textbf {\bibinfo {volume} {128}},\ \bibinfo {pages} {244302} (\bibinfo {year} {2022})}\BibitemShut {NoStop}%
\bibitem [{\citenamefont {B\"{a}uerle}\ \emph {et~al.}(2018)\citenamefont {B\"{a}uerle}, \citenamefont {Christian~Glattli}, \citenamefont {Meunier}, \citenamefont {Portier}, \citenamefont {Roche}, \citenamefont {Roulleau}, \citenamefont {Takada},\ and\ \citenamefont {Waintal}}]{Buerle2018}%
  \BibitemOpen
  \bibfield  {author} {\bibinfo {author} {\bibfnamefont {C.}~\bibnamefont {B\"{a}uerle}}, \bibinfo {author} {\bibfnamefont {D.}~\bibnamefont {Christian~Glattli}}, \bibinfo {author} {\bibfnamefont {T.}~\bibnamefont {Meunier}}, \bibinfo {author} {\bibfnamefont {F.}~\bibnamefont {Portier}}, \bibinfo {author} {\bibfnamefont {P.}~\bibnamefont {Roche}}, \bibinfo {author} {\bibfnamefont {P.}~\bibnamefont {Roulleau}}, \bibinfo {author} {\bibfnamefont {S.}~\bibnamefont {Takada}}, \ and\ \bibinfo {author} {\bibfnamefont {X.}~\bibnamefont {Waintal}},\ }\href {\doibase 10.1088/1361-6633/aaa98a} {\bibfield  {journal} {\bibinfo  {journal} {Reports on Progress in Physics}\ }\textbf {\bibinfo {volume} {81}},\ \bibinfo {pages} {056503} (\bibinfo {year} {2018})}\BibitemShut {NoStop}%
\bibitem [{\citenamefont {Fève}\ \emph {et~al.}(2007)\citenamefont {Fève}, \citenamefont {Mahé}, \citenamefont {Berroir}, \citenamefont {Kontos}, \citenamefont {Plaçais}, \citenamefont {Glattli}, \citenamefont {Cavanna}, \citenamefont {Etienne},\ and\ \citenamefont {Jin}}]{Feve2007}%
  \BibitemOpen
  \bibfield  {author} {\bibinfo {author} {\bibfnamefont {G.}~\bibnamefont {Fève}}, \bibinfo {author} {\bibfnamefont {A.}~\bibnamefont {Mahé}}, \bibinfo {author} {\bibfnamefont {J.-M.}\ \bibnamefont {Berroir}}, \bibinfo {author} {\bibfnamefont {T.}~\bibnamefont {Kontos}}, \bibinfo {author} {\bibfnamefont {B.}~\bibnamefont {Plaçais}}, \bibinfo {author} {\bibfnamefont {D.~C.}\ \bibnamefont {Glattli}}, \bibinfo {author} {\bibfnamefont {A.}~\bibnamefont {Cavanna}}, \bibinfo {author} {\bibfnamefont {B.}~\bibnamefont {Etienne}}, \ and\ \bibinfo {author} {\bibfnamefont {Y.}~\bibnamefont {Jin}},\ }\href {\doibase 10.1126/science.1141243} {\bibfield  {journal} {\bibinfo  {journal} {Science}\ }\textbf {\bibinfo {volume} {316}},\ \bibinfo {pages} {1169–1172} (\bibinfo {year} {2007})}\BibitemShut {NoStop}%
\bibitem [{\citenamefont {Lanzillotti-Kimura}\ \emph {et~al.}(2011)\citenamefont {Lanzillotti-Kimura}, \citenamefont {Fainstein}, \citenamefont {Perrin},\ and\ \citenamefont {Jusserand}}]{PhysRevB.84.064307}%
  \BibitemOpen
  \bibfield  {author} {\bibinfo {author} {\bibfnamefont {N.~D.}\ \bibnamefont {Lanzillotti-Kimura}}, \bibinfo {author} {\bibfnamefont {A.}~\bibnamefont {Fainstein}}, \bibinfo {author} {\bibfnamefont {B.}~\bibnamefont {Perrin}}, \ and\ \bibinfo {author} {\bibfnamefont {B.}~\bibnamefont {Jusserand}},\ }\href {\doibase 10.1103/PhysRevB.84.064307} {\bibfield  {journal} {\bibinfo  {journal} {Phys. Rev. B}\ }\textbf {\bibinfo {volume} {84}},\ \bibinfo {pages} {064307} (\bibinfo {year} {2011})}\BibitemShut {NoStop}%
\bibitem [{\citenamefont {Eisfeld}\ and\ \citenamefont {Renk}(1979)}]{10.1063/1.90855}%
  \BibitemOpen
  \bibfield  {author} {\bibinfo {author} {\bibfnamefont {W.}~\bibnamefont {Eisfeld}}\ and\ \bibinfo {author} {\bibfnamefont {K.~F.}\ \bibnamefont {Renk}},\ }\href {\doibase 10.1063/1.90855} {\bibfield  {journal} {\bibinfo  {journal} {Applied Physics Letters}\ }\textbf {\bibinfo {volume} {34}},\ \bibinfo {pages} {481} (\bibinfo {year} {1979})}\BibitemShut {NoStop}%
\bibitem [{\citenamefont {Giorgianni}\ \emph {et~al.}(2022)\citenamefont {Giorgianni}, \citenamefont {Udina}, \citenamefont {Cea}, \citenamefont {Paris}, \citenamefont {Caputo}, \citenamefont {Radovic}, \citenamefont {Boie}, \citenamefont {Sakai}, \citenamefont {Schneider},\ and\ \citenamefont {Johnson}}]{pump-probe-1}%
  \BibitemOpen
  \bibfield  {author} {\bibinfo {author} {\bibfnamefont {F.}~\bibnamefont {Giorgianni}}, \bibinfo {author} {\bibfnamefont {M.}~\bibnamefont {Udina}}, \bibinfo {author} {\bibfnamefont {T.}~\bibnamefont {Cea}}, \bibinfo {author} {\bibfnamefont {E.}~\bibnamefont {Paris}}, \bibinfo {author} {\bibfnamefont {M.}~\bibnamefont {Caputo}}, \bibinfo {author} {\bibfnamefont {M.}~\bibnamefont {Radovic}}, \bibinfo {author} {\bibfnamefont {L.}~\bibnamefont {Boie}}, \bibinfo {author} {\bibfnamefont {J.}~\bibnamefont {Sakai}}, \bibinfo {author} {\bibfnamefont {C.~W.}\ \bibnamefont {Schneider}}, \ and\ \bibinfo {author} {\bibfnamefont {S.~L.}\ \bibnamefont {Johnson}},\ }\href {\doibase 10.1038/s42005-022-00882-7} {\bibfield  {journal} {\bibinfo  {journal} {Communications Physics}\ }\textbf {\bibinfo {volume} {5}} (\bibinfo {year} {2022}),\ 10.1038/s42005-022-00882-7}\BibitemShut {NoStop}%
\bibitem [{\citenamefont {Yoon}\ \emph {et~al.}(2023{\natexlab{a}})\citenamefont {Yoon}, \citenamefont {Lu}, \citenamefont {Uzundal}, \citenamefont {Qi}, \citenamefont {Zhao}, \citenamefont {Chen}, \citenamefont {Feng}, \citenamefont {Watanabe}, \citenamefont {Taniguchi}, \citenamefont {Crommie},\ and\ \citenamefont {Wang}}]{https://doi.org/10.48550/arxiv.2310.04939}%
  \BibitemOpen
  \bibfield  {author} {\bibinfo {author} {\bibfnamefont {Y.}~\bibnamefont {Yoon}}, \bibinfo {author} {\bibfnamefont {Z.}~\bibnamefont {Lu}}, \bibinfo {author} {\bibfnamefont {C.}~\bibnamefont {Uzundal}}, \bibinfo {author} {\bibfnamefont {R.}~\bibnamefont {Qi}}, \bibinfo {author} {\bibfnamefont {W.}~\bibnamefont {Zhao}}, \bibinfo {author} {\bibfnamefont {S.}~\bibnamefont {Chen}}, \bibinfo {author} {\bibfnamefont {Q.}~\bibnamefont {Feng}}, \bibinfo {author} {\bibfnamefont {K.}~\bibnamefont {Watanabe}}, \bibinfo {author} {\bibfnamefont {T.}~\bibnamefont {Taniguchi}}, \bibinfo {author} {\bibfnamefont {M.~F.}\ \bibnamefont {Crommie}}, \ and\ \bibinfo {author} {\bibfnamefont {F.}~\bibnamefont {Wang}},\ }\href {\doibase 10.48550/ARXIV.2310.04939} {\enquote {\bibinfo {title} {Terahertz phonon engineering and spectroscopy with van der waals heterostructures},}\ } (\bibinfo {year} {2023}{\natexlab{a}})\BibitemShut {NoStop}%
\bibitem [{\citenamefont {Yoon}\ \emph {et~al.}(2023{\natexlab{b}})\citenamefont {Yoon}, \citenamefont {Lu}, \citenamefont {Uzundal}, \citenamefont {Qi}, \citenamefont {Zhao}, \citenamefont {Chen}, \citenamefont {Feng}, \citenamefont {Watanabe}, \citenamefont {Taniguchi}, \citenamefont {Crommie},\ and\ \citenamefont {Wang}}]{Yoon2023}%
  \BibitemOpen
  \bibfield  {author} {\bibinfo {author} {\bibfnamefont {Y.}~\bibnamefont {Yoon}}, \bibinfo {author} {\bibfnamefont {Z.}~\bibnamefont {Lu}}, \bibinfo {author} {\bibfnamefont {C.}~\bibnamefont {Uzundal}}, \bibinfo {author} {\bibfnamefont {R.}~\bibnamefont {Qi}}, \bibinfo {author} {\bibfnamefont {W.}~\bibnamefont {Zhao}}, \bibinfo {author} {\bibfnamefont {S.}~\bibnamefont {Chen}}, \bibinfo {author} {\bibfnamefont {Q.}~\bibnamefont {Feng}}, \bibinfo {author} {\bibfnamefont {K.}~\bibnamefont {Watanabe}}, \bibinfo {author} {\bibfnamefont {T.}~\bibnamefont {Taniguchi}}, \bibinfo {author} {\bibfnamefont {M.~F.}\ \bibnamefont {Crommie}}, \ and\ \bibinfo {author} {\bibfnamefont {F.}~\bibnamefont {Wang}},\ }\href {\doibase 10.48550/ARXIV.2310.04939} {\enquote {\bibinfo {title} {Terahertz phonon engineering and spectroscopy with van der waals heterostructures},}\ } (\bibinfo {year} {2023}{\natexlab{b}})\BibitemShut {NoStop}%
\bibitem [{\citenamefont {Sanders}\ \emph {et~al.}(2009)\citenamefont {Sanders}, \citenamefont {Stanton}, \citenamefont {Kim}, \citenamefont {Yee}, \citenamefont {Lim}, \citenamefont {H\'aroz}, \citenamefont {Booshehri}, \citenamefont {Kono},\ and\ \citenamefont {Saito}}]{PhysRevB.79.205434}%
  \BibitemOpen
  \bibfield  {author} {\bibinfo {author} {\bibfnamefont {G.~D.}\ \bibnamefont {Sanders}}, \bibinfo {author} {\bibfnamefont {C.~J.}\ \bibnamefont {Stanton}}, \bibinfo {author} {\bibfnamefont {J.-H.}\ \bibnamefont {Kim}}, \bibinfo {author} {\bibfnamefont {K.-J.}\ \bibnamefont {Yee}}, \bibinfo {author} {\bibfnamefont {Y.-S.}\ \bibnamefont {Lim}}, \bibinfo {author} {\bibfnamefont {E.~H.}\ \bibnamefont {H\'aroz}}, \bibinfo {author} {\bibfnamefont {L.~G.}\ \bibnamefont {Booshehri}}, \bibinfo {author} {\bibfnamefont {J.}~\bibnamefont {Kono}}, \ and\ \bibinfo {author} {\bibfnamefont {R.}~\bibnamefont {Saito}},\ }\href {\doibase 10.1103/PhysRevB.79.205434} {\bibfield  {journal} {\bibinfo  {journal} {Phys. Rev. B}\ }\textbf {\bibinfo {volume} {79}},\ \bibinfo {pages} {205434} (\bibinfo {year} {2009})}\BibitemShut {NoStop}%
\bibitem [{\citenamefont {Fathi}(2011)}]{cnt-review}%
  \BibitemOpen
  \bibfield  {author} {\bibinfo {author} {\bibfnamefont {D.}~\bibnamefont {Fathi}},\ }\href {\doibase 10.1155/2011/471241} {\bibfield  {journal} {\bibinfo  {journal} {Journal of Nanotechnology}\ }\textbf {\bibinfo {volume} {2011}} (\bibinfo {year} {2011}),\ 10.1155/2011/471241}\BibitemShut {NoStop}%
\bibitem [{\citenamefont {Kane}\ and\ \citenamefont {Mele}(1997)}]{PhysRevLett.78.1932}%
  \BibitemOpen
  \bibfield  {author} {\bibinfo {author} {\bibfnamefont {C.~L.}\ \bibnamefont {Kane}}\ and\ \bibinfo {author} {\bibfnamefont {E.~J.}\ \bibnamefont {Mele}},\ }\href {\doibase 10.1103/PhysRevLett.78.1932} {\bibfield  {journal} {\bibinfo  {journal} {Phys. Rev. Lett.}\ }\textbf {\bibinfo {volume} {78}},\ \bibinfo {pages} {1932} (\bibinfo {year} {1997})}\BibitemShut {NoStop}%
\bibitem [{\citenamefont {Saito}\ \emph {et~al.}(1992)\citenamefont {Saito}, \citenamefont {Fujita}, \citenamefont {Dresselhaus},\ and\ \citenamefont {Dresselhaus}}]{PhysRevB.46.1804}%
  \BibitemOpen
  \bibfield  {author} {\bibinfo {author} {\bibfnamefont {R.}~\bibnamefont {Saito}}, \bibinfo {author} {\bibfnamefont {M.}~\bibnamefont {Fujita}}, \bibinfo {author} {\bibfnamefont {G.}~\bibnamefont {Dresselhaus}}, \ and\ \bibinfo {author} {\bibfnamefont {M.~S.}\ \bibnamefont {Dresselhaus}},\ }\href {\doibase 10.1103/PhysRevB.46.1804} {\bibfield  {journal} {\bibinfo  {journal} {Phys. Rev. B}\ }\textbf {\bibinfo {volume} {46}},\ \bibinfo {pages} {1804} (\bibinfo {year} {1992})}\BibitemShut {NoStop}%
\bibitem [{\citenamefont {Mintmire}\ \emph {et~al.}(1992)\citenamefont {Mintmire}, \citenamefont {Dunlap},\ and\ \citenamefont {White}}]{PhysRevLett.68.631}%
  \BibitemOpen
  \bibfield  {author} {\bibinfo {author} {\bibfnamefont {J.~W.}\ \bibnamefont {Mintmire}}, \bibinfo {author} {\bibfnamefont {B.~I.}\ \bibnamefont {Dunlap}}, \ and\ \bibinfo {author} {\bibfnamefont {C.~T.}\ \bibnamefont {White}},\ }\href {\doibase 10.1103/PhysRevLett.68.631} {\bibfield  {journal} {\bibinfo  {journal} {Phys. Rev. Lett.}\ }\textbf {\bibinfo {volume} {68}},\ \bibinfo {pages} {631} (\bibinfo {year} {1992})}\BibitemShut {NoStop}%
\bibitem [{\citenamefont {Hamada}\ \emph {et~al.}(1992)\citenamefont {Hamada}, \citenamefont {Sawada},\ and\ \citenamefont {Oshiyama}}]{PhysRevLett.68.1579}%
  \BibitemOpen
  \bibfield  {author} {\bibinfo {author} {\bibfnamefont {N.}~\bibnamefont {Hamada}}, \bibinfo {author} {\bibfnamefont {S.-i.}\ \bibnamefont {Sawada}}, \ and\ \bibinfo {author} {\bibfnamefont {A.}~\bibnamefont {Oshiyama}},\ }\href {\doibase 10.1103/PhysRevLett.68.1579} {\bibfield  {journal} {\bibinfo  {journal} {Phys. Rev. Lett.}\ }\textbf {\bibinfo {volume} {68}},\ \bibinfo {pages} {1579} (\bibinfo {year} {1992})}\BibitemShut {NoStop}%
\bibitem [{Sup()}]{SupplementalMaterials}%
  \BibitemOpen
  \href@noop {} {}\bibinfo {note} {See Supplemental Material at [url] for details of derivations and numerical calculations, which includes Refs. \cite{Sato2019MicroscopicGraphene,PhysRevLett.73.902}.}\BibitemShut {Stop}%
\bibitem [{\citenamefont {von Oppen}\ \emph {et~al.}(2009)\citenamefont {von Oppen}, \citenamefont {Guinea},\ and\ \citenamefont {Mariani}}]{PhysRevB.80.075420}%
  \BibitemOpen
  \bibfield  {author} {\bibinfo {author} {\bibfnamefont {F.}~\bibnamefont {von Oppen}}, \bibinfo {author} {\bibfnamefont {F.}~\bibnamefont {Guinea}}, \ and\ \bibinfo {author} {\bibfnamefont {E.}~\bibnamefont {Mariani}},\ }\href {\doibase 10.1103/PhysRevB.80.075420} {\bibfield  {journal} {\bibinfo  {journal} {Phys. Rev. B}\ }\textbf {\bibinfo {volume} {80}},\ \bibinfo {pages} {075420} (\bibinfo {year} {2009})}\BibitemShut {NoStop}%
\bibitem [{Lon()}]{Longitudinal}%
  \BibitemOpen
  \href@noop {} {}\bibinfo {note} {While we consider coherent longitudinal phonon modes, coherent transverse modes should induce similar behavior.}\BibitemShut {Stop}%
\bibitem [{Com()}]{Commensurate}%
  \BibitemOpen
  \href@noop {} {}\bibinfo {note} {For simplicity, we assume that the wavelength of the coherent phonon mode is commensurate with the periodicity of the CNT along the tube axis.}\BibitemShut {Stop}%
\bibitem [{\citenamefont {Rudner}\ and\ \citenamefont {Lindner}(2020)}]{floquethandbook}%
  \BibitemOpen
  \bibfield  {author} {\bibinfo {author} {\bibfnamefont {M.~S.}\ \bibnamefont {Rudner}}\ and\ \bibinfo {author} {\bibfnamefont {N.~H.}\ \bibnamefont {Lindner}},\ }\href {\doibase 10.48550/ARXIV.2003.08252} {\enquote {\bibinfo {title} {The floquet engineer's handbook},}\ } (\bibinfo {year} {2020})\BibitemShut {NoStop}%
\bibitem [{\citenamefont {Esin}\ \emph {et~al.}(2018)\citenamefont {Esin}, \citenamefont {Rudner}, \citenamefont {Refael},\ and\ \citenamefont {Lindner}}]{transport}%
  \BibitemOpen
  \bibfield  {author} {\bibinfo {author} {\bibfnamefont {I.}~\bibnamefont {Esin}}, \bibinfo {author} {\bibfnamefont {M.~S.}\ \bibnamefont {Rudner}}, \bibinfo {author} {\bibfnamefont {G.}~\bibnamefont {Refael}}, \ and\ \bibinfo {author} {\bibfnamefont {N.~H.}\ \bibnamefont {Lindner}},\ }\href {\doibase 10.1103/PhysRevB.97.245401} {\bibfield  {journal} {\bibinfo  {journal} {Phys. Rev. B}\ }\textbf {\bibinfo {volume} {97}},\ \bibinfo {pages} {245401} (\bibinfo {year} {2018})}\BibitemShut {NoStop}%
\bibitem [{Spi()}]{SpinDegen}%
  \BibitemOpen
  \href@noop {} {}\bibinfo {note} {The factor of two in Eq. (\ref{eq:current-floquet}) accounts for spin degeneracy.}\BibitemShut {Stop}%
\bibitem [{\citenamefont {Zhao}\ and\ \citenamefont {Mazumdar}(2004)}]{PhysRevLett.93.157402}%
  \BibitemOpen
  \bibfield  {author} {\bibinfo {author} {\bibfnamefont {H.}~\bibnamefont {Zhao}}\ and\ \bibinfo {author} {\bibfnamefont {S.}~\bibnamefont {Mazumdar}},\ }\href {\doibase 10.1103/PhysRevLett.93.157402} {\bibfield  {journal} {\bibinfo  {journal} {Phys. Rev. Lett.}\ }\textbf {\bibinfo {volume} {93}},\ \bibinfo {pages} {157402} (\bibinfo {year} {2004})}\BibitemShut {NoStop}%
\bibitem [{\citenamefont {Malic}\ \emph {et~al.}(2008)\citenamefont {Malic}, \citenamefont {Hirtschulz}, \citenamefont {Milde}, \citenamefont {Richter}, \citenamefont {Maultzsch}, \citenamefont {Reich},\ and\ \citenamefont {Knorr}}]{coulomb-cnt}%
  \BibitemOpen
  \bibfield  {author} {\bibinfo {author} {\bibfnamefont {E.}~\bibnamefont {Malic}}, \bibinfo {author} {\bibfnamefont {M.}~\bibnamefont {Hirtschulz}}, \bibinfo {author} {\bibfnamefont {F.}~\bibnamefont {Milde}}, \bibinfo {author} {\bibfnamefont {M.}~\bibnamefont {Richter}}, \bibinfo {author} {\bibfnamefont {J.}~\bibnamefont {Maultzsch}}, \bibinfo {author} {\bibfnamefont {S.}~\bibnamefont {Reich}}, \ and\ \bibinfo {author} {\bibfnamefont {A.}~\bibnamefont {Knorr}},\ }\href {\doibase 10.1002/pssb.200879612} {\bibfield  {journal} {\bibinfo  {journal} {physica status solidi (b)}\ }\textbf {\bibinfo {volume} {245}},\ \bibinfo {pages} {2155 } (\bibinfo {year} {2008})}\BibitemShut {NoStop}%
\bibitem [{\citenamefont {Seetharam}\ \emph {et~al.}(2019)\citenamefont {Seetharam}, \citenamefont {Bardyn}, \citenamefont {Lindner}, \citenamefont {Rudner},\ and\ \citenamefont {Refael}}]{fbe_adv}%
  \BibitemOpen
  \bibfield  {author} {\bibinfo {author} {\bibfnamefont {K.~I.}\ \bibnamefont {Seetharam}}, \bibinfo {author} {\bibfnamefont {C.-E.}\ \bibnamefont {Bardyn}}, \bibinfo {author} {\bibfnamefont {N.~H.}\ \bibnamefont {Lindner}}, \bibinfo {author} {\bibfnamefont {M.~S.}\ \bibnamefont {Rudner}}, \ and\ \bibinfo {author} {\bibfnamefont {G.}~\bibnamefont {Refael}},\ }\href {\doibase 10.1103/PhysRevB.99.014307} {\bibfield  {journal} {\bibinfo  {journal} {Phys. Rev. B}\ }\textbf {\bibinfo {volume} {99}},\ \bibinfo {pages} {014307} (\bibinfo {year} {2019})}\BibitemShut {NoStop}%
\bibitem [{\citenamefont {Seetharam}\ \emph {et~al.}(2015)\citenamefont {Seetharam}, \citenamefont {Bardyn}, \citenamefont {Lindner}, \citenamefont {Rudner},\ and\ \citenamefont {Refael}}]{fbe_orig}%
  \BibitemOpen
  \bibfield  {author} {\bibinfo {author} {\bibfnamefont {K.~I.}\ \bibnamefont {Seetharam}}, \bibinfo {author} {\bibfnamefont {C.-E.}\ \bibnamefont {Bardyn}}, \bibinfo {author} {\bibfnamefont {N.~H.}\ \bibnamefont {Lindner}}, \bibinfo {author} {\bibfnamefont {M.~S.}\ \bibnamefont {Rudner}}, \ and\ \bibinfo {author} {\bibfnamefont {G.}~\bibnamefont {Refael}},\ }\href {\doibase 10.1103/PhysRevX.5.041050} {\bibfield  {journal} {\bibinfo  {journal} {Phys. Rev. X}\ }\textbf {\bibinfo {volume} {5}},\ \bibinfo {pages} {041050} (\bibinfo {year} {2015})}\BibitemShut {NoStop}%
\bibitem [{\citenamefont {Genske}\ and\ \citenamefont {Rosch}(2015)}]{PhysRevA.92.062108}%
  \BibitemOpen
  \bibfield  {author} {\bibinfo {author} {\bibfnamefont {M.}~\bibnamefont {Genske}}\ and\ \bibinfo {author} {\bibfnamefont {A.}~\bibnamefont {Rosch}},\ }\href {\doibase 10.1103/PhysRevA.92.062108} {\bibfield  {journal} {\bibinfo  {journal} {Phys. Rev. A}\ }\textbf {\bibinfo {volume} {92}},\ \bibinfo {pages} {062108} (\bibinfo {year} {2015})}\BibitemShut {NoStop}%
\bibitem [{\citenamefont {Hone}\ \emph {et~al.}(2009)\citenamefont {Hone}, \citenamefont {Ketzmerick},\ and\ \citenamefont {Kohn}}]{PhysRevE.79.051129}%
  \BibitemOpen
  \bibfield  {author} {\bibinfo {author} {\bibfnamefont {D.~W.}\ \bibnamefont {Hone}}, \bibinfo {author} {\bibfnamefont {R.}~\bibnamefont {Ketzmerick}}, \ and\ \bibinfo {author} {\bibfnamefont {W.}~\bibnamefont {Kohn}},\ }\href {\doibase 10.1103/PhysRevE.79.051129} {\bibfield  {journal} {\bibinfo  {journal} {Phys. Rev. E}\ }\textbf {\bibinfo {volume} {79}},\ \bibinfo {pages} {051129} (\bibinfo {year} {2009})}\BibitemShut {NoStop}%
\bibitem [{\citenamefont {Kohn}(2001)}]{Kohn2001PeriodicThermodynamics}%
  \BibitemOpen
  \bibfield  {author} {\bibinfo {author} {\bibfnamefont {W.}~\bibnamefont {Kohn}},\ }\href {\doibase 10.1023/A:1010327828445} {\bibfield  {journal} {\bibinfo  {journal} {Journal of Statistical Physics}\ }\textbf {\bibinfo {volume} {103}},\ \bibinfo {pages} {417} (\bibinfo {year} {2001})}\BibitemShut {NoStop}%
\bibitem [{\citenamefont {Esin}\ \emph {et~al.}(2021)\citenamefont {Esin}, \citenamefont {Gupta}, \citenamefont {Berg}, \citenamefont {Rudner},\ and\ \citenamefont {Lindner}}]{gyro}%
  \BibitemOpen
  \bibfield  {author} {\bibinfo {author} {\bibfnamefont {I.}~\bibnamefont {Esin}}, \bibinfo {author} {\bibfnamefont {G.}~\bibnamefont {Gupta}}, \bibinfo {author} {\bibfnamefont {E.}~\bibnamefont {Berg}}, \bibinfo {author} {\bibfnamefont {M.}~\bibnamefont {Rudner}}, \ and\ \bibinfo {author} {\bibfnamefont {N.}~\bibnamefont {Lindner}},\ }\href {\doibase 10.1038/s41467-021-25511-9} {\bibfield  {journal} {\bibinfo  {journal} {Nature Communications}\ }\textbf {\bibinfo {volume} {12}} (\bibinfo {year} {2021}),\ 10.1038/s41467-021-25511-9}\BibitemShut {NoStop}%
\bibitem [{\citenamefont {Khrapak}(2020)}]{PhysRevResearch.2.012040}%
  \BibitemOpen
  \bibfield  {author} {\bibinfo {author} {\bibfnamefont {S.~A.}\ \bibnamefont {Khrapak}},\ }\href {\doibase 10.1103/PhysRevResearch.2.012040} {\bibfield  {journal} {\bibinfo  {journal} {Phys. Rev. Res.}\ }\textbf {\bibinfo {volume} {2}},\ \bibinfo {pages} {012040} (\bibinfo {year} {2020})}\BibitemShut {NoStop}%
\bibitem [{\citenamefont {Lindemann}(1910)}]{lindemann-orig}%
  \BibitemOpen
  \bibfield  {author} {\bibinfo {author} {\bibfnamefont {F.}~\bibnamefont {Lindemann}},\ }\href@noop {} {\bibfield  {journal} {\bibinfo  {journal} {Z. Phys.}\ }\textbf {\bibinfo {volume} {11}},\ \bibinfo {pages} {609} (\bibinfo {year} {1910})}\BibitemShut {NoStop}%
\bibitem [{\citenamefont {Thouless}(1983)}]{PhysRevB.27.6083}%
  \BibitemOpen
  \bibfield  {author} {\bibinfo {author} {\bibfnamefont {D.~J.}\ \bibnamefont {Thouless}},\ }\href {\doibase 10.1103/PhysRevB.27.6083} {\bibfield  {journal} {\bibinfo  {journal} {Phys. Rev. B}\ }\textbf {\bibinfo {volume} {27}},\ \bibinfo {pages} {6083} (\bibinfo {year} {1983})}\BibitemShut {NoStop}%
\bibitem [{\citenamefont {Talyanskii}\ \emph {et~al.}(2001)\citenamefont {Talyanskii}, \citenamefont {Novikov}, \citenamefont {Simons},\ and\ \citenamefont {Levitov}}]{PhysRevLett.87.276802}%
  \BibitemOpen
  \bibfield  {author} {\bibinfo {author} {\bibfnamefont {V.~I.}\ \bibnamefont {Talyanskii}}, \bibinfo {author} {\bibfnamefont {D.~S.}\ \bibnamefont {Novikov}}, \bibinfo {author} {\bibfnamefont {B.~D.}\ \bibnamefont {Simons}}, \ and\ \bibinfo {author} {\bibfnamefont {L.~S.}\ \bibnamefont {Levitov}},\ }\href {\doibase 10.1103/PhysRevLett.87.276802} {\bibfield  {journal} {\bibinfo  {journal} {Phys. Rev. Lett.}\ }\textbf {\bibinfo {volume} {87}},\ \bibinfo {pages} {276802} (\bibinfo {year} {2001})}\BibitemShut {NoStop}%
\bibitem [{\citenamefont {Leek}\ \emph {et~al.}(2005)\citenamefont {Leek}, \citenamefont {Buitelaar}, \citenamefont {Talyanskii}, \citenamefont {Smith}, \citenamefont {Anderson}, \citenamefont {Jones}, \citenamefont {Wei},\ and\ \citenamefont {Cobden}}]{PhysRevLett.95.256802}%
  \BibitemOpen
  \bibfield  {author} {\bibinfo {author} {\bibfnamefont {P.~J.}\ \bibnamefont {Leek}}, \bibinfo {author} {\bibfnamefont {M.~R.}\ \bibnamefont {Buitelaar}}, \bibinfo {author} {\bibfnamefont {V.~I.}\ \bibnamefont {Talyanskii}}, \bibinfo {author} {\bibfnamefont {C.~G.}\ \bibnamefont {Smith}}, \bibinfo {author} {\bibfnamefont {D.}~\bibnamefont {Anderson}}, \bibinfo {author} {\bibfnamefont {G.~A.~C.}\ \bibnamefont {Jones}}, \bibinfo {author} {\bibfnamefont {J.}~\bibnamefont {Wei}}, \ and\ \bibinfo {author} {\bibfnamefont {D.~H.}\ \bibnamefont {Cobden}},\ }\href {\doibase 10.1103/PhysRevLett.95.256802} {\bibfield  {journal} {\bibinfo  {journal} {Phys. Rev. Lett.}\ }\textbf {\bibinfo {volume} {95}},\ \bibinfo {pages} {256802} (\bibinfo {year} {2005})}\BibitemShut {NoStop}%
\bibitem [{\citenamefont {Buitelaar}\ \emph {et~al.}(2006)\citenamefont {Buitelaar}, \citenamefont {Leek}, \citenamefont {Talyanskii}, \citenamefont {Smith}, \citenamefont {Anderson}, \citenamefont {Jones}, \citenamefont {Wei},\ and\ \citenamefont {Cobden}}]{Buitelaar2006}%
  \BibitemOpen
  \bibfield  {author} {\bibinfo {author} {\bibfnamefont {M.~R.}\ \bibnamefont {Buitelaar}}, \bibinfo {author} {\bibfnamefont {P.~J.}\ \bibnamefont {Leek}}, \bibinfo {author} {\bibfnamefont {V.~I.}\ \bibnamefont {Talyanskii}}, \bibinfo {author} {\bibfnamefont {C.~G.}\ \bibnamefont {Smith}}, \bibinfo {author} {\bibfnamefont {D.}~\bibnamefont {Anderson}}, \bibinfo {author} {\bibfnamefont {G.~A.~C.}\ \bibnamefont {Jones}}, \bibinfo {author} {\bibfnamefont {J.}~\bibnamefont {Wei}}, \ and\ \bibinfo {author} {\bibfnamefont {D.~H.}\ \bibnamefont {Cobden}},\ }\href {\doibase 10.1088/0268-1242/21/11/s10} {\bibfield  {journal} {\bibinfo  {journal} {Semiconductor Science and Technology}\ }\textbf {\bibinfo {volume} {21}},\ \bibinfo {pages} {S69–S77} (\bibinfo {year} {2006})}\BibitemShut {NoStop}%
\bibitem [{\citenamefont {Ahlers}\ \emph {et~al.}(2004)\citenamefont {Ahlers}, \citenamefont {Fletcher}, \citenamefont {Ebbecke},\ and\ \citenamefont {Janssen}}]{Ahlers2004}%
  \BibitemOpen
  \bibfield  {author} {\bibinfo {author} {\bibfnamefont {F.}~\bibnamefont {Ahlers}}, \bibinfo {author} {\bibfnamefont {N.}~\bibnamefont {Fletcher}}, \bibinfo {author} {\bibfnamefont {J.}~\bibnamefont {Ebbecke}}, \ and\ \bibinfo {author} {\bibfnamefont {T.}~\bibnamefont {Janssen}},\ }\href {\doibase 10.1016/j.cap.2004.01.012} {\bibfield  {journal} {\bibinfo  {journal} {Current Applied Physics}\ }\textbf {\bibinfo {volume} {4}},\ \bibinfo {pages} {529–533} (\bibinfo {year} {2004})}\BibitemShut {NoStop}%
\bibitem [{\citenamefont {Ebbecke}\ \emph {et~al.}(2004)\citenamefont {Ebbecke}, \citenamefont {Strobl},\ and\ \citenamefont {Wixforth}}]{PhysRevB.70.233401}%
  \BibitemOpen
  \bibfield  {author} {\bibinfo {author} {\bibfnamefont {J.}~\bibnamefont {Ebbecke}}, \bibinfo {author} {\bibfnamefont {C.~J.}\ \bibnamefont {Strobl}}, \ and\ \bibinfo {author} {\bibfnamefont {A.}~\bibnamefont {Wixforth}},\ }\href {\doibase 10.1103/PhysRevB.70.233401} {\bibfield  {journal} {\bibinfo  {journal} {Phys. Rev. B}\ }\textbf {\bibinfo {volume} {70}},\ \bibinfo {pages} {233401} (\bibinfo {year} {2004})}\BibitemShut {NoStop}%
\bibitem [{\citenamefont {Fletcher}\ \emph {et~al.}(2003)\citenamefont {Fletcher}, \citenamefont {Ebbecke}, \citenamefont {Janssen}, \citenamefont {Ahlers}, \citenamefont {Pepper}, \citenamefont {Beere},\ and\ \citenamefont {Ritchie}}]{PhysRevB.68.245310}%
  \BibitemOpen
  \bibfield  {author} {\bibinfo {author} {\bibfnamefont {N.~E.}\ \bibnamefont {Fletcher}}, \bibinfo {author} {\bibfnamefont {J.}~\bibnamefont {Ebbecke}}, \bibinfo {author} {\bibfnamefont {T.~J. B.~M.}\ \bibnamefont {Janssen}}, \bibinfo {author} {\bibfnamefont {F.~J.}\ \bibnamefont {Ahlers}}, \bibinfo {author} {\bibfnamefont {M.}~\bibnamefont {Pepper}}, \bibinfo {author} {\bibfnamefont {H.~E.}\ \bibnamefont {Beere}}, \ and\ \bibinfo {author} {\bibfnamefont {D.~A.}\ \bibnamefont {Ritchie}},\ }\href {\doibase 10.1103/PhysRevB.68.245310} {\bibfield  {journal} {\bibinfo  {journal} {Phys. Rev. B}\ }\textbf {\bibinfo {volume} {68}},\ \bibinfo {pages} {245310} (\bibinfo {year} {2003})}\BibitemShut {NoStop}%
\bibitem [{\citenamefont {Kundu}\ \emph {et~al.}(2020)\citenamefont {Kundu}, \citenamefont {Rudner}, \citenamefont {Berg},\ and\ \citenamefont {Lindner}}]{PhysRevB.101.041403}%
  \BibitemOpen
  \bibfield  {author} {\bibinfo {author} {\bibfnamefont {A.}~\bibnamefont {Kundu}}, \bibinfo {author} {\bibfnamefont {M.}~\bibnamefont {Rudner}}, \bibinfo {author} {\bibfnamefont {E.}~\bibnamefont {Berg}}, \ and\ \bibinfo {author} {\bibfnamefont {N.~H.}\ \bibnamefont {Lindner}},\ }\href {\doibase 10.1103/PhysRevB.101.041403} {\bibfield  {journal} {\bibinfo  {journal} {Phys. Rev. B}\ }\textbf {\bibinfo {volume} {101}},\ \bibinfo {pages} {041403} (\bibinfo {year} {2020})}\BibitemShut {NoStop}%
\bibitem [{\citenamefont {Shu}\ \emph {et~al.}(2023)\citenamefont {Shu}, \citenamefont {Zhang},\ and\ \citenamefont {Sun}}]{one-way-transport}%
  \BibitemOpen
  \bibfield  {author} {\bibinfo {author} {\bibfnamefont {C.}~\bibnamefont {Shu}}, \bibinfo {author} {\bibfnamefont {K.}~\bibnamefont {Zhang}}, \ and\ \bibinfo {author} {\bibfnamefont {K.}~\bibnamefont {Sun}},\ }\href {\doibase 10.48550/ARXIV.2306.10000} {\enquote {\bibinfo {title} {Loss-induced universal one-way transport in periodically driven systems},}\ } (\bibinfo {year} {2023})\BibitemShut {NoStop}%
\bibitem [{\citenamefont {Sato}\ \emph {et~al.}(2019)\citenamefont {Sato}, \citenamefont {McIver}, \citenamefont {Nuske}, \citenamefont {Tang}, \citenamefont {Jotzu}, \citenamefont {Schulte}, \citenamefont {H\"ubener}, \citenamefont {De~Giovannini}, \citenamefont {Mathey}, \citenamefont {Sentef}, \citenamefont {Cavalleri},\ and\ \citenamefont {Rubio}}]{Sato2019MicroscopicGraphene}%
  \BibitemOpen
  \bibfield  {author} {\bibinfo {author} {\bibfnamefont {S.~A.}\ \bibnamefont {Sato}}, \bibinfo {author} {\bibfnamefont {J.~W.}\ \bibnamefont {McIver}}, \bibinfo {author} {\bibfnamefont {M.}~\bibnamefont {Nuske}}, \bibinfo {author} {\bibfnamefont {P.}~\bibnamefont {Tang}}, \bibinfo {author} {\bibfnamefont {G.}~\bibnamefont {Jotzu}}, \bibinfo {author} {\bibfnamefont {B.}~\bibnamefont {Schulte}}, \bibinfo {author} {\bibfnamefont {H.}~\bibnamefont {H\"ubener}}, \bibinfo {author} {\bibfnamefont {U.}~\bibnamefont {De~Giovannini}}, \bibinfo {author} {\bibfnamefont {L.}~\bibnamefont {Mathey}}, \bibinfo {author} {\bibfnamefont {M.~A.}\ \bibnamefont {Sentef}}, \bibinfo {author} {\bibfnamefont {A.}~\bibnamefont {Cavalleri}}, \ and\ \bibinfo {author} {\bibfnamefont {A.}~\bibnamefont {Rubio}},\ }\href {\doibase 10.1103/PhysRevB.99.214302} {\bibfield  {journal} {\bibinfo  {journal} {Phys. Rev. B}\ }\textbf {\bibinfo {volume} {99}},\ \bibinfo {pages} {214302} (\bibinfo {year} {2019})}\BibitemShut {NoStop}%
\bibitem [{\citenamefont {Meier}\ \emph {et~al.}(1994)\citenamefont {Meier}, \citenamefont {von Plessen}, \citenamefont {Thomas},\ and\ \citenamefont {Koch}}]{PhysRevLett.73.902}%
  \BibitemOpen
  \bibfield  {author} {\bibinfo {author} {\bibfnamefont {T.}~\bibnamefont {Meier}}, \bibinfo {author} {\bibfnamefont {G.}~\bibnamefont {von Plessen}}, \bibinfo {author} {\bibfnamefont {P.}~\bibnamefont {Thomas}}, \ and\ \bibinfo {author} {\bibfnamefont {S.~W.}\ \bibnamefont {Koch}},\ }\href {\doibase 10.1103/PhysRevLett.73.902} {\bibfield  {journal} {\bibinfo  {journal} {Phys. Rev. Lett.}\ }\textbf {\bibinfo {volume} {73}},\ \bibinfo {pages} {902} (\bibinfo {year} {1994})}\BibitemShut {NoStop}%
\end{thebibliography}%


\begin{thebibliography}{15}%
\makeatletter
\providecommand \@ifxundefined [1]{%
 \@ifx{#1\undefined}
}%
\providecommand \@ifnum [1]{%
 \ifnum #1\expandafter \@firstoftwo
 \else \expandafter \@secondoftwo
 \fi
}%
\providecommand \@ifx [1]{%
 \ifx #1\expandafter \@firstoftwo
 \else \expandafter \@secondoftwo
 \fi
}%
\providecommand \natexlab [1]{#1}%
\providecommand \enquote  [1]{``#1''}%
\providecommand \bibnamefont  [1]{#1}%
\providecommand \bibfnamefont [1]{#1}%
\providecommand \citenamefont [1]{#1}%
\providecommand \href@noop [0]{\@secondoftwo}%
\providecommand \href [0]{\begingroup \@sanitize@url \@href}%
\providecommand \@href[1]{\@@startlink{#1}\@@href}%
\providecommand \@@href[1]{\endgroup#1\@@endlink}%
\providecommand \@sanitize@url [0]{\catcode `\\12\catcode `\$12\catcode `\&12\catcode `\#12\catcode `\^12\catcode `\_12\catcode `\%12\relax}%
\providecommand \@@startlink[1]{}%
\providecommand \@@endlink[0]{}%
\providecommand \url  [0]{\begingroup\@sanitize@url \@url }%
\providecommand \@url [1]{\endgroup\@href {#1}{\urlprefix }}%
\providecommand \urlprefix  [0]{URL }%
\providecommand \Eprint [0]{\href }%
\providecommand \doibase [0]{http://dx.doi.org/}%
\providecommand \selectlanguage [0]{\@gobble}%
\providecommand \bibinfo  [0]{\@secondoftwo}%
\providecommand \bibfield  [0]{\@secondoftwo}%
\providecommand \translation [1]{[#1]}%
\providecommand \BibitemOpen [0]{}%
\providecommand \bibitemStop [0]{}%
\providecommand \bibitemNoStop [0]{.\EOS\space}%
\providecommand \EOS [0]{\spacefactor3000\relax}%
\providecommand \BibitemShut  [1]{\csname bibitem#1\endcsname}%
\let\auto@bib@innerbib\@empty
\bibitem [{\citenamefont {Talyanskii}\ \emph {et~al.}(2001)\citenamefont {Talyanskii}, \citenamefont {Novikov}, \citenamefont {Simons},\ and\ \citenamefont {Levitov}}]{PhysRevLett.87.276802}%
  \BibitemOpen
  \bibfield  {author} {\bibinfo {author} {\bibfnamefont {V.~I.}\ \bibnamefont {Talyanskii}}, \bibinfo {author} {\bibfnamefont {D.~S.}\ \bibnamefont {Novikov}}, \bibinfo {author} {\bibfnamefont {B.~D.}\ \bibnamefont {Simons}}, \ and\ \bibinfo {author} {\bibfnamefont {L.~S.}\ \bibnamefont {Levitov}},\ }\href {\doibase 10.1103/PhysRevLett.87.276802} {\bibfield  {journal} {\bibinfo  {journal} {Phys. Rev. Lett.}\ }\textbf {\bibinfo {volume} {87}},\ \bibinfo {pages} {276802} (\bibinfo {year} {2001})}\BibitemShut {NoStop}%
\bibitem [{\citenamefont {Leek}\ \emph {et~al.}(2005)\citenamefont {Leek}, \citenamefont {Buitelaar}, \citenamefont {Talyanskii}, \citenamefont {Smith}, \citenamefont {Anderson}, \citenamefont {Jones}, \citenamefont {Wei},\ and\ \citenamefont {Cobden}}]{PhysRevLett.95.256802}%
  \BibitemOpen
  \bibfield  {author} {\bibinfo {author} {\bibfnamefont {P.~J.}\ \bibnamefont {Leek}}, \bibinfo {author} {\bibfnamefont {M.~R.}\ \bibnamefont {Buitelaar}}, \bibinfo {author} {\bibfnamefont {V.~I.}\ \bibnamefont {Talyanskii}}, \bibinfo {author} {\bibfnamefont {C.~G.}\ \bibnamefont {Smith}}, \bibinfo {author} {\bibfnamefont {D.}~\bibnamefont {Anderson}}, \bibinfo {author} {\bibfnamefont {G.~A.~C.}\ \bibnamefont {Jones}}, \bibinfo {author} {\bibfnamefont {J.}~\bibnamefont {Wei}}, \ and\ \bibinfo {author} {\bibfnamefont {D.~H.}\ \bibnamefont {Cobden}},\ }\href {\doibase 10.1103/PhysRevLett.95.256802} {\bibfield  {journal} {\bibinfo  {journal} {Phys. Rev. Lett.}\ }\textbf {\bibinfo {volume} {95}},\ \bibinfo {pages} {256802} (\bibinfo {year} {2005})}\BibitemShut {NoStop}%
\bibitem [{\citenamefont {Buitelaar}\ \emph {et~al.}(2006)\citenamefont {Buitelaar}, \citenamefont {Leek}, \citenamefont {Talyanskii}, \citenamefont {Smith}, \citenamefont {Anderson}, \citenamefont {Jones}, \citenamefont {Wei},\ and\ \citenamefont {Cobden}}]{Buitelaar2006}%
  \BibitemOpen
  \bibfield  {author} {\bibinfo {author} {\bibfnamefont {M.~R.}\ \bibnamefont {Buitelaar}}, \bibinfo {author} {\bibfnamefont {P.~J.}\ \bibnamefont {Leek}}, \bibinfo {author} {\bibfnamefont {V.~I.}\ \bibnamefont {Talyanskii}}, \bibinfo {author} {\bibfnamefont {C.~G.}\ \bibnamefont {Smith}}, \bibinfo {author} {\bibfnamefont {D.}~\bibnamefont {Anderson}}, \bibinfo {author} {\bibfnamefont {G.~A.~C.}\ \bibnamefont {Jones}}, \bibinfo {author} {\bibfnamefont {J.}~\bibnamefont {Wei}}, \ and\ \bibinfo {author} {\bibfnamefont {D.~H.}\ \bibnamefont {Cobden}},\ }\href {\doibase 10.1088/0268-1242/21/11/s10} {\bibfield  {journal} {\bibinfo  {journal} {Semiconductor Science and Technology}\ }\textbf {\bibinfo {volume} {21}},\ \bibinfo {pages} {S69–S77} (\bibinfo {year} {2006})}\BibitemShut {NoStop}%
\bibitem [{\citenamefont {Ahlers}\ \emph {et~al.}(2004)\citenamefont {Ahlers}, \citenamefont {Fletcher}, \citenamefont {Ebbecke},\ and\ \citenamefont {Janssen}}]{Ahlers2004}%
  \BibitemOpen
  \bibfield  {author} {\bibinfo {author} {\bibfnamefont {F.}~\bibnamefont {Ahlers}}, \bibinfo {author} {\bibfnamefont {N.}~\bibnamefont {Fletcher}}, \bibinfo {author} {\bibfnamefont {J.}~\bibnamefont {Ebbecke}}, \ and\ \bibinfo {author} {\bibfnamefont {T.}~\bibnamefont {Janssen}},\ }\href {\doibase 10.1016/j.cap.2004.01.012} {\bibfield  {journal} {\bibinfo  {journal} {Current Applied Physics}\ }\textbf {\bibinfo {volume} {4}},\ \bibinfo {pages} {529–533} (\bibinfo {year} {2004})}\BibitemShut {NoStop}%
\bibitem [{\citenamefont {Ebbecke}\ \emph {et~al.}(2004)\citenamefont {Ebbecke}, \citenamefont {Strobl},\ and\ \citenamefont {Wixforth}}]{PhysRevB.70.233401}%
  \BibitemOpen
  \bibfield  {author} {\bibinfo {author} {\bibfnamefont {J.}~\bibnamefont {Ebbecke}}, \bibinfo {author} {\bibfnamefont {C.~J.}\ \bibnamefont {Strobl}}, \ and\ \bibinfo {author} {\bibfnamefont {A.}~\bibnamefont {Wixforth}},\ }\href {\doibase 10.1103/PhysRevB.70.233401} {\bibfield  {journal} {\bibinfo  {journal} {Phys. Rev. B}\ }\textbf {\bibinfo {volume} {70}},\ \bibinfo {pages} {233401} (\bibinfo {year} {2004})}\BibitemShut {NoStop}%
\bibitem [{\citenamefont {Fletcher}\ \emph {et~al.}(2003)\citenamefont {Fletcher}, \citenamefont {Ebbecke}, \citenamefont {Janssen}, \citenamefont {Ahlers}, \citenamefont {Pepper}, \citenamefont {Beere},\ and\ \citenamefont {Ritchie}}]{PhysRevB.68.245310}%
  \BibitemOpen
  \bibfield  {author} {\bibinfo {author} {\bibfnamefont {N.~E.}\ \bibnamefont {Fletcher}}, \bibinfo {author} {\bibfnamefont {J.}~\bibnamefont {Ebbecke}}, \bibinfo {author} {\bibfnamefont {T.~J. B.~M.}\ \bibnamefont {Janssen}}, \bibinfo {author} {\bibfnamefont {F.~J.}\ \bibnamefont {Ahlers}}, \bibinfo {author} {\bibfnamefont {M.}~\bibnamefont {Pepper}}, \bibinfo {author} {\bibfnamefont {H.~E.}\ \bibnamefont {Beere}}, \ and\ \bibinfo {author} {\bibfnamefont {D.~A.}\ \bibnamefont {Ritchie}},\ }\href {\doibase 10.1103/PhysRevB.68.245310} {\bibfield  {journal} {\bibinfo  {journal} {Phys. Rev. B}\ }\textbf {\bibinfo {volume} {68}},\ \bibinfo {pages} {245310} (\bibinfo {year} {2003})}\BibitemShut {NoStop}%
\bibitem [{\citenamefont {Sato}\ \emph {et~al.}(2019)\citenamefont {Sato}, \citenamefont {McIver}, \citenamefont {Nuske}, \citenamefont {Tang}, \citenamefont {Jotzu}, \citenamefont {Schulte}, \citenamefont {H\"ubener}, \citenamefont {De~Giovannini}, \citenamefont {Mathey}, \citenamefont {Sentef}, \citenamefont {Cavalleri},\ and\ \citenamefont {Rubio}}]{Sato2019MicroscopicGraphene}%
  \BibitemOpen
  \bibfield  {author} {\bibinfo {author} {\bibfnamefont {S.~A.}\ \bibnamefont {Sato}}, \bibinfo {author} {\bibfnamefont {J.~W.}\ \bibnamefont {McIver}}, \bibinfo {author} {\bibfnamefont {M.}~\bibnamefont {Nuske}}, \bibinfo {author} {\bibfnamefont {P.}~\bibnamefont {Tang}}, \bibinfo {author} {\bibfnamefont {G.}~\bibnamefont {Jotzu}}, \bibinfo {author} {\bibfnamefont {B.}~\bibnamefont {Schulte}}, \bibinfo {author} {\bibfnamefont {H.}~\bibnamefont {H\"ubener}}, \bibinfo {author} {\bibfnamefont {U.}~\bibnamefont {De~Giovannini}}, \bibinfo {author} {\bibfnamefont {L.}~\bibnamefont {Mathey}}, \bibinfo {author} {\bibfnamefont {M.~A.}\ \bibnamefont {Sentef}}, \bibinfo {author} {\bibfnamefont {A.}~\bibnamefont {Cavalleri}}, \ and\ \bibinfo {author} {\bibfnamefont {A.}~\bibnamefont {Rubio}},\ }\href {\doibase 10.1103/PhysRevB.99.214302} {\bibfield  {journal} {\bibinfo  {journal} {Phys. Rev. B}\ }\textbf {\bibinfo {volume} {99}},\ \bibinfo {pages} {214302} (\bibinfo {year} {2019})}\BibitemShut {NoStop}%
\bibitem [{\citenamefont {Meier}\ \emph {et~al.}(1994)\citenamefont {Meier}, \citenamefont {von Plessen}, \citenamefont {Thomas},\ and\ \citenamefont {Koch}}]{PhysRevLett.73.902}%
  \BibitemOpen
  \bibfield  {author} {\bibinfo {author} {\bibfnamefont {T.}~\bibnamefont {Meier}}, \bibinfo {author} {\bibfnamefont {G.}~\bibnamefont {von Plessen}}, \bibinfo {author} {\bibfnamefont {P.}~\bibnamefont {Thomas}}, \ and\ \bibinfo {author} {\bibfnamefont {S.~W.}\ \bibnamefont {Koch}},\ }\href {\doibase 10.1103/PhysRevLett.73.902} {\bibfield  {journal} {\bibinfo  {journal} {Phys. Rev. Lett.}\ }\textbf {\bibinfo {volume} {73}},\ \bibinfo {pages} {902} (\bibinfo {year} {1994})}\BibitemShut {NoStop}%
\bibitem [{\citenamefont {Sanders}\ \emph {et~al.}(2009)\citenamefont {Sanders}, \citenamefont {Stanton}, \citenamefont {Kim}, \citenamefont {Yee}, \citenamefont {Lim}, \citenamefont {H\'aroz}, \citenamefont {Booshehri}, \citenamefont {Kono},\ and\ \citenamefont {Saito}}]{PhysRevB.79.205434}%
  \BibitemOpen
  \bibfield  {author} {\bibinfo {author} {\bibfnamefont {G.~D.}\ \bibnamefont {Sanders}}, \bibinfo {author} {\bibfnamefont {C.~J.}\ \bibnamefont {Stanton}}, \bibinfo {author} {\bibfnamefont {J.-H.}\ \bibnamefont {Kim}}, \bibinfo {author} {\bibfnamefont {K.-J.}\ \bibnamefont {Yee}}, \bibinfo {author} {\bibfnamefont {Y.-S.}\ \bibnamefont {Lim}}, \bibinfo {author} {\bibfnamefont {E.~H.}\ \bibnamefont {H\'aroz}}, \bibinfo {author} {\bibfnamefont {L.~G.}\ \bibnamefont {Booshehri}}, \bibinfo {author} {\bibfnamefont {J.}~\bibnamefont {Kono}}, \ and\ \bibinfo {author} {\bibfnamefont {R.}~\bibnamefont {Saito}},\ }\href {\doibase 10.1103/PhysRevB.79.205434} {\bibfield  {journal} {\bibinfo  {journal} {Phys. Rev. B}\ }\textbf {\bibinfo {volume} {79}},\ \bibinfo {pages} {205434} (\bibinfo {year} {2009})}\BibitemShut {NoStop}%
\bibitem [{\citenamefont {Fathi}(2011)}]{cnt-review}%
  \BibitemOpen
  \bibfield  {author} {\bibinfo {author} {\bibfnamefont {D.}~\bibnamefont {Fathi}},\ }\href {\doibase 10.1155/2011/471241} {\bibfield  {journal} {\bibinfo  {journal} {Journal of Nanotechnology}\ }\textbf {\bibinfo {volume} {2011}} (\bibinfo {year} {2011}),\ 10.1155/2011/471241}\BibitemShut {NoStop}%
\bibitem [{\citenamefont {Kane}\ and\ \citenamefont {Mele}(1997)}]{PhysRevLett.78.1932}%
  \BibitemOpen
  \bibfield  {author} {\bibinfo {author} {\bibfnamefont {C.~L.}\ \bibnamefont {Kane}}\ and\ \bibinfo {author} {\bibfnamefont {E.~J.}\ \bibnamefont {Mele}},\ }\href {\doibase 10.1103/PhysRevLett.78.1932} {\bibfield  {journal} {\bibinfo  {journal} {Phys. Rev. Lett.}\ }\textbf {\bibinfo {volume} {78}},\ \bibinfo {pages} {1932} (\bibinfo {year} {1997})}\BibitemShut {NoStop}%
\bibitem [{\citenamefont {Saito}\ \emph {et~al.}(1992)\citenamefont {Saito}, \citenamefont {Fujita}, \citenamefont {Dresselhaus},\ and\ \citenamefont {Dresselhaus}}]{PhysRevB.46.1804}%
  \BibitemOpen
  \bibfield  {author} {\bibinfo {author} {\bibfnamefont {R.}~\bibnamefont {Saito}}, \bibinfo {author} {\bibfnamefont {M.}~\bibnamefont {Fujita}}, \bibinfo {author} {\bibfnamefont {G.}~\bibnamefont {Dresselhaus}}, \ and\ \bibinfo {author} {\bibfnamefont {M.~S.}\ \bibnamefont {Dresselhaus}},\ }\href {\doibase 10.1103/PhysRevB.46.1804} {\bibfield  {journal} {\bibinfo  {journal} {Phys. Rev. B}\ }\textbf {\bibinfo {volume} {46}},\ \bibinfo {pages} {1804} (\bibinfo {year} {1992})}\BibitemShut {NoStop}%
\bibitem [{\citenamefont {Mintmire}\ \emph {et~al.}(1992)\citenamefont {Mintmire}, \citenamefont {Dunlap},\ and\ \citenamefont {White}}]{PhysRevLett.68.631}%
  \BibitemOpen
  \bibfield  {author} {\bibinfo {author} {\bibfnamefont {J.~W.}\ \bibnamefont {Mintmire}}, \bibinfo {author} {\bibfnamefont {B.~I.}\ \bibnamefont {Dunlap}}, \ and\ \bibinfo {author} {\bibfnamefont {C.~T.}\ \bibnamefont {White}},\ }\href {\doibase 10.1103/PhysRevLett.68.631} {\bibfield  {journal} {\bibinfo  {journal} {Phys. Rev. Lett.}\ }\textbf {\bibinfo {volume} {68}},\ \bibinfo {pages} {631} (\bibinfo {year} {1992})}\BibitemShut {NoStop}%
\bibitem [{\citenamefont {Hamada}\ \emph {et~al.}(1992)\citenamefont {Hamada}, \citenamefont {Sawada},\ and\ \citenamefont {Oshiyama}}]{PhysRevLett.68.1579}%
  \BibitemOpen
  \bibfield  {author} {\bibinfo {author} {\bibfnamefont {N.}~\bibnamefont {Hamada}}, \bibinfo {author} {\bibfnamefont {S.-i.}\ \bibnamefont {Sawada}}, \ and\ \bibinfo {author} {\bibfnamefont {A.}~\bibnamefont {Oshiyama}},\ }\href {\doibase 10.1103/PhysRevLett.68.1579} {\bibfield  {journal} {\bibinfo  {journal} {Phys. Rev. Lett.}\ }\textbf {\bibinfo {volume} {68}},\ \bibinfo {pages} {1579} (\bibinfo {year} {1992})}\BibitemShut {NoStop}%
\bibitem [{\citenamefont {Saito}\ \emph {et~al.}(1998)\citenamefont {Saito}, \citenamefont {Takeya}, \citenamefont {Kimura}, \citenamefont {Dresselhaus},\ and\ \citenamefont {Dresselhaus}}]{PhysRevB.57.4145}%
  \BibitemOpen
  \bibfield  {author} {\bibinfo {author} {\bibfnamefont {R.}~\bibnamefont {Saito}}, \bibinfo {author} {\bibfnamefont {T.}~\bibnamefont {Takeya}}, \bibinfo {author} {\bibfnamefont {T.}~\bibnamefont {Kimura}}, \bibinfo {author} {\bibfnamefont {G.}~\bibnamefont {Dresselhaus}}, \ and\ \bibinfo {author} {\bibfnamefont {M.~S.}\ \bibnamefont {Dresselhaus}},\ }\href {\doibase 10.1103/PhysRevB.57.4145} {\bibfield  {journal} {\bibinfo  {journal} {Phys. Rev. B}\ }\textbf {\bibinfo {volume} {57}},\ \bibinfo {pages} {4145} (\bibinfo {year} {1998})}\BibitemShut {NoStop}%
\end{thebibliography}%

\end{document}


\widetext
\onecolumngrid

\begin{center}
\textbf{\large Supplemental Material:
\\Quantized Acoustoelectric Floquet Effect in Quantum Nanowires}
\\[0.4ex] Christopher Yang, Will Hunt, and Gil Refael, Iliya Esin
\end{center}
\par
\setcounter{page}{1}
\twocolumngrid

\setcounter{equation}{0}
\setcounter{figure}{0}
\setcounter{table}{0}
\setcounter{page}{1}
\makeatletter
\renewcommand{\theequation}{S\arabic{equation}}
\renewcommand{\thefigure}{S\arabic{figure}}
\setcounter{secnumdepth}{4}

\def\thefootnote{*}\footnotetext{These authors contributed equally to this work}

\section{Effective Description for Adiabatic Drives}
In this section, we derive an effective model for the steady states and quantized transport in 1D systems driven adiabatically by coherent phonons, in the regime $\Delta \gg \hbar\omega$. Systems adiabatically driven by GHz-frequency surface acoustic phonon waves have been observed experimentally to host a quantized acousto-electric current response \cite{PhysRevLett.87.276802,PhysRevLett.95.256802,Buitelaar2006,Ahlers2004,PhysRevB.70.233401,PhysRevB.68.245310}. 

We first consider the time-averaged current
\begin{equation} \label{eq:current}
    J = \frac{2e}{\hbar} \frac{1}{T} \int_0^{T} dt \int_0^q \frac{dk}{2\pi} \ \text{Tr} \left[\frac{\partial H_0(k,t)}{\partial k} \rho(k,t) \right]
\end{equation}
where the factor of two accounts for spin degeneracy, $T = 2\pi/\omega$, ${\rho}(k, t)$ is the density matrix for an electron with crystal momentum $k$, and $\hat{H}_0(t)=\int_0^{q} dk/(2\pi) \hat{\boldsymbol{\psi}}^\dagger_k H_0(k,t)\hat{\boldsymbol{\psi}}_k$ [see main text for definition]. The actual value of $\rho(k,t)$ is controlled by the free propagation Hamiltonian, $\hat{H}_0(t)$, collisions with incoherent phonon modes, described by $\hat{H}_b(t)$, and electron-electron interactions. We assume that the incoherent phonons are in a low-temperature thermal equilibrium, serving as a low-temperature heat bath for the electrons. In the regime where the characteristic relaxation is faster than the drive frequency, $1/\tau \gg \hbar\omega$, we approximate the relaxation dynamics of the electrons in the CNT due to scattering processes by the Lindbladian $\mathcal{L}\{ \rho(k,t) \} \equiv -1/\tau [\rho(k,t) - \rho^{\text{eq}}(k,t)]$. Here, the electrons relax to a thermal equilibrium state in the instantaneous basis of $\hat{H}_0(t)$, denoted by $\rho^{\text{eq}}(k,t)$. The time evolution of the density matrix can then be described by the master equation \cite{Sato2019MicroscopicGraphene,PhysRevLett.73.902}
\begin{equation} \label{eq:rellinb}
    \dot{\rho}(k,t) = \frac{i}{\hbar} [\rho(k,t) , H_0(k,t)] + \mathcal{L}\{ \rho(k,t) \}.
\end{equation}
In the limit where coupling strength of the coherent electrons via the vector potential $\hat{\boldsymbol{\mathcal{A}}}_{\text{ph}}^{(\lambda)}(\boldsymbol{r},t)$ is much smaller than the separation between the lowest conduction and highest valence bands, i.e., $\hbar v_F \beta a^{-1} \ll \hbar v_F \delta k$ [see definitions of $\delta k$ and $\beta$ in the main text], we can further approximate $V(\boldsymbol{r},t) \approx {\Delta} \cos(qr-\omega t)$, where $V(\boldsymbol{r},t) \equiv \langle \hat{V}^{(\lambda_0)}(\boldsymbol{r},t) \rangle$, $\boldsymbol{r} = r \hat{\boldsymbol{k}}_{\parallel}$ is the spatial position along the tube axis of the CNT. We also approximate $E_+(k) \approx \hbar^2 k^2 / (2m) + \hbar v_F \delta k$ where $m$ is the effective mass at the band bottom [see Fig. 1(c) in the main text]. For details of the derivation, see Sec. \ref{sec:effh}.

\begin{figure}
    \centering    \includegraphics[width=0.98\linewidth]{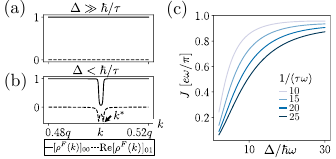}
    \caption{(a) Steady state occupation $[\rho^F(k)]_{00}$ of the $\alpha = 0$ Floquet band and steady state coherence $[\rho^F(k)]_{01}$ at a strong drive amplitude ${\Delta} \tau/\hbar = 10$ vs. crystal momentum $k$, following Eqs. (\ref{eq:diagtoy}) and (\ref{eq:cohtoy}). The coherences are negligible and the $\alpha = 0$ band is fully-occupied. (b) Same as (a) but for a weak drive amplitude ${\Delta} \tau/\hbar = 0.007$ where the occupatons resemble the equilibrium distribution: the coherences are large and the $\alpha = 0$ is empty in a small region $q/2 < k < k^*$, where $k^*$ is the momentum at which the Floquet gap $\Delta$ opens [see Fig. 1(c) in the main text]. (c) Steady state current $J$ as a function of ${\Delta}/\hbar\omega$. The current response is nearly quantized when ${\Delta} \gg \hbar/\tau, \hbar\omega$. }
    \label{fig:lindblad-results}
\end{figure}

To solve the Lindblad master equation analytically, we truncate the eigenvalue relation, Eq. (1) in the main text, for the Floquet spectrum around $\alpha = 0, 1$ so as to ignore contributions of order $[\Delta/(\hbar^2 q^2 / 2m)]^2$ or higher. We work in the Floquet basis in which the density matrix is given by $\rho^F(k,t)$, where the density operator is expressed as $\hat{\rho}(k,t) = \sum_{\alpha\beta} [\rho^F(k,t) ]_{\alpha\beta} \hat{f}_{k\alpha}^{\dagger}(t) \hat{f}_{k\beta}(t)$ and $\hat{f}_{k\alpha}^{\dagger}(t)$ creates the Floquet eigenstate $|\psi_{k\alpha}(\boldsymbol{r}, t) \rangle$. The steady state density matrix $\rho^F(k)$ satisfies $\dot{\rho}^F(k) = 0$ and, following Eq. (\ref{eq:current}), leads to the current $J = 2e \int_0^q {dk}/{2\pi} \  \text{Tr}[J^F(k)  \rho^F(k)]$, where $[J^F(k)]_{\alpha\alpha} = \hbar^{-1} d\varepsilon_{k\alpha}/dk$ and, for momenta $0 \leq k \leq q$, one can approximate $[J^F(k)]_{01} = [J^F(k)]_{10} \approx -{(4 \hbar q {\Delta}}/{m) [2{\Delta}^2 + (\hbar\omega + 2\delta)^2]^{-1/2}}$.

To determine the conditions under which the current is quantized, i.e., $\rho^F(k) \approx \rho^{\text{opt}}(k)$, where $[\rho^{\text{opt}}(k)]_{00} = 1 - [\rho^{\text{opt}}(k)]_{11} = 1$ and $[\rho^{\text{opt}}(k)]_{01} = [\rho^{\text{opt}}(k)]_{10} = 0$, we analyze the dependence of $\rho^F(k)$ on the Floquet gap ${\Delta}$. For momenta $0 \leq k \leq q$,
\begin{equation} \label{eq:diagtoy}
    [\rho^F(k)]_{00} \approx \frac{1}{2} + \frac{2 + \delta \hbar\omega / (\delta^2 + {\Delta}^2)}{2 \{(\delta^2 + {\Delta}^2)[4\Delta^2+(\hbar\omega-  2\delta)^2]\}^{1/2}}
\end{equation}
where $\delta = \hbar^2 kq/2m + \hbar^2 q^2/4m$ and $[\rho^F(k)]_{11} = 1- [\rho^F(k)]_{00}$. The real component of the steady state coherence is given by
\begin{equation} \label{eq:cohtoy}
    \text{Re} [\rho^F(k)]_{01} \approx  \frac{{\Delta}\hbar\omega/\{(\delta^2 + {\Delta}^2)[4\Delta^2+(\hbar\omega-  2\delta)^2]\}^{1/2} }{2 [1 + 4(\tau \Delta/\hbar)^2 + \tau^2 ( \hbar\omega - 2\delta)^2 /\hbar^2]\}}.
\end{equation}
The imaginary component of the coherence does not contribute to the current. (See Sec. \ref{sec:ssdetails} for details of the derivation.) The electronic occupation $[\rho^F(k)]_{00}$ and coherence $\text{Re} [\rho^F(k)]_{01}$ are largest for momenta $q/2 < k < k^*$ near the Floquet gap where $\delta \approx \hbar\omega / 2$ and $k^*$ is the momentum at which the Floquet gap opens [see Fig. 1(c) in the main text]. The density matrix approaches the optimal distribution $\rho^F(k) \to \rho^{\text{opt}}(k)$ in the strong drive limit ${\Delta}/\hbar\omega, {\Delta}\tau/\hbar \to \infty$, where $[\rho^F(k)]_{00} \to 1$ and $\text{Re} [\rho^F(k)]_{01} \to 0$ for all $k$. In contrast, in the low intensity limit ${\Delta} \to 0$, $[\rho^F(k)]_{00} \to [1 - \text{sign}(\hbar\omega - 2 \delta)] / 2$ and $[\rho^F(k)]_{01} \to 0$, which recovers the equilibrium Fermi-Dirac distribution. 

We plot the corresponding steady state occupations $[\rho^F(k)]_{00}$ and coherences $[\rho^F(k)]_{01}$ in Fig. \ref{fig:lindblad-results}(a)-(b) for a strong (${\Delta} \gg \hbar/\tau, \hbar\omega$) and weak (${\Delta} < \hbar/\tau$) drive amplitude, respectively. Here, we use a large $\hbar^2q^2/2m \approx 250 \hbar\omega$ and $1/(\tau\omega) = 10$. Fig. \ref{fig:lindblad-results}(c) shows the steady state current $J$ as a function of drive amplitude ${\Delta}$ for various characteristic relaxation rates $\hbar/\tau \gg \hbar\omega$. Consistent with the analysis above, the current approaches its quantized value and $\rho^F(k) \to \rho^{\text{opt}}(k)$ in the regime ${\Delta} \gg \hbar/\tau, \hbar \omega$. 

\section{Effective Hamiltonian} \label{sec:effh}
In this section, we derive the effective Hamiltonian $V(\boldsymbol{r},t) \approx \Delta \cos(qr-\omega t)$ which describes electronic coupling to the coherent phonon drive.

To obtain the effective drive Hamiltonian, we work in the eigenbasis of the electronic Hamiltonian 
\begin{equation}
    H_{\text{e}}(k) = \hbar v_F \boldsymbol{k} \cdot \sigma ,
\end{equation}
where $\boldsymbol{k} = \delta k \hat{\boldsymbol{k}}_{\perp} + k \hat{\boldsymbol{k}}_{\parallel}$ [see the full definition in the main text]. Let $U$ be the matrix that diagonalizes $H_{\text{e}}(k)$, where
\begin{equation}
    U^{-1} H_{\text{e}}(k) U = \begin{pmatrix}
        -\hbar v_F |\boldsymbol{k}| & 0 \\ 0 & \hbar v_F |\boldsymbol{k}|
    \end{pmatrix}.
\end{equation}
In the limit near $k = 0$, one can show that
\begin{equation}
    U^{-1} H_{\text{e}}(k) U \approx \begin{pmatrix}
        E_{-}(k) & 0 \\ 0 & E_{+}(k)
    \end{pmatrix}
\end{equation}
where $E_{\pm}(k) = \hbar^2 k^2/(2m) \pm \hbar v_F \delta k$. The electronic coupling to a coherent phonon mode is described by the Hamiltonian
\begin{equation}
    V(\boldsymbol{r}, t) = \hbar v_F \langle \hat{\boldsymbol{\mathcal{A}}}_{\text{ph}}^{(\lambda_0)}(\boldsymbol{r},t) \rangle  \cdot \boldsymbol{\sigma} + \langle \hat{\phi}_{\text{ph}}^{(\lambda_0)}(\boldsymbol{r},t)\rangle.
\end{equation}
Note that
\begin{equation}
    [\langle \hat{\boldsymbol{\mathcal{A}}}_{\text{ph}}^{(\lambda_0)}(\boldsymbol{r},t) \rangle ]_x = \frac{\sqrt{3} \beta}{2a} [(\hat{k}_{\parallel}^y)^2 - (\hat{k}_{\parallel}^x)^2] q u \sin (qr - \omega t)
\end{equation}
\begin{equation}
    [\langle \hat{\boldsymbol{\mathcal{A}}}_{\text{ph}}^{(\lambda_0)}(\boldsymbol{r},t) \rangle ]_y = \frac{\sqrt{3} \beta}{2a} (-2 \hat{k}_{\parallel}^x \hat{k}_{\parallel}^y) q u \sin (qr - \omega t)
\end{equation}
and
\begin{equation}
    \langle \hat{\phi}_{\text{ph}}^{(\lambda_0)}(\boldsymbol{r},t)\rangle = D qu \sin(qr-\omega t) I.
\end{equation}
To transform $V(\boldsymbol{r},t)$ into the diagonal basis of $H_{\text{e}}(k)$, first note that
\begin{equation} \label{eq:transformvec}
\begin{split}
    &U^{-1} [\hbar v_F \langle \hat{\boldsymbol{\mathcal{A}}}_{\text{ph}}^{(\lambda_0)}({r},t) \rangle  \cdot \boldsymbol{\sigma}] U\\
    &= \begin{pmatrix}
        - \langle \hat{\boldsymbol{\mathcal{A}}}_{\text{ph}}^{(\lambda_0)}(\boldsymbol{r},t) \rangle  \cdot {\boldsymbol{k}}/|\boldsymbol{k}| & - i\langle \hat{\boldsymbol{\mathcal{A}}}_{\text{ph}}^{(\lambda_0)}(\boldsymbol{r},t) \rangle  \times {\boldsymbol{k}}/|\boldsymbol{k}| \\
        i\langle \hat{\boldsymbol{\mathcal{A}}}_{\text{ph}}^{(\lambda_0)}(\boldsymbol{r},t) \rangle  \times {\boldsymbol{k}}/|\boldsymbol{k}| & \langle \hat{\boldsymbol{\mathcal{A}}}_{\text{ph}}^{(\lambda_0)}(\boldsymbol{r},t) \rangle  \cdot {\boldsymbol{k}}/|\boldsymbol{k}|
    \end{pmatrix}.
\end{split}
\end{equation}
In a semiconducting nanotube, near the band extrema, $k \ll \delta k$, so $\boldsymbol{k}$ and $\hat{\boldsymbol{k}}_{\perp}$ are roughly parallel and 
\begin{equation}
    \langle \hat{\boldsymbol{\mathcal{A}}}_{\text{ph}}^{(\lambda_0)}({r},t) \rangle  \cdot {\boldsymbol{k}}/|\boldsymbol{k}| \approx \hat{k}_{\parallel}^y \hbar v_F \frac{\sqrt{3}\beta}{2a} qu \sin(qr-\omega t) .
\end{equation}
When $\hbar v_F \beta a^{-1} \ll \hbar v_F \delta k$, the off-diagonal components of Eq. (\ref{eq:transformvec}) are negligible. In this limit, we finally obtain an effective Hamiltonian for the lowest conduction band of the CNT,
\begin{equation}
\begin{split}
   &[U^{-1} H_0(k,t) U]_{00} \\
   &\approx  E_+(k) + \left[\hat{k}_{\parallel}^y \hbar v_F \frac{\sqrt{3}\beta}{2a} + D\right] qu \sin(qr-\omega t).
\end{split}
\end{equation}
Up to a phase shift in time, the second term corresponds to the effective Hamiltonian $V(\boldsymbol{r},t) \approx \Delta \cos(qr-\omega t)$ used in the main text [see discussion below Eq. (\ref{eq:rellinb})].

\section{Details of the Linblad Master Equation} \label{sec:ssdetails}

In this section, we detail the analytical analysis used to determine the steady state density matrix $\rho^F(k)$ via the Linblad master equation. 

Let us first derive an approximate Floquet Hamiltonian to the driven system. Adapting Eq. (1) in the main text to the effective Hamiltonian $V(\boldsymbol{r},t) \approx \Delta \cos(qr-\omega t)$, we find that
\begin{equation} \label{eq:freq-ladder}
\begin{split}
    i\hbar \partial_t &|\phi_{k\alpha}\rangle = H_{\text{e}}(k+nq) |\phi_{k\alpha}^{(n)}\rangle \\
    &+ \frac{1}{2} \Delta e^{i\omega t} |\phi_{k\alpha}^{(n+1)}\rangle + \frac{1}{2} \Delta e^{-i\omega t} |\phi_{k\alpha}^{(n-1)}\rangle.
\end{split}
\end{equation}
To simplify the analytical treatment, we focus on the range $k \in [0, q]$ and consider only the two closely lying bands in this range, described by $H_{\text{e}}(k)$ and $H_{\text{e}}(k-q)$. Here, $H_{\text{e}}(k)$ and $H_{\text{e}}(k-q)$ are separated at most by energy $\hbar^2 q^2 / 2m$ and are separated from all other bands by at least energy $\hbar^2 q^2 / 2m$. Thus, in the limit $\hbar^2 q^2 / 2m \gg \Delta$, we keep only the $|\phi_{k\alpha}^{(0)}\rangle$ and $|\phi_{k\alpha}^{(1)}\rangle$ harmonics in Eq. (\ref{eq:freq-ladder}), neglecting corrections of order $[\Delta/(\hbar^2 q^2 / 2m)]^2$ or higher. We obtain the approximate Schrödinger equation
\begin{equation}
    i \hbar \partial_t \begin{pmatrix} |\phi_{k\alpha}^{(0)}\rangle \\ |\phi_{k\alpha}^{(1)}\rangle \end{pmatrix} \approx \tilde{H}_{\text{eff}}(k,t) \begin{pmatrix} |\phi_{k\alpha}^{(0)}\rangle \\ |\phi_{k\alpha}^{(1)}\rangle \end{pmatrix},
\end{equation}
where 
\begin{equation}
    \tilde{H}_{\text{eff}}(k,t) \equiv \begin{pmatrix} k^2/2m & \Delta e^{i\omega t} \\ \Delta e^{-i\omega t} & (k+ q)^2 / 2m \end{pmatrix}.
\end{equation}
The components of the single-particle reduced density matrix ${\rho}(k,t)$ may be written in terms of the harmonics as $[\rho(k,t)]_{mn} = \langle \phi_{k\alpha}^{(m)} | \hat{\rho} | \phi_{k\alpha}^{(n)} \rangle$ for $m, n = 0, 1$, where $\hat{\rho}$ is the density operator.

Let us begin by describing the instantaneous eigenbasis of the Hamiltonian $\tilde{H}_{\text{eff}}(k,t)$ and the equilibrium distribution $\rho^{\text{eq}}(k,t)$ that enters the Linbladian $\mathcal{L}\{\rho(k,t) \}$ [see definition above Eq. (\ref{eq:rellinb})]. For simplicity of notation, we define $\epsilon = \hbar^2 k^2/2m$ and $\delta = \hbar^2 kq/2m + \hbar^2 q^2/4m$. We diagonalize $\tilde{H}_{\text{eff}}(k,t)$ via a  transformation matrix $M(t)$ to obtain the instantaneous eigenbasis of $\tilde{H}_{\text{eff}}(k,t)$
\begin{equation}
\mathcal{H}(k) \equiv M(k,t)^{\dagger} \tilde{H}_{\text{eff}}(k,t) M(k,t) = \begin{pmatrix}
E_{-}(k) & 0 \\
0 & E_{+}(k)
\end{pmatrix} 
\end{equation}
where the instantaneous eigenenergies are given by $E_{-}(k) = \epsilon + \delta - \sqrt{\delta^2+\Delta^2}$ and $E_{+}(k) = \epsilon + \delta + \sqrt{\delta^2+\Delta^2}$. In this specific case, the instantaneous eigenenergies happen to be time-independent, simplifying the analytic evaluation of the Linblad master equation. In this instantaneous eigenbasis, we can define the instantaneous equilibrium distribution
\begin{equation}
    \rho^{\text{eq}}({k}) = \begin{pmatrix}
f(E_-(k)) & 0 \\
0 & f(E_+(k))
\end{pmatrix}
\end{equation}
where $f(E) = \theta(E -\mu)$ and $\mu = \hbar^2 q^2 / (8m)$ at zero temperature and optimal doping considered in the main text.

To analytically determine the Floquet eigenstates, we begin with the Kernel $\tilde{H}_{\text{eff}}(k,t) - i\hbar \partial_t$ and perform the rotating wave transformation to obtain $H_{\text{RW}}(k) \equiv R(t)^{\dagger} [\tilde{H}_{\text{eff}}(k,t) - i\hbar \partial_t] R(t)$, where
\begin{equation}
    R(t) = \begin{pmatrix}
        e^{-i\omega t} & 0 \\
        0 & 1
    \end{pmatrix}.
\end{equation}
Diagonalizing $H_{\text{RW}}(k)$ via the transformation matrix $T(k)$, we obtain the Floquet Hamiltonian
\begin{equation}
\begin{split}
    H_F(k) &= T(k)^{\dagger} H_{\text{RW}}(k) T(k) \\
    &=  \begin{pmatrix}
b - \sqrt{c^2+d^2} & 0 \\
0 & b + \sqrt{c^2+d^2}
\end{pmatrix}
\end{split}
\end{equation}
where
\begin{equation}
b \equiv \epsilon + \delta - \frac{\hbar\omega}{2},
\end{equation}
\begin{equation}
c \equiv -\sqrt{\delta^2+\Delta^2} - \frac{\hbar\omega \delta}{2\sqrt{\delta^2+\Delta^2}} ,
\end{equation}
and
\begin{equation}
d \equiv \frac{\hbar\omega \Delta}{2\sqrt{\delta^2+\Delta^2}}.
\end{equation}

Inserting the instantaneous equilibrium distribution $\rho^{\text{eq}}(k)$ and setting $\dot{\rho}^F(k,t) = 0$ into the Linblad master equation [see Eq. (\ref{eq:rellinb})], we obtain the steady state occupation
\begin{equation}
[{\rho}^F_k]_{00} = N_{-}^2
\end{equation}
and off-diagonal coherence
\begin{equation}
[{\rho}^F(k)]_{01} = \frac{2N_{+}N_{-} [2+ i(4\tau\sqrt{c^2+d^2})]}{4+16\tau^2(c^2+d^2)}
\end{equation}
where
\begin{equation}
    N_{\pm} = \frac{c\pm\sqrt{c^2+d^2}}{\sqrt{d^2 + (c\pm\sqrt{c^2 + d^2})^2}}.
\end{equation}
Note that $[{\rho}^F(k)]_{11} = 1- [{\rho}^F(k)]_{00}$ and $[{\rho}^F(k)]_{10} = ([{\rho}^F(k)]_{01})^*$. Upon algebraic manupulation, one finds Eqs. (\ref{eq:diagtoy}) and (\ref{eq:cohtoy}).

Lastly, we note that the current operator $J^F(k)$ used in the main text [see definition above Eq. (\ref{eq:diagtoy})] is derived using the transformation matrices defined above.
\begin{equation}
    J^F(k) = T(k)^{\dagger} R(t)^{\dagger} M(k,t)^{\dagger} \frac{1}{\hbar} \frac{\partial  \tilde{H}_{\text{eff}}(k,t)}{\partial k}  M(k,t) R(t) T(k).
\end{equation}

\section{Details of the Carbon Nanotube Hamiltonian}
The carbon nanotube (CNT) structure is described by the chirality indices $(m, n)$ \cite{PhysRevB.79.205434, cnt-review, PhysRevLett.78.1932,PhysRevB.46.1804,PhysRevLett.68.631,PhysRevLett.68.1579}. Along the axis of the CNT, the lattice is periodic under translations by $\boldsymbol{T} = t_1 \boldsymbol{a}_1 + t_2 \boldsymbol{a}_2$, where $t_1 = (2m+n) / d_R$, $t_2 = -(2n+m)/d_R$, and $d_R = \text{gcd}(2n+m, 2m+n)$, resulting in enlarged CNT unit cells that each contain $N = 2(n^2 + nm + m^2) / d_R$ graphene unit cells. Here, $\boldsymbol{a}_1 = \boldsymbol{\delta}_1 - \boldsymbol{\delta}_3$ and $\boldsymbol{a}_2 = \boldsymbol{\delta}_2 - \boldsymbol{\delta}_3$ are the primitive lattice vectors the graphene layer, where $\boldsymbol{\delta}_j = a/\sqrt{3} ( \sin (2 \pi j/3), \cos (2 \pi j/3))$ and $a = 0.246 \ \mathrm{nm}$. Along the circumference of the tube, the electron and phonon momenta acquire discrete values, while the momenta remain approximately continuous along the tube axis for a long CNT. The possible momenta can be expressed as $\boldsymbol{k} = s {\boldsymbol{k}}_{\perp} + k \hat{\boldsymbol{k}}_{\parallel}$, where $s = 0, 1, \hdots, N$ and $k \in [-\pi/|\boldsymbol{T}|, \pi/|\boldsymbol{T}|]$, with momentum vectors $\boldsymbol{k}_{\perp} = (-t_2 \boldsymbol{b}_1 + t_1 \boldsymbol{b}_2)$ and $\boldsymbol{k}_{\parallel} = (m \boldsymbol{b}_1 - n \boldsymbol{b}_2) / N$, and unit vectors $\boldsymbol{\hat{k}}_{\perp} = \boldsymbol{k}_{\perp} / |\boldsymbol{k}_{\perp}|$ and $\boldsymbol{\hat{k}}_{\parallel} = \boldsymbol{k}_{\parallel} / |\boldsymbol{k}_{\parallel}|$. We use $\boldsymbol{b}_1$ and $\boldsymbol{b}_2$ to denote the reciprocal lattice vectors of the graphene layer ($\boldsymbol{a}_i \cdot \boldsymbol{b}_j = 2\pi \delta_{ij}$). The family of possible momenta for each value of $s$ represents a linear path, or `cut,' along the Brillouin zone of the monolayer graphene. Now, the Hamiltonian for the CNT is of a block-diagonal form, with $N$ blocks each corresponding to the Hamiltonian $\hat{H}_{s}(\boldsymbol{k}) = \hat{H}_{\text{g}}(s {\boldsymbol{k}}_{\perp} + k \hat{\boldsymbol{k}}_{\parallel})$ along a cut $s = 0, \hdots N$, where the single-particle tight-binding Hamiltonian for a monolayer graphene sheet with nearest-neighbor hopping is given by
\begin{equation}
    \hat{H}_{\text{g}} = \sum_{\boldsymbol{k}} \boldsymbol{\hat{\psi}}_{\boldsymbol{k}}^{\dagger} \begin{pmatrix} 0 & h^*(\boldsymbol{k}) \\ h(\boldsymbol{k}) & 0 \end{pmatrix} \boldsymbol{\hat{\psi}}_{\boldsymbol{k}}.
\end{equation}
Here, $\boldsymbol{\hat{\psi}}_{\boldsymbol{k}} \equiv \begin{pmatrix} \hat{\psi}_{\boldsymbol{k}, A} & \hat{\psi}_{\boldsymbol{k}, B} \end{pmatrix}$ where $\hat{\psi}_{\boldsymbol{k}, x}^\dagger$ creates a fermion of crystal momentum $\boldsymbol{k}$ on sublattice $x = A,B$ of the graphene sheet, $h(\boldsymbol{k}) = h \sum_j e^{i\boldsymbol{k} \cdot \boldsymbol{\delta}_j}$, and $h = 2.8 \ \mathrm{eV}$. 

The lowest-energy conduction and highest-energy valence bands are indexed by $s_{\text{m}} = \text{argmin}_{s} \min_k | s {\boldsymbol{k}}_{\perp} + k \hat{\boldsymbol{k}}_{\parallel} - \boldsymbol{K}|$, where $\boldsymbol{K}$ is the momentum of the Dirac $K$ point in the monolayer graphene Brillouin zone. The lowest energy electronic state along the cut $s_m$ has momentum $k_m = \text{argmin}_{k} | s_m {\boldsymbol{k}}_{\perp} + k \hat{\boldsymbol{k}}_{\parallel} - \boldsymbol{K}|$. Near the momentum $k_m$, the Hamiltonian can be approximated by the Dirac Hamiltonian shown in Eq. (1) of the main text, where $\delta k = | s_m {\boldsymbol{k}}_{\perp} + k_m \hat{\boldsymbol{k}}_{\parallel} - \boldsymbol{K}|$.

\section{Microscopic Hamiltonian for Electron-Phonon Interactions}
In this section, we provide the full expression for the microscopic Hamiltonian $\hat{H}_{\text{b}}(t)$ accounting for electronic coupling to incoherent bath phonon modes. The low-energy longitudinal acoustic phonons dominate the electron-phonon scattering near the Fermi surface of the lightly-doped CNT. Upon writing the displacement operators $\hat{\boldsymbol{u}}^{(\lambda)}(\boldsymbol{r},t)$ for such phonon modes in terms of bath acoustic phonon creation operators $\hat{b}^\dagger_{p}$, we derive the effective Hamiltonian \cite{PhysRevB.57.4145}: 
\begin{equation} \label{eq:h-el-ph}
    \hat{H}_{\text{b}} = \int \frac{dkdq}{(2\pi)^2} {M}_{k,p}  \hat{c}_{{k}+p,+}^{\dagger} \hat{c}_{k,+}(\hat{b}^{\dagger}_{p} + \hat{b}_{-p}) + \text{h.c.} 
\end{equation}
where ${M}_{k, p} = {D \sqrt{\hbar c_{\text{ph}} {p}}}/({\sqrt{2A\rho}c_{\text{ph}}}) \mathcal{W}_{k,p}$ with unit cell area $A = \sqrt{3} N a^2/2$, and graphene density $\rho$.

\section{Floquet Boltzmann Equation} \label{sec:sup-fbe}
In this section, we present the full expressions for the electron-phonon  and electron-electron collision integrals, discretized on a 1D momentum grid of $N$ points. The electron-phonon collision integral is given by
\begin{widetext}
\begin{align} \label{eq:el-ph-i}
\begin{split}
    I^{\text{b}}_{{k}\alpha}&[\{ F_{{k}\alpha} \}] =  \frac{2\pi}{\hbar} \frac{1}{N} \sum_{{{k}'\in \mathrm{BZ}}} \sum_{\alpha'} \sum_{s} \sum_n |\mathcal{G}_{{k}\alpha}^{{k}'\alpha'}(n)|^2 \frac{1}{\hbar c_{\text{ph}}} \times \\
    & \times  (\{-\mathcal{N}(\varepsilon_{{k}'\alpha'} - \varepsilon_{{k}\alpha}) F_{{k}\alpha} (1-F_{{k}'\alpha'}) \delta(k'-k + n q +k_s) \\
    & \hspace{2cm} +[1+\mathcal{N}(\varepsilon_{{k}'\alpha'} - \varepsilon_{{k}\alpha})](1-F_{{k}\alpha}) F_{{k}'\alpha'} \delta(k'-k + n q -k_s) \} \theta(\varepsilon_{{k}'\alpha'} - \varepsilon_{{k}\alpha}) \\
    &\hspace{1cm} + \{-[1+\mathcal{N}(\varepsilon_{{k}'\alpha'} - \varepsilon_{{k}\alpha})] F_{{k}\alpha} (1-F_{{k}'\alpha'}) \delta(k'-k + n q -k_s) \\
    & \hspace{2cm} +\mathcal{N}(\varepsilon_{{k}'\alpha'} - \varepsilon_{{k}\alpha})(1-F_{{k}\alpha}) F_{{k}'\alpha'} \delta(k'-k + n q +k_s) \} \theta(\varepsilon_{{k}'\alpha'} - \varepsilon_{{k}\alpha}) \\
\end{split}
\end{align}
\begin{equation} \label{eq:el-ph-m}
    \Ch{\mathcal{G}_{{k}\alpha}^{{k}'\alpha'}(n) =  \frac{1}{\sqrt{A}} \frac{D \hbar k_s}{\sqrt{2\rho \hbar|\varepsilon_{{k}'\alpha'} - \varepsilon_{{k}\alpha}|}}   \sum_m \langle \phi^{n+m}_{{k}'\alpha'} | +, {k}' \rangle \mathcal{W}_{{k},k'-k}\langle +, {k} | \phi^m_{{k}\alpha} \rangle}
\end{equation}
\end{widetext}
where $\rho = \Ch{1.52} \times 10^{-6} \ \mathrm{kg/m^2}$ is the 2D density of the graphene layers, $D$ is the deformation potential, $k_s$ satisfies $\hbar c_{\text{ph}} |s \boldsymbol{k}_{\perp} + k_s \boldsymbol{\hat{k}}_{\parallel}| = \varepsilon_{{k}'\alpha'} - \varepsilon_{{k}\alpha}$, and $\mathcal{N}(\varepsilon) = 1/(e^{\varepsilon/k_B T} - 1)$ is the Bose-Einstein distribution for incoherent phonons maintained in thermal equilibrium at temperature $T$. The electron-electron collision integral is given by
\begin{widetext}
\begin{align} \label{eq:el-el-i}
\begin{split}
    I^{\text{ee}}_{{k}\alpha}[\{ F_{{k}\alpha} \}] = \frac{4\pi}{\hbar} &\frac{1}{N^2} \sum_{{k}_2 \in \mathrm{BZ}} \sum_{{k}_3 \in \mathrm{BZ}} \sum_{\alpha_2, \alpha_3, \alpha_4} \sum_n  | \mathcal{V}_{({k},\alpha), ({k}_2,\alpha_2)}^{({k}_3,\alpha_3),({k}_1+{k}_2-{k}_3,\alpha_4)}(n) |^2 \times \\
    &\times \delta(\varepsilon_{{k}\alpha} +  \varepsilon_{{k}_2\alpha_2} - \varepsilon_{{k}_3\alpha_3} - \varepsilon_{{k} + {k}_2-{k}_3,\alpha_4} + n\hbar \Omega)  \times\\
    & \times\left[ (1-F_{{k}\alpha})(1-F_{{k}_2\alpha_2}) F_{{k}_3 \alpha_3} F_{{k}_1+{k}_2-{k}_3,\alpha_4} - F_{{k}\alpha}F_{{k}_2\alpha_2} (1-F_{{k}_3 \alpha_3}) (1-F_{{k}_1+{k}_2-{k}_3,\alpha_4} )\right]
\end{split}
\end{align}
\begin{align}
\begin{split}
    \mathcal{V}_{({k},\alpha), ({k}_2,\alpha_2)}^{({k}_3,\alpha_3),({k}_1+{k}_2-{k}_3,\alpha_4)}(n) = \sum_{n_2,n_3,n_4} & V({{k}_2 -{k}_3}) \Ch{\mathcal{W}_{{k}_1,k_3 - k_2} \mathcal{W}_{{k}_2,-(k_3 - k_2)}} \langle \phi^{n-n_2 + n_3 + n_4}_{{k} \alpha} | +, {k} \rangle  \langle \phi^{n_2}_{{k}_2 \alpha_2} | +, {k}_2 \rangle \times \\ 
    & \times \langle +, {k}_3 | \phi^{n_3}_{{k}_3 \alpha_3} \rangle \langle +, {k}_4 | \phi^{n_4}_{{k} + {k}_2 - {k}_3, \alpha_4} \rangle.
\end{split}
\end{align}
\end{widetext}
To solve for the steady-state, we use the Newton-Raphson algorithm to find the roots $\partial_t F_{{k}\alpha} = 0$ of the FBE. We set the doping of the system by adding the Lagrange multiplier term $\lambda (\sum_{{k}\alpha} F_{{k}\alpha} -N k_F / q)$ with large constant $\lambda$ to the FBE.

\begin{figure}
    \centering    \includegraphics[width=0.98\linewidth]{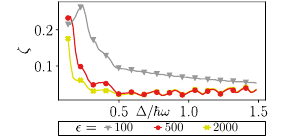}
    \caption{The ratio $\zeta$ of the maximum interband scattering rate to the Floquet band energy separation vs. Floquet gap $\Delta$ for various dielectric constants $\epsilon$. Note that $\zeta \ll 1$, so steady state coherences are suppressed. }
    \label{fig:sfig2}
\end{figure}

\section{Validity of Floquet Boltzmann Equation and Steady State Scattering Times}
The interband scattering rates $1/\tau^{\text{ph}}_{k\alpha\alpha'}$ and $1/\tau^{\text{el}}_{k\alpha\alpha'}$ are given by
\begin{widetext}
\begin{align} \label{eq:el-ph-scat-time}
\begin{split}
    \frac{1}{\tau^{\text{ph}}_{{k}\alpha\alpha'}}&=  \frac{2\pi}{\hbar} \frac{1}{N} \sum_{{{k}'\in \mathrm{BZ}}} \sum_{s} \sum_n |\mathcal{G}_{{k}\alpha}^{{k}'\alpha'}(n)|^2 \frac{1}{\hbar c_{\text{ph}}} \times \\
    & \times  +\mathcal{N}(\varepsilon_{{k}'\alpha'} - \varepsilon_{{k}\alpha}) (1-F_{{k}'\alpha'}) \delta(k'-k + n q +k_s) +[1+\mathcal{N}(\varepsilon_{{k}'\alpha'} - \varepsilon_{{k}\alpha})] (1-F_{{k}'\alpha'}) \delta(k'-k + n q -k_s)  \\
\end{split}
\end{align}
and
\begin{align} \label{eq:el-el-scat-time}
\begin{split}
     \frac{1}{\tau^{\text{el}}_{{k}\alpha\alpha'}} = \frac{4\pi}{\hbar} &\frac{1}{N^2} \sum_{{k}_2 \in \mathrm{BZ}} \sum_{{k}_3 \in \mathrm{BZ}} \sum_{\substack{\alpha_2, \alpha_3, \alpha_4 \\ \alpha_3 = \alpha' \text{ or } \alpha_4 = \alpha'}} \sum_n \sum_{{G}} | \mathcal{V}_{({k},\alpha), ({k}_2,\alpha_2)}^{({k}_3,\alpha_3),({k}_1+{k}_2-{k}_3,\alpha_4)}(n,{G}) |^2 \times \\
    &\times \delta(\varepsilon_{{k}\alpha} +  \varepsilon_{{k}_2\alpha_2} - \varepsilon_{{k}_3\alpha_3} - \varepsilon_{{k} + {k}_2-{k}_3,\alpha_4} + n\hbar \Omega) F_{{k}_2\alpha_2} (1-F_{{k}_3 \alpha_3}) (1-F_{{k}_1+{k}_2-{k}_3,\alpha_4} ).
\end{split}
\end{align}
\end{widetext}
To verify that the steady state coherences are suppressed and that the Floquet Boltzmann equation is valid, we check that $\zeta \ll 1$, where $\zeta \equiv \max_{k,\alpha,\alpha'} \hbar / (\tau^{\text{tot}}_{k\alpha\alpha'} |\varepsilon_{k\alpha} - \varepsilon_{k\alpha'}|)$, where $1/ \tau^{\text{tot}}_{k\alpha\alpha'} \equiv 1/ \tau^{\text{ph}}_{k\alpha\alpha'} + 1/ \tau^{\text{el}}_{k\alpha\alpha'}$ [see full definition in the main text]. Fig. \ref{fig:sfig2} verifies that $\zeta \ll 1$ as a function of drive amplitude.

\bibliographystyle{apsrev4-1}
\bibliography{references_use}